\documentclass[]{rsos}

\begin{document}

\title{Post-main-sequence planetary system evolution}

\author{Dimitri Veras$^{1}$}

\address{$^{1}$Department of Physics,
University of Warwick,
Coventry CV4 7AL,
UK}

\subject{Extrasolar planets, Astrophysics, Solar system}

\keywords{Dynamics, White dwarfs, Giant branch stars, Pulsars, Asteroids, Formation}

\corres{Dimitri Veras\\
\email{d.veras@warwick.ac.uk}}

\begin{abstract}
The fates of planetary systems provide unassailable insights
into their formation and represent rich cross-disciplinary
dynamical laboratories.  Mounting observations of post-main-sequence
planetary systems necessitate a complementary level of 
theoretical scrutiny.  Here, I review the diverse dynamical 
processes which affect planets, asteroids, comets and pebbles
as their parent stars evolve into giant branch, white dwarf
and neutron stars.
This reference provides a foundation for the interpretation and 
modelling of currently known systems and upcoming discoveries.
\end{abstract}


\begin{fmtext}

\section{Introduction} \label{secIntro}

Decades of unsuccessful attempts to find
planets around other Sun-like stars
preceded the unexpected 1992 discovery 
of planetary bodies orbiting a pulsar 
\cite{wolfra1992,wolszczan1994}.  The three planets 
around the millisecond pulsar PSR B1257+12 were the first 
confidently reported extrasolar planets to withstand enduring scrutiny
due to their well-constrained masses and orbits.
However, a retrospective historical analysis reveals even more surprises.
We now know that the eponymous celestial body that Adriaan van Maanen 
observed in the late 1910s \cite{vanmaanen1917,vanmaanen1919} is an isolated 
white dwarf (WD) with a metal-enriched atmosphere: direct evidence
for the accretion of planetary remnants.

These pioneering discoveries of planetary material around or in 
post-main-sequence
(post-MS) stars, although exciting, represented a poor harbinger for how the
field of exoplanetary science has since matured.  The first viable hints of
exoplanets found around MS stars ($\gamma$ Cephei Ab and HD 114762 b) 
\cite{cametal1988,latetal1989}, in 1987-1989, 
were not promulgated as such due to uncertainties in the interpretation of 
the observations and the inability to place upper bounds on the companion masses. 
A confident detection of an MS exoplanet emerged with 
the 1995 discovery of
51 Pegasi b \cite{mayque1995}, followed quickly by 70 Virginis b \cite{marbut1996}
and 47 Ursae Majoris b \cite{butmar1996}, although 
\end{fmtext} \maketitle 
\begin{figure}[!h]
\centering\includegraphics[width=5.5in]{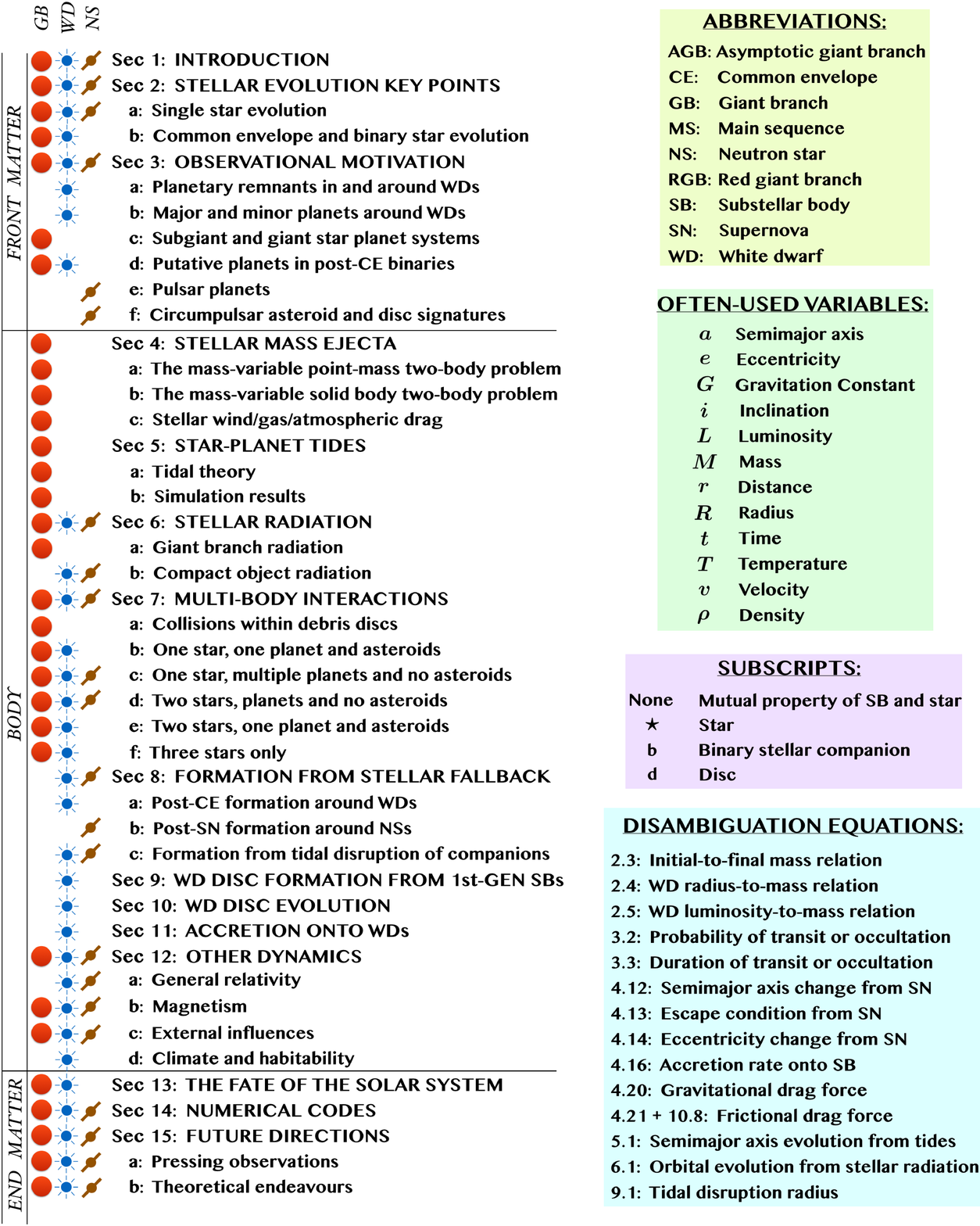}
\caption{Paper outline and nomenclature.  
Some section titles are abbreviated
to save space. Variables 
not listed here are described in situ, and usually
contain descriptive subscripts and/or superscripts.
The important abbreviation ``substellar body'' (SB) 
can refer to, for example, a brown dwarf, planet, moon, 
asteroid, comet or pebble.  ``Disambiguation equations'' refer to relations
that have appeared in multiple different forms in the literature.  
In this paper, these other forms are referenced in the
text that surrounds these 
equations, so that readers can decide which form is best 
to use (or newly derive) for their purposes. Overdots 
always refer to time derivatives. The expression $<>$ refers to 
averaged quantities.
}
\label{FigTOC}
\end{figure}
\begin{figure}[!h]
\centering\includegraphics[width=5.4in]{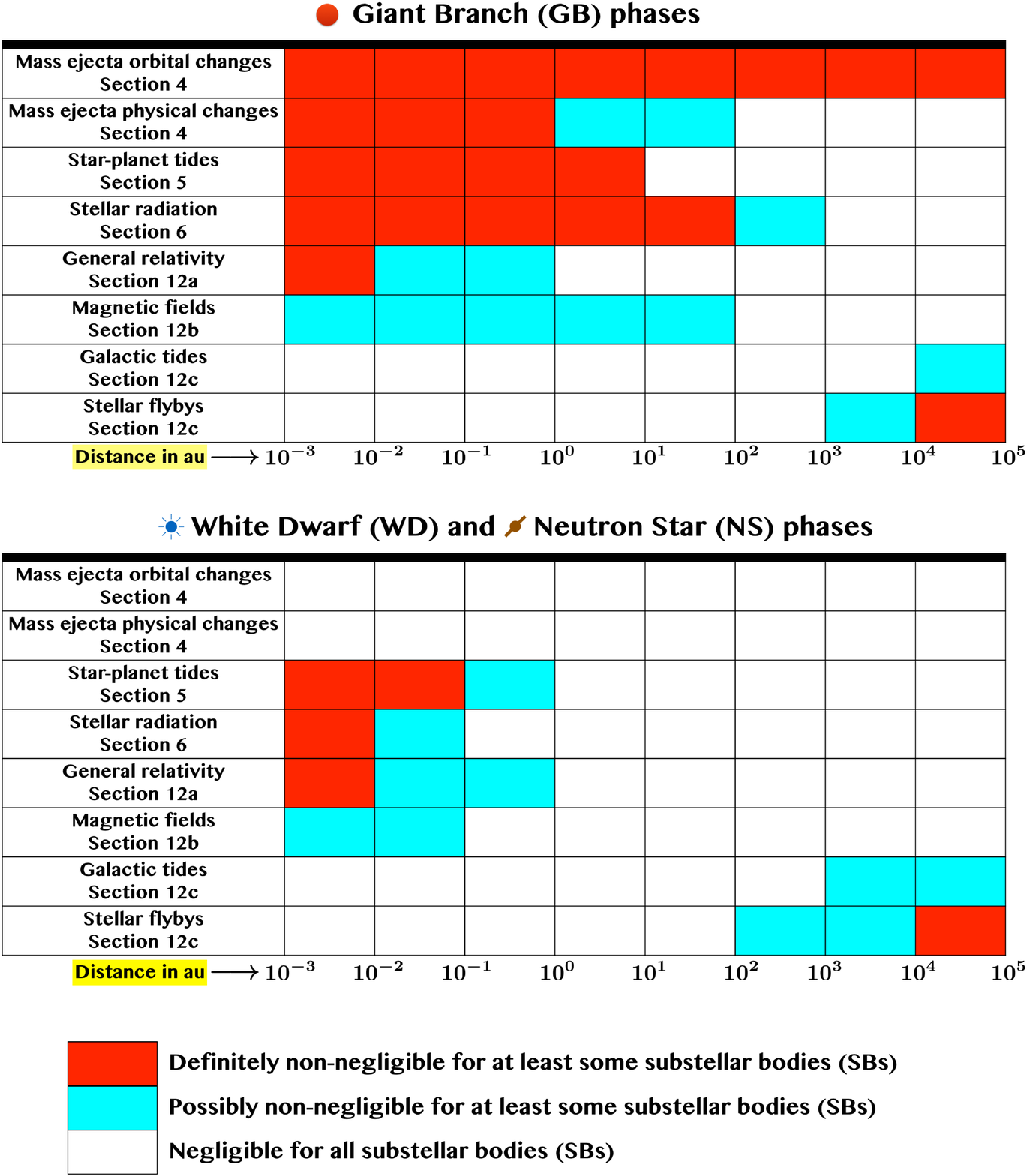}
\caption{Important forces in post-MS systems.  These charts
represent just a first point of reference.  Every system should be
treated on a case-by-case basis.  Magnetic fields include those of
both the star and the SB, and external effects are less penetrative
in the GB phases because they are relatively short.}
\label{FigForces}
\end{figure}
\noindent{}in all cases the mass
degeneracy remained.  These planets ushered in
a burgeoning and flourishing era of astrophysics.
Now, two decades later, our planet inventory numbers in 
the thousands; over 90\% of all known 
exoplanets orbit MS stars that will eventually become WDs,
and WDs will eventually become the most common stars in the Milky Way.

Nevertheless, major uncertainties linger.  MS exoplanet 
detection techniques currently provide minimal inferences about the bulk chemical
composition of exoplanetary material.  How planets form and dynamically
settle into their observed states remains unanswered and represents 
a vigorous area of active research.  Calls for a better understanding
of post-MS evolution arise from MS discoveries of planets near the end
of their lives \cite{liletal2014} and a desire to inform planet formation
models \cite{caretal2012}.  Direct observation of MS smaller
bodies, such as exo-asteroids, exo-comets or exo-moons, remains tantalizingly
out-of-reach, except in a handful of cases 
\cite{lagetal1990,meletal2001,welmon2013,kieetal2014a,kieetal2014b}.

\begin{table}[!h]
\caption{Some notable post-main-sequence planetary systems. $^1$Potentially polluted with lithium. $^2$First WD debris disc. $^3$Polluted with 17 different metals. $^4$Planet orbits the MS star. $^5$Multiple circumbinary planets. $^6$First confirmed exoplanetary system. $^7$First confirmed circumbinary planet. $^8$Disc probably eccentric and axisymmetric. $^9$Planet at several thousand au. $^{10}$Only WD with transiting SBs, a disc and pollution. $^{11}$Highly-variable WD disc. $^{12}$First polluted WD.}
\label{PopularSys}
\begin{tabular}{ccccc}
\hline
Name & Type & See Sections & Notes \\
\hline
BD+48 740       & GB star with possible pollution& \ref{secOM}\ref{secOMgiant}\ref{secOMgiantGB} & 1 \\ 
G 29-38          & WD with disc and pollution     & \ref{secOM}\ref{secOMWDpoll}\ref{WDDebrisDiscs}  & 2  \\
GD 362          & WD with disc and pollution     & \ref{secOM}\ref{secOMWDpoll}\ref{WDDebrisDiscs}   & 3  \\
GJ 86           & Binary WD-MS with planet       & \ref{secOM}\ref{secOMWDplan}\ref{secOribitingComp}   & 4  \\
NN Ser          & Binary WD-MS with planets       & \ref{secOM}\ref{secOMPCEB}, \ref{secMulti}\ref{secMulti2sNp0a}\ref{secMulti2sNp0aWOS}, \ref{secMulti}\ref{secMulti2sNp0a}\ref{secMulti2sNp0aWS}, \ref{secFall}\ref{secFallCE}, \ref{secIdeas}\ref{secIdeasPres}\ref{secIdeasPresMon}    & 5  \\
PSR B1257+12    & Pulsar with planets            & \ref{secIntro}, \ref{secOM}\ref{secOMpulplan}, \ref{secMulti}\ref{secMulti1sNp0a}\ref{secMulti1sNp0aPul}, \ref{secFall}\ref{secFallSN}, \ref{secFall}\ref{secFallTides}   & 6  \\
PSR B1620-26    & Binary pulsar-WD with planet   & \ref{secOM}\ref{secOMWDplan}\ref{secOribitingWDs}, \ref{secMulti}\ref{secMulti2sNp0a}\ref{secMulti2sNp0aWOS}   & 7  \\
SDSS J1228+1040 & WD with disc and pollution     & \ref{secOM}\ref{secOMWDpoll}\ref{WDDebrisDiscs}, \ref{secWDevol}, \ref{secIdeas}\ref{secIdeasPres}\ref{secIdeasPresNew}   & 8  \\
WD 0806-661     & WD with planet                 & \ref{secOM}\ref{secOMWDplan}\ref{secOribitingWDs}   & 9  \\
WD 1145+017     & WD with asteroids, disc, and pollution   &  \ref{secOM}\ref{secOMWDplan}\ref{secOribitingWDs}, \ref{secIdeas}\ref{secIdeasPres}\ref{secIdeasPresMon}   & 10  \\
WD J0959-0200   & WD with disc and pollution     & \ref{secOM}\ref{secOMWDpoll}\ref{WDDebrisDiscs}, \ref{secWDevol}, \ref{secIdeas}\ref{secIdeasPres}\ref{secIdeasPresMon}   & 11  \\
vMa2            & WD with pollution              & \ref{secIntro}, \ref{secOM}\ref{secOMWDpoll}\ref{WDpollInd}   & 12  \\
\hline
\end{tabular}
\end{table}

Post-MS planetary system investigations help alleviate
these uncertainties, particularly with escalating
observations of exoplanetary remnants in WD systems.
Unlike for pulsar systems, planetary signatures 
are common in and around WD stars. The exquisite chemical 
constraints on rocky planetesimals
that are gleaned from WD atmospheric abundance studies 
is covered in detail by the review of \cite{juryou2014}, and is not a
focus of this article.  Similarly, I do not focus on the revealing
observational aspects of the nearly forty debris discs orbiting WDs, 
a topic recently reviewed by \cite{farihi2016}.

Instead, I place into context and describe the complex and varied 
dynamical processes that influence planetary bodies
after the star has turned off of the MS.  I attempt to touch upon all
theoretical aspects of post-MS planetary science, although my focus is on 
the giant branch (GB) and WD phases of stellar evolution.  The vital
inclusion of bodies smaller than planets -- e.g. exo-asteroids and 
exo-comets -- in this review highlights 
both the necessity of incorporating Solar system constraints and models
and the interdisciplinary nature of post-MS planetary science.

\subsection{Article layout}

I begin by providing a visual table of contents in Fig. \ref{FigTOC},
which includes handy references for the abbreviations
and commonly-used variables in this article. 
I use the abbreviation ``SB'' (``substellar body'' or ``smaller body'') 
extensively in the text and equations; constraining relations to just
one of planets,
asteroids, comets or pebbles is too restrictive for 
the strikingly diverse field of post-MS planetary science.
The term also includes brown dwarfs, for which many physical
relations presented here also apply.  ``Planetary systems'' is defined 
as systems which include SBs.  The ``disambiguation'' equations
identified in Fig. \ref{FigTOC} refer to relations that have appeared in multiple
different forms in the previous post-MS planetary literature; I attempt to consolidate
these references.  In Fig. \ref{FigForces} I characterise distances from the star
in which various forces are important, or might be important.  This figure
may be used as a guide when modelling a particular system or set of
systems.  Table \ref{PopularSys} lists some notable post-MS planetary 
systems, along with brief descriptions and pointers to where they are mentioned
in the text.

My deliberately basic treatment of introductory material
(stellar evolution and observations from Secs. \ref{secSE}-\ref{secOM})
is intended to provide the necessary background for subsequent
sections, and not meant to emulate an in-depth synopsis.
The body of the article (Secs. \ref{secMass}-\ref{secOther}) provides more detail
on the dynamical aspects of post-MS planetary science.
This review concludes with brief comments on the fate of the
Solar system (Sec. \ref{secFate}), a hopefully helpful summary of 
the numerical codes that have or may be used in theoretical 
investigations (Sec. \ref{secCodes}) and a promising outlook on the future of this
science (Sec. \ref{secIdeas}), with guidance for how upcoming observations can 
maximize scientific return.

\section{Stellar evolution key points} \label{secSE}

The infrangible link between SBs and their host star
is highlighted during post-MS evolution, and requires one to understand
the star's temporal evolution. My treatment below is purposefully 
simplified to provide the necessary information
for post-MS planetary system studies; for more detail, see e.g. 
\cite{iben2013a,iben2013b}.

\subsection{Single star evolution} \label{secSEsingle}

\subsubsection{Main sequence}

MS evolution is important because it provides the historical context and
initial conditions for dedicated post-MS studies.  MS stars quiescently 
burn hydrogen to produce helium in their cores, and do lose mass through winds 
according to Eq. 4 of \cite{wooetal2002} and Eq. 9
of \cite{zenetal2010}.  The Sun currently loses mass at a rate of about 
$2.4 \times 10^{-14} M_{\odot}$/yr (pg. 15 of \cite{vial2013}).  The MS
lifetime is sensitively dependent on the initial value of $M_{\star}^{({\rm MS})}$
and less so on the star's metallicity $Z_{\star}$. This lifetime decreases drastically
(by two orders of magnitude, from about 10 Gyr to 0.1 Gyr) as the initial 
mass increases from $1 M_{\odot}$ to $6M_{\odot}$ (see Fig. 5 of \cite{veretal2013}).

\subsubsection{Giant branches}

All stars experience the ``red giant branch'' (RGB) phase, when 
hydrogen in the core is exhausted and the remaining hydrogen burns in a contracting 
shell as the envelope expands.  The extent of convection in the star increases, potentially ``dredging-up'' already-burnt matter.
Eventually core temperatures become high enough to burn helium.  For stars with 
$M_{\rm star}^{({\rm MS})} < 2.0 M_{\odot}$, helium ignition sets off so-called ``helium flashes''.
This value of $2.0 M_{\odot}$ represents a key transition mass; the duration and charactervof the mass loss changes markedly when crossing this threshold.
After the core helium is exhausted, a helium-burning shell is formed.
At this point, the star is said to have begun evolving on the ``asymptotic giant branch'' (AGB).
Another expansion of convection may then cause a ``second dredge-up''.
When, during the AGB, the helium-burning shell reaches the hydrogen outer envelope, 
a different type 
of helium flash occurs (denoted a ``thermal pulse''), one which emits a sudden burst 
of luminosity and mass.  
This event, which can occur many times, also causes a sudden increase and 
then drop in stellar radius
(see Fig. 3 of \cite{wicetal2010}).  
Therefore, AGB thermal pulses literally cause the star to pulsate.  Changes in
the star's convective properties during this violent time may also allow for a ``third dredge-up'' 
to then occur.

\begin{figure}[!h]
\centering\includegraphics[width=3.0in]{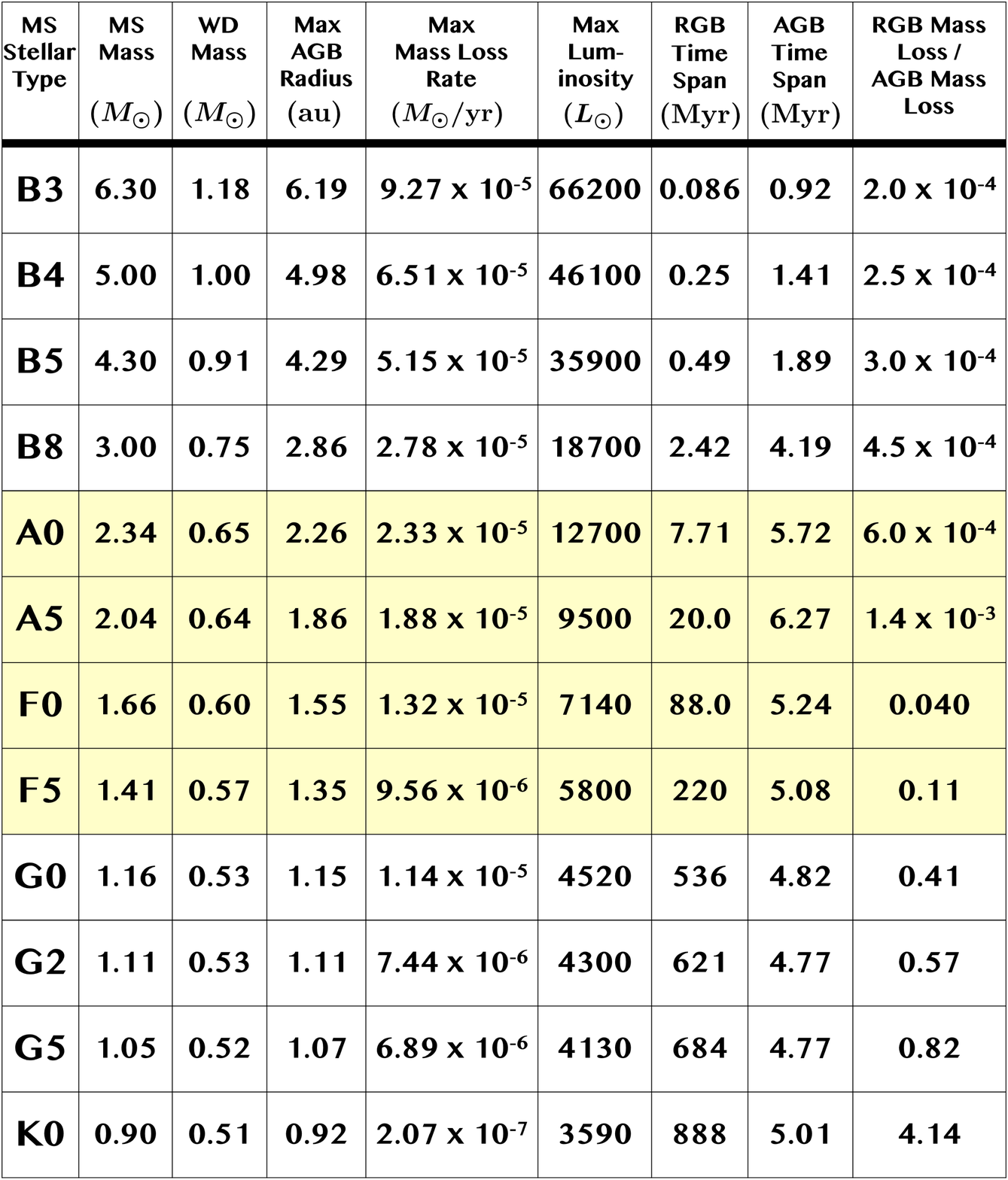}
\caption{Useful values for twelve different stellar evolution tracks.
I mapped the first column to the second by using
Appendix B of \cite{gray2008}, and then created the remaining columns by
using the \textsc{SSE} code \cite{huretal2000} by assuming its default values
(which includes Solar metallicity). The four highlighted rows roughly represent the
range of the most common progenitor stars for the present-day WD population in the
Milky Way.}
\label{FigSpec}
\end{figure}

During both the RGB and AGB phases, the star undergoes significant 
mass loss (up to 80\%), radius variability (up to about 10 au, from an initial value 
of $10^{-3}-10^{-2}$ au), and luminosity variability (up to many tens of thousand 
times the MS value) regardless of the extent of the pulses.  Fig. \ref{FigSpec} 
provides representative values;
the highlighted rows indicate the most typical progenitors for the currently observed
Milky Way WD population.  These changes along the GB phases may completely
transform a planetary system; indeed linking WD and MS planetary systems is a goal and a
challenge, and may also help constrain stellar evolution. Unfortunately,
identifying the dominant mechanisms responsible for mass loss -- both isotropic
and anisotropic -- on the RGB and AGB continues to prove difficult.

{\bf RGB MASS LOSS}
\ \ \ \
On the RGB, mass-loss is traditionally parameterised by the Reimers formula, a series of
proportionalities that was later calibrated by \cite{kudrei1978} and recently improved upon
by \cite{schcun2005} to finally give

\begin{eqnarray}
\frac{dM_{\star}^{({\rm RGB})}}{dt}
&=&
8 \times 10^{-14} \frac{M_{\odot}}{{\rm yr}} \ 
\left( \frac{L_{\star}^{({\rm RGB})}}{L_{\odot}} \right)
\left( \frac{R_{\star}^{({\rm RGB})}}{R_{\odot}} \right)
\left( \frac{M_{\star}^{({\rm RGB})}}{M_{\odot}} \right)^{-1}
\nonumber
\\
&\times&
\left( \frac{T_{\star}^{({\rm RGB})}}{4000 \ {\rm K}} \right)^{\frac{7}{2}}
\left[1 + 2.3 \times 10^{-4} \left( \frac{g_{\star}^{({\rm RGB})}}{g_{\odot}} \right)^{-1}  \right]
.
\label{eqtruere}
\end{eqnarray}

\noindent{}where $g$ refers to surface gravity.  Traditional formulations of Eq. \ref{eqtruere}, 
which are still widely used, do not include the final two terms, and have a leading coefficient
of $2 \times 10^{-13} M_{\odot}$/yr.

{\bf AGB MASS LOSS}
\ \ \ \
Applying the Reimers formula on the AGB can produce significantly erroneous results 
(Fig 13 of \cite{musvil2012}).  Instead, during this phase a different prescription  
is often applied, whose formulation \cite{vaswoo1993} has stood the test of ongoing
observations and can be found, e.g. in Eqs. 2-3 of \cite{bonwya2010}.  
Accompanying each AGB pulse is a variation in mass loss of potentially a few orders
of magnitude, a phenomenon now claimed to have been observed \cite{maeetal2012}. 
At the final stage of the AGB -- ``the tip'' of the AGB -- the wind is particularly powerful
and is known as the ``superwind'' (e.g. \cite{lagzij2008}).  A star's peak mass loss rate 
typically occurs during the superwind unless the AGB phase is nonexistent or negligible.

{\bf GB MASS EJECTA SPEED}
\ \ \ \
The speed at which mass is ejected is generally a function of the internal properties
of the star and the location of ejection.  One simplified estimate that may be useful 
for post-MS studies is \cite{owoetal1998}

\begin{equation}
v_{\rm wind} = 
\sqrt{ 
\left( 
\frac{2GM_{\star}}{R_{\star}}
\right)
\left( 
1
-
\frac{{\left[v_{\star}^{({\rm rot})}\right]}^2 R_{\star}}{G M_{\star}}
\sin^2{\theta}
\right)
},
\end{equation}

\noindent{}where $\theta$ is the stellar co-latitude and $v_{\star}^{({\rm rot})}$ is the 
stellar rotational speed at the equator.

{\bf GB CHANGES FROM SB INGESTION}
\ \ \ \
If a large SB such as a brown dwarf of planet is ingested during the GB phases,
two significant events might result: a enhancement of lithium in the photosphere
and spin-up of the star.  The former was predicted in 1967 by 
\cite{alexander1967}. \cite{adaetal2012} claimed that SB accretion onto 
stars can increase their Li surface abundance for a few Myr.
However, an enhanced abundance of Li in GB stars could also indicate
dredge-up by the Cameron-Fowler mechanism, mixing through tides, or thermohaline
or magneto-thermohaline processes.  Therefore, a planetary origin interpretation
for Li overabundance remains degenerate.

Several investigations have considered how a GB star spins up due to SB accretion.
In his Eq. 1, \cite{massarotti2008} computed the change in the star's rotational speed.
He suggested that a population of GB fast-rotators due to planet ingestion would be
detectable if the speed increased by at least 2 km/s.  \cite{caretal2009} found 
that a few percent of the known population of exoplanets (at the time) could 
create rapid rotators, where rapid is defined as having a rotational speed 
larger than about $10$ km/s. 

SB ingestion may cause other changes, such as enhanced mass loss 
\cite{sieliv1999,beaetal2011} and displacement on the Hertzsprung-Russel 
diagram \cite{soker1998}. The presence or ingestion of a SB 
could be a reason \cite{beasok2011}
why some GB stars prematurely lose their entire envelope before
core fusion of helium begins.  These stars are known as
``extreme horizontal branch stars'', which are also known as ``hot subdwarfs''
or ``sdB'' stars (see \cite{heber2009} for a review specifically of these
types of stars).  This SB ingestion explanation is particularly
relevant for hot subdwarfs with no known stellar binary companions.
When modelling AGB envelopes for this or other purposes,
one may use a power-law density profile (see, e.g. Sec. 2 \cite{harsok1994});  
\cite{wilwu2005} instead gave a more complex form in their Eq. 5.

\subsubsection{White dwarfs}

For stars with $M_{\star}^{({\rm MS})} \lesssim 8M_{\odot}$, after the GB 
envelope is completely blown away the remaining core becomes a WD 
\cite{huretal2000,altetal2010,koester2013}.  In the Milky Way, about 
about 95\% of all stars will become WDs \cite{altetal2010}. The term 
``white'' in WD originates from the notion that the majority of 
known WDs are hotter than the Sun \cite{altetal2010}.  
The expelled material photoionizes and
the resulting observed structure, which might not have any relation to planets whatsoever,
is confusingly termed a ``planetary nebula''.  Although the expelled material will encounter
remnant planets and asteroids, few investigations so far have tried to link these nebulae with planets.
Even the link between nebula morphology and stellar configurations remains uncertain \cite{demsok2011}
although SBs which are at least as massive as planets, as well as stellar-mass companions, 
are thought to play a significant role in shaping and driving the nebulae 
(e.g. \cite{soker1996,norbla2006}).  

The time elapsed since the moment an AGB star becomes a WD is denoted the ``cooling age''
because the WD is in a state of monotonic cooling (as nuclear burning has now stopped). 
The term cooling age allows one to distinguish from the total age of the star, which
includes its previous evolutionary phases. Although some investigations refer to 
planetary nebula or ``post-AGB'' as the name of a separate stellar evolutionary phase 
\cite{villiv2007,soker2008},
I do not, and assume that the transition from AGB to WD contains no other evolutionary phase.

{\bf WD DESIGNATIONS}
\ \ \ \
WDs have and continue to be characterised observationally by the dominant spectral 
absorption lines in their atmospheres.  These designations \cite{sioetal1983} 
include ``D'', which stands for 
degenerate, ``A'', for hydrogen rich, ``B'' for helium rich, ``Z'' for metal-rich
(metals are elements heavier than helium), and ``H'' for magnetic.  About 80-85\% of 
the WD population are DA WDs \cite{altetal2010,koester2013}.  Non-DA WDs probably 
lost their hydrogen in a relatively late-occurring shell flash.

{\bf WD COMPOSITION}
\ \ \ \
The composition of the WD cores are some combination of carbon (from the burning of helium), 
oxygen (from the burning of carbon) and rarely neon (from the burning of oxygen).  
The vast majority of WD cores contain carbon and oxygen because they are not hot enough
to host copious quantities of oxygen and neon. Only trace amounts of other metals should exist.

{\bf WD MASS}
\ \ \ \
The initial mass function combined with the current age
of the Galaxy has conspired to yield a present-day distribution of WD
masses according to Fig. 2 of \cite{altetal2010} and Figs. 8, 10 and 
11 of \cite{kleetal2013}. These figures indicate a unimodal distribution that is 
peaked at about $0.6 M_{\odot}$ and contains a long tail at masses higher than 
$0.8 M_{\odot}$.  This distribution also conforms with a previous large (348 objects)
survey \cite{lieetal2005}, where $0.4M_{\odot}$ and  $0.8M_{\odot}$ values are
considered to be ``low-mass'' and ``high-mass'' \cite{livetal2005}.
In principle, WD masses can range up to about $1.4M_{\odot}$.
Only single stars with $M_{\star}^{({\rm MS})} \gtrsim 0.8M_{\odot}$ could have already 
become WDs, and hence single WDs must have masses that satisfy 
$M_{\star}^{({\rm WD})} \gtrsim 0.4 M_{\odot}$.  For comparable or lower mass single WDs, 
perhaps substellar companions could 
have stripped away some of this mass during the CE phase \cite{neltau1998}.

How the mass of a WD is related to its progenitor MS mass represents an extensive field
of study characterized by the ``initial-to-final-mass relation''. Observationally, this
relation is often determined with WDs which are members of stellar clusters whose
ages are well-constrained.  However,
the relation is dependent on stellar metallicity, and in particular the chemistry
of individual stars.  Ignoring those dependencies, some relations used in the
post-MS planetary literature include Eq. 6 of \cite{debetal2012} (originally 
from \cite{wiletal2009}), Eq. 9 of
\cite{frehan2014} (originally from \cite{wooetal2002}), and Eq. 6 of 
\cite{shatho2013} (originally from \cite{kaletal2008}).  

One study which did evaluate how the initial-final mass relationship is
a function of metallicity is \cite{menetal2008}.  They found that metallicity can change
the final mass by $0.4M_{\odot}$, a potentially alarming variation
given the difference between a ``low-mass'' WD ($0.4 M_{\odot}$) 
and a ``high-mass'' WD ($0.8 M_{\odot}$).  \cite{menetal2008} also provided in 
their Appendix potentially useful WD-MS mass relations as a function of metallicity,
for 
$Z_{\star} = \left[0.0001, 0.0003, 0.001, 0.004, 0.01, 0.02, 0.03, 0.04, 0.05, 0.06, 0.08, 0.1\right]$.
For Solar metallicity ($Z_{\star} = Z_{\odot} =0.02$) and any star that will 
become a WD for $0.8M_{\odot} < M_{\star}^{({\rm MS})} < 6.0M_{\odot}$, they found

\[
\frac{M_{\star}^{({\rm WD})}}{M_{\odot}} =
{\rm min}
  \bigg[
  0.572-0.046\frac{M_{\star}^{({\rm MS})}}{M_{\odot}}  +0.0288\left(\frac{M_{\star}^{({\rm MS})}}{M_{\odot}}\right)^2,
\]

\begin{equation}
\ \ \ \ \ \ \ \ 
\ \ \ \ \ \ \ \ 
  1.153-0.242\frac{M_{\star}^{({\rm MS})}}{M_{\odot}}  +0.0409\left(\frac{M_{\star}^{({\rm MS})}}{M_{\odot}}\right)^2
  \bigg].
\label{eqIFMR}
\end{equation}

{\bf WD RADIUS}
\ \ \ \
Usefully for modellers, the radius of the WD can be estimated 
entirely in terms of $M_{\star}^{({\rm WD})}$ with explicit formulae.
A particularly compact but broad approximation is
$R_{\star}^{({\rm WD})}/R_{\odot} \sim 10^{-2} (M_{\star}^{({\rm WD})}/M_{\odot})^{-1/3}$.
Alternatively, more accurate formulae 
-- which are derived assuming that the WD temperature is zero --
that are within a few
percent of one another are from Eq. 15 of \cite{verrap1988},
and, as shown below, Eqs. 27-28 of \cite{nauenberg1972}:

\begin{equation}
\frac{R_{\star}^{(\rm WD)}}{R_{\odot}}
\approx
0.0127 
\left( \frac{M_{\star}^{(\rm WD)}}{M_{\odot}} \right)^{-1/3}
\sqrt{
1 - 0.607
\left( \frac{M_{\star}^{(\rm WD)}}{M_{\odot}} \right)^{4/3}
}
.
\label{eqWDR1}
\end{equation}

\noindent{}From Eq. \ref{eqWDR1}, I obtain a canonical WD radius of 
8750 km $= 0.0126R_{\odot}$, assuming the fiducial value of 
$M_{\star}^{(\rm WD)} = 0.6M_{\odot}$.

{\bf WD LUMINOSITY}
\ \ \ \
The luminosity of WDs can be estimated in multiple ways. A rough approximation
that does not include dependencies on stellar mass or metallicity is
from Eq. 8 of \cite{stoetal2015}, which is originally from \cite{altetal2009}:  
$L_{\star} = L(t_{\rm cool}=0)
\times [t_{\rm cool}/10^5 \ {\rm yr}]^{-1.25}$, where $t_{\rm cool}$
is the WD cooling age.
I include these dependencies by combining the prescription originally from 
\cite{mestel1952} with expressions used in post-MS planetary contexts from Eq. 6 
of \cite{bonwya2010} and Eq. 5 of \cite{verDisc2015} to obtain

\begin{equation}
L_{\star}^{(\rm WD)} = 3.26 L_{\odot} 
\left( \frac{M_{\star}^{(\rm WD)}}{0.6M_{\odot}} \right)
\left( \frac{Z_{\star}}{0.02} \right)^{0.4}
\left(0.1 + \frac{t_{\rm cool}}{\rm Myr} \right)^{-1.18}
,
\end{equation}

\noindent{}where $Z_{\star}$ is the assumed-to-be-fixed stellar metallicity.
Depending on the WD cooling age, the star's luminosity can
range from about $10^{3}L_{\odot}$ to $10^{-5}L_{\odot}$.
\noindent{}This formula also applies only until ``crystallisation''
sets in, which occurs for $T_{\star}^{({\rm WD})} \lesssim 6000-8000$ K \cite{metetal2004}.

\subsubsection{Neutron stars}

For stars with $M_{\star}^{(\rm MS)} \gtrsim 8M_{\odot}$, the end of the AGB
phase results in an explosion: a core collapse plus an 
outwardly-expanding shockwave that nearly 
instantaneously (with velocities of about $10^3-10^4$ km/s) expels the envelope and causes
the star to lose at least half of its mass.  This event is a supernova (SN).
Any asymmetry in the SN will cause a velocity ``kick''.
The remaining stellar core becomes either a neutron star (NS) or a black hole.
Of most relevance to post-MS planetary science are pulsars, which are an august
class of NSs that represent
precise, stable and reliable clocks.

Although NSs and WDs are together grouped as ``compact stars'', NSs are much
more compact, with radii on the order of 10 km.  NS masses are greater than
those of WD stars. Typically $M_{\star}^{({\rm NS})} \ge 1.4 M_{\odot}$.
NSs cool much faster than WDs, with a decreasing luminosity which can be
modelled by (pg. 30 of \cite{shateu1983})

\begin{equation}
L_{\star}^{({\rm NS})}  = 0.02 L_{\odot} 
\left( \frac{M_{\star}^{({\rm NS})}}{M_{\odot}} \right)^{2/3}
\left[
\frac{{\rm max}\left(t_{\rm cool}, 0.1 {\rm Myr} \right)}
{{\rm Myr}}
\right]
,
\end{equation}

\noindent{}where $t_{\rm cool}$ represents the NS cooling time in this context.

Millisecond pulsars have rotational periods on the order of ms.  They are
thought to have been spun up by accretion, and are hence said to be ``recycled''.
\cite{milham2001} argued that the presence of planets around millisecond pulsars
can constrain the evolutionary history of the star.  In particular,
they posed that the moon-sized SB around the millisecond pulsar
PSR B1257+12 demonstrates how that particular star is not recycled by (i) favouring
a second-generation formation scenario for the SB (see Sec. \ref{secFall}), and (ii)
suggesting that the formation cannot have occurred during an accretional event
nor in a post-spin-up disc.  They claimed that the moon-sized SB must have 
formed, post-SN, around the star as is with its current rotational 
frequency and magnetic moment.

\subsection{Common envelope and binary star evolution} \label{secSEbinary}

Stellar binary systems are important because they represent several tens of percent
of all Milky Way stellar systems.  The presence of a stellar binary companion 
can significantly complicate the evolution if the mutual separation is within 
a few tens of au.  Both star-star tides and the formation of a 
``common envelope'' (CE) can alter the fate otherwise predicted from single-star evolution.
\cite{ivaetal2013} reviewed the theoretical work performed on and the 
physical understanding of CEs; see their Fig. 1 for some illustrative
evolutionary track examples. \cite{taaric2010} provided a shorter, simulation-based 
review of the topic.

A CE is a collection of mass that envelopes either (i) two stars, or, 
(ii) one star and one large SB like a 
giant planet.  In both cases, as the smaller binary component spirals into the 
larger one, the former transfers energy to the
envelope.  The transfer efficiency is
a major unknown in the theory of stellar evolution.  The smaller 
companion may blow off the
CE by depositing a sufficient amount of energy in the envelope during inspiral.
Relevant equations describing this process include Eq. 17 of \cite{noretal2010},
Eqs. 2-5 of \cite{schdre2014}, and Eq. 8 of \cite{soker2015}.  
A more complete treatment that takes into
account shock propagation and rotation may be found in Eqs. 6-25 of \cite{schdre2014};
also see the earlier work by \cite{soker1993}.
The more massive the companion, and the more extended the envelope,
the more likely ejection will occur.  
The speed of infall within the CE may be expressed generally
as a quartic equation in terms of the radial velocity (Eq. 9 of
\cite{norbla2006}), but only if the SB's tangential velocity is known,
as well as the accretion rate onto the SB.  Equation 1 of 
\cite{neltau1998} approximates the final post-CE separation after
inspiral.

Even without a CE, the interaction between both stars might dynamically
excite any SBs in that system, particularly when one or both stars
leave the MS. Both stars might be similar enough in age (and hence MS mass) to undergo
coupled GB mass loss. Section 5.2 of \cite{verevawyatou2014} quantified
this possibility, and finds that the MS masses of both components must roughly
lie within 10\% of one another in order for both to simultaneously lose
mass during their AGB phases.

\section{Observational motivation} \label{secOM}

Post-MS planetary systems provide multiple insights that are
not available from MS planetary systems, including: (i) substantive 
access to surface and interior SB chemistry, (ii) a way to link SB fate and 
formation, (iii) different constraints on tidal, mass-losing and radiative
processes, and (iv) the environments to allow for detections
of extreme SBs.  The agents for all this insight come from GB, WD
and NS planetary systems.  Overall, the total number of WD
remnant planetary systems is of the same order ($\sim 1000$) as
MS planetary systems, and about one order of magnitude more
than GB planetary systems.
The number of remnant planetary systems around NSs is a few.

\subsection{Planetary remnants in and around WDs} \label{secOMWDpoll}

Fragments and constituents of disrupted SBs that were planets, asteroids, 
moons, comets, boulders and pebbles observationally manifest 
themselves in the atmospheres of WDs and the
debris discs which surround WDs.
The mounting evidence for and growing importance of both topics is respectively 
highlighted in the recent reviews of 
\cite{juryou2014,farihi2016}. Here, I devote just one subsection to the 
observational aspects of each topic.

\subsubsection{WD atmospheric pollution}  \label{WDpollInd}

Because WDs are roughly the size of the Earth but contain approximately 
the mass of the Sun,
WDs are about $10^5$ as dense as the Sun.  Consequently, due to gravitational
settling, WD atmospheres quickly separate
light elements from heavy elements \cite{schatzman1945}, causing the latter 
to sink as oil would in water.  
This stratification of WD atmospheres by atomic weight provides a tabula rasa
upon which any ingested contaminants conspicuously stand out -- as long as we detect
them before they sink.

{\bf COMPOSITION OF INTRINSIC WD PHOTOSPHERE}
\ \ \ \
The chemical composition of the atmosphere 
is dependent on (1) how the WD evolved from the GB phase, and (2) the WD's cooling age.  
At the end of the AGB phase, the star's photosphere becomes either hydrogen-rich (DA), 
helium-rich (DB), or a mixed hydrogen-helium composition (DAB, DBA). The link between
spectral type and composition is sometimes not so clear, as more literally a DA WD
refers to a WD whose strongest absorption features arise from H, and similarly 
a DB WD has the strongest absorption features arising from He.  If the cooling age 
is within a few tens of Myr, then the WD
is hot enough to still contain heavy elements in the photosphere.  These elements
are said to be ``radiatively levitated''.  
For cooling ages between tens of Myr and
about 500 Myr, the atmosphere consists of hydrogen and/or helium only.  For cooling
ages greater than 500 Myr, some carbon -- but usually only carbon -- 
from the core may be dredged 
up onto the atmosphere. Effectively then, WDs which are older than a few tens of Myr 
and do not accrete anything have atmospheres which are composed
of some combination of hydrogen, helium and carbon only.  

{\bf COMPOSITION OF POLLUTED WD PHOTOSPHERE}
\ \ \ \
Yet, we have now detected a total of 18 metals heavier than carbon in WDs
with $30$ Myr $\lesssim t_{\rm cool} \lesssim 500$ Myr.
These metals, which are said to ``pollute'' the WD (thereby adding a ``Z'' designation
to its spectral class), are, with atomic number, O(8),
Na(11), Mg(12), Al(13), Si(14), P(15), S(16), Ca(20), Sc(21), Ti(22), V(23), Cr(24),
Mn(25), Fe(26), Co(27), Ni(28), Cu(29) and Sr(38).  Although N(7) has
not been directly detected, there are published upper limits for that chemical element.  
These metals include 
rock-forming elements (Si, Fe, Mg, O), refractory lithophiles (Ca, Al, Ti), 
volatile elements (C, N, P, S) and siderophiles (Cr, Mn, S, Ni).
The first polluted WD (Van Maanen 2, or vMa 2), discovered
in the late 1910s, contains observable Ca, which happens to be the strongest signature
in WD spectra (Mg is next) \cite{vanmaanen1917,vanmaanen1919}.  Only about 90 years later,
with the availability of high resolution spectroscopy from the Hubble Space Telescope,
plus ground-based observations with Keck, VLT, HST and SDSS,
did the floodgates open with the detection of 17 of the above metals all within the same WD
(GD 362) \cite{zucetal2007}.  A steady stream of highly metal-polluted WDs
has now revealed unique, detailed and exquisite chemical signatures 
(e.g. \cite{gaeetal2012,kleetal2010,kleetal2011,juretal2012,xuetal2013,xuetal2014,wiletal2015}).
Two notable cases \cite{faretal2013b,radetal2015} include a high-enough level of oxygen
to indicate that the origin of the pollution consisted of water, a possibility 
envisaged by e.g. \cite{jurxu2010}.

\begin{figure}[!t]
\centering\includegraphics[width=4.0in]{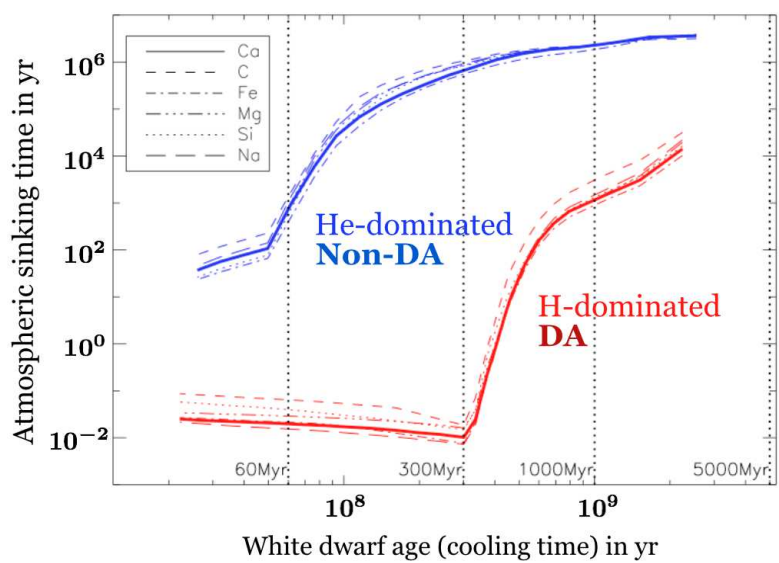}
\caption{Cosmetically enhanced version of Fig. 1 of
\cite{wyaetal2014}.  Shown are the shrinking times of six
metals in WD atmospheres.  These times are orders 
of magnitude less than the WD cooling ages.  The sinking 
timescales of DA WDs younger than about 300 Myr are days-to-weeks.
}
\label{FigWya}
\end{figure}

{\bf PLANETARY ORIGIN OF POLLUTION}
\ \ \ \
A common explanation for the presence of all these metals is accretion of
remnant planetary material.  The now-overwhelming evidence includes: 
(i) the presence of accompanying debris discs 
(see Sec. \ref{secOM}\ref{secOMWDpoll}\ref{WDDebrisDiscs}),
(ii) SBs caught in the act of disintegrating around a polluted WD
(see Sec. \ref{secOM}\ref{secOMWDplan}\ref{secOribitingWDs})
(iii) chemical abundances which resemble the bulk Earth to zeroth order 
(see e.g. \cite{juryou2014}), 
(iv) a variety of chemical signatures that are comparable to the diversity seen 
across Solar system meteorite families (see e.g. \cite{juryou2014}),
(v) the debunking of the possibility of accretion from the interstellar medium 
(see, e.g. Eqs. 2-6 and Table 3 of \cite{faretal2010}),
and
(vi) the fraction of polluted WD systems, which is 25\%--50\% 
\cite{zucetal2003,zucetal2010,koeetal2014} and hence roughly commensurate
with estimates of Milky Way MS planet-hosting systems \cite{casetal2012}.
This last point is particularly remarkable because
metals heavier than carbon will sink 
(or ``diffuse'', ``settle'' or ``sediment'') through the convection or ``mixing'' 
zone quickly (see Fig. \ref{FigWya}):
in days or weeks for DA WDs younger than about 300 Myr and within 
Myrs for DB WDs.
In all cases, the
sinking times are orders of magnitude shorter than the WD cooling age.
Therefore we should always expect to detect heavy metal pollution at 
a level well-under 0.1\%.

\begin{figure}[!t]
\centering\includegraphics[width=5.0in]{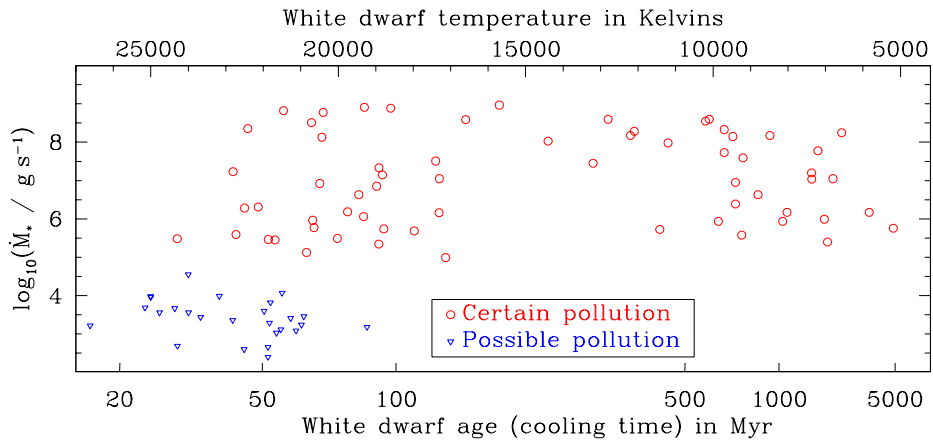}
\caption{Cosmetically reconstructed version of the top panel of
Fig. 8 of \cite{koeetal2014}.  The blue downward triangles refer
to upper limits.  The plot illustrates that accretion rate appears
to be a flat function of WD cooling age: pollution occurs at similar
rates for young and old WDs. 
}
\label{FigAccr}
\end{figure}

Because the percentage is actually 25\%-50\%, then for most DA WDs (which represent 
about 80\% of all WDs \cite{kleetal2013}) the accretion is ongoing right now.
The accretion occurs at similar levels along all detectable WD cooling ages 
(up to about 5 Gyr;
see Fig. \ref{FigAccr}), highlighting an important challenge for theorists: what 
planetary architectures can generate comparably high levels of accretion at such 
late ages?  (the first polluted
WD, vMa 2, is relatively ``old'', with $t_{\rm cool} = 3$ Gyr).

{\bf IMPLICATIONS FOR PLANETARY CHEMISTRY AND SURFACES/INTERIORS}
\ \ \ \
For the foreseeable future, the only reliable way to study the chemistry of SBs 
will be through spectroscopic observations of their tidally disrupted 
remnants in WD atmospheres.  
Samples from Solar system meteorites, comets and planets (including the Earth) allow
us to make direct connections to chemical element distributions in WD atmospheres.
For example, we know that an overabundance of S, Cr, Fe and Ni indicates melting
and perhaps differentiation \cite{gaeetal2012}.  Signatures of core and crust formation 
are imbued in the ratio of iron to siderophiles or refractory lithophiles.  Also, in
particular, Fe-rich cores, Fe-poor mantles or Al-rich crusts may all be distinguished
\cite{juretal2014}.  A carbon-to-oxygen ratio $\gtrsim 0.8$ would 
result in drastically different physical setup than the Solar system's \cite{moretal2014}.
For more details, see \cite{juryou2014}.

{\bf IMPLICATIONS FOR PLANETARY STATISTICS}
\ \ \ \
{Because polluted WDs signify planetary systems, these stars can be used
to probe characteristics of the Galactic exoplanet population.  
\cite{zuckerman2014} considered the population of polluted WDs which
are in wide binary systems, and concluded that a comparable fraction of
both single-star and wide binary-star systems with $r_{\rm b} \gtrsim 1000$ au 
host planets. For $r_{\rm b} \lesssim 1000$ au, however, the binary planet-hosting
fraction is less, implying perhaps that in these cases the binary companion suppresses
planet formation or more easily creates dynamical instability.}

\begin{figure}[!h]
\centering\includegraphics[width=4.0in]{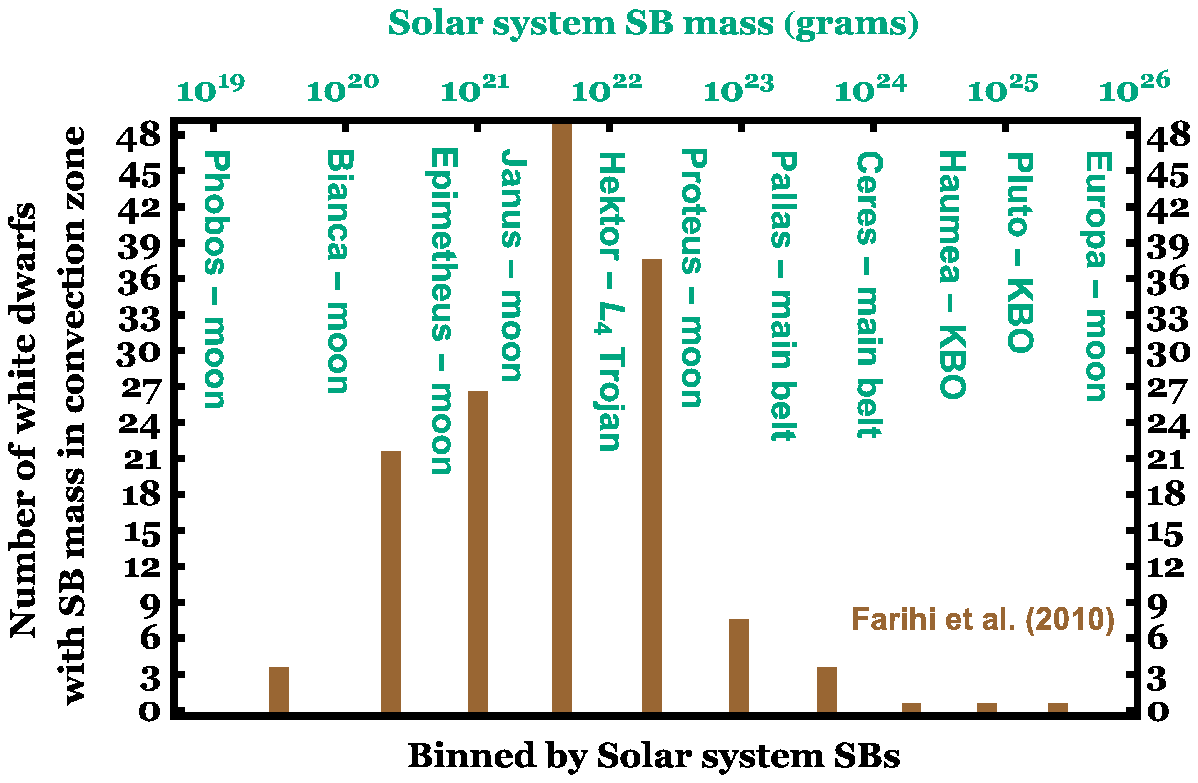}
\centering\includegraphics[width=4.0in]{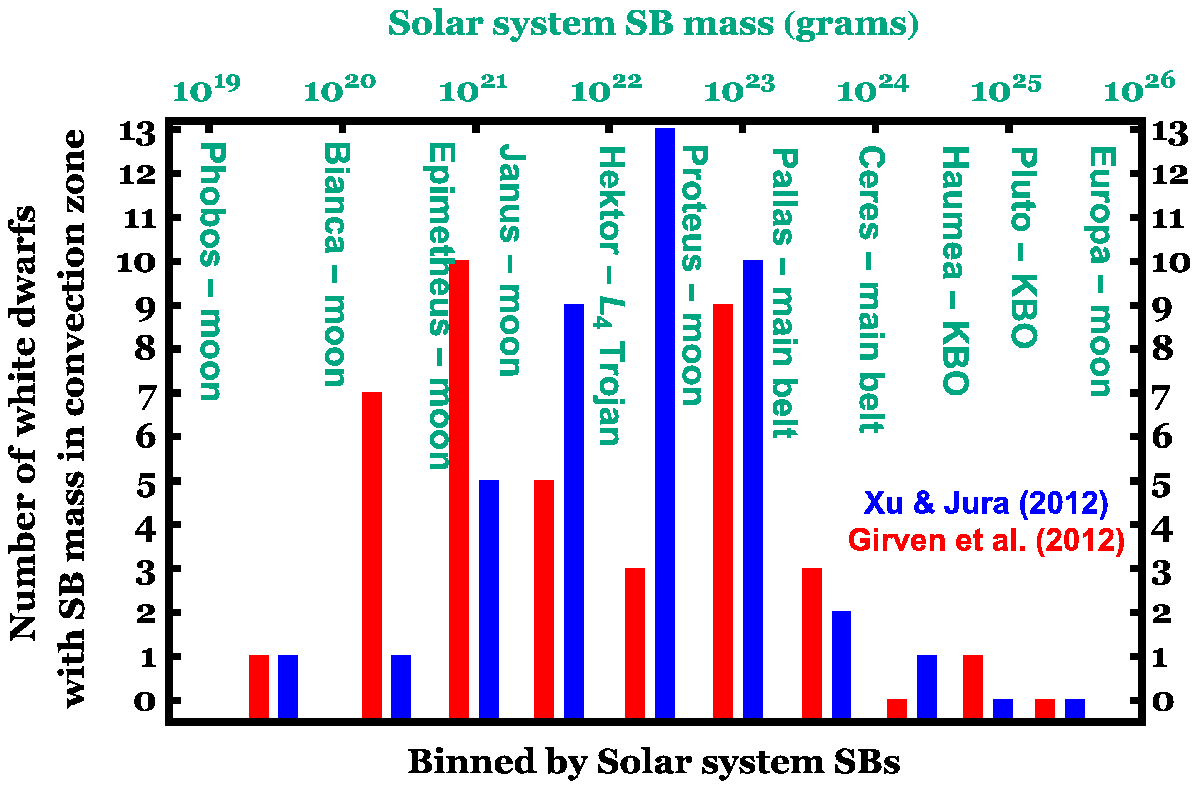}
\caption{Histograms of the accumulated mass of rocky 
substellar bodies that were accreted onto
white dwarfs during the last Myr or so, including
both detections and limiting values.  
Differently coloured bars refer to
three different WD samples 
(brown: data from \cite{faretal2010} assuming that Ca represents 
1.6\% of the mass of the accreted bodies, similar to the corresponding
mass fraction of the bulk Earth -- see Table 3 of \cite{mcdonough2000};
blue: data displayed in Fig. 12 of \cite{xujur2012}; 
red: data displayed in Fig. 9 of \cite{giretal2012}).
The panels are separated according to sample size (see $y$-axes).
For observational subtleties associated with the data, see the corresponding
papers. 
The bin sizes are according
to the Solar system objects displayed in green, with masses given on the top axis.
This plot demonstrates that pollution may arise from a wide variety of objects.}
\label{FigConv}
\end{figure}

{\bf ACCUMULATED METAL MASS IN THE CONVECTION ZONE}
\ \ \ \
For DB WDs, the situation is different.  Their convection zones are deep enough to hold
a record of all remnant planetary accretion over the last Myr or so.  This feature allows
us to estimate lower bounds for the total amount accreted over this timescale.  
Fig. \ref{FigConv} illustrates the amount 
of mass in metals,
in terms of Solar system asteroids, moons and one L$_4$ Jupiter Trojan, that has been
accreted in two different WD samples.

{\bf OTHER CONSTRAINTS}
\ \ \ \
Some metal-polluted WDs are magnetic (denoted with an ``H'' in the spectral classification).  
At least 10 DZH WDs harbour magnetic fields of $0.5-10$ MG \cite{holetal2015}, although 
this preliminary work indicates this number is likely to be double or triple.   The 
theoretical implications of magnetic fields have previously only briefly been touched upon 
(Sec. \ref{secOther}\ref{secOtherMagnetism}).

Further, hydrogen abundance in WD atmospheres, although not considered a ``pollutant'', 
nevertheless might provide an important constraint on pollution.  Because hydrogen does 
not diffuse out of a WD atmosphere, this chemical element represents a 
permanent historical tracer of accretion throughout the WD lifetime 
(even if the WD's spectral type changes as a function of time).
Accretion analyses and interpretations, however, must assume that 
the WD begins life with a certain amount of primordial H.  
This accretion arises  
from a combination of the interstellar medium, asteroids, comets and any planets.  
Of these, comets -- 
and in particular exo-Oort cloud comets -- might provide the greatest amount of this hydrogen
through ices.
Consequently, linking WD hydrogen content with cooling age may help determine the accretion
rate of exo-Oort cloud comets soon after the WD is formed \cite{stoetal2015} and over time
\cite{vershagan2014}.  Fig. 5 of \cite{radetal2015} illustrates how WD
hydrogen mass appears to be a steadily increasing function of cooling age, and increases
at a rate far greater than realistic estimates of accretion from the interstellar medium.

\subsubsection{WD debris discs}  \label{WDDebrisDiscs}

Debris discs have been detected orbiting nearly 40 polluted WDs.
The first disc discovered orbits the WD Giclas 29-38 (commonly known as G 29-38) 
\cite{zucbec1987} in 1987.  Nearly two decades passed before the second disc, 
orbiting GD 362 \cite{becetal2005,kiletal2005}, prompted rapid progress.
No confidently reported debris disc around a single unpolluted
WD exists, suggesting the link between pollution and discs is strong.
At least a few percent and up to 100\% of all WDs host discs
\cite{faretal2009,baretal2012,beretal2014}.  The lower limit for the Galactic
population is based on actually observed discs, whereas the one-to-one
potential correspondence between pollution and the presence of a disc
is based on most discs likely being too faint to detect.  Although
observational sensitivities allow pollution to be discovered in
WDs with $t_{\rm cool}$ as high as about 5 Gyr, discs are difficult to detect
for $t_{\rm cool} > 0.5$ Gyr \cite{beretal2014}.  \cite{farihi2016} recently
summarized observations of these discs.  See also Table
1 of \cite{faretal2009}, Table 1 of \cite{xujur2012} and Table 2 of \cite{rafgar2012}
for some details of dust-only discs found before 2012.

{\bf DETECTION CONSTRAINTS}
All these discs are dusty, and dust comprises the major if not sole component.
Consequently, the detection and characterization of the discs rely on 
modeling spectral 
energy distributions with a signature (``excess'' with respect to the flux
from the WD) in the infrared and a total
flux, $\mathcal{F}$, prescription that is given in Eq. 3 of \cite{jura2003}:

\begin{equation}
\mathcal{F} \approx 12\pi^{\frac{1}{3}}
\cos{\left(i_{\rm LOS}\right)}
\frac{R_{\star}^2}{D^2} 
\left( 
\frac{2 k_{\rm B} T_{\star}}{3 h \nu}
\right)^{\frac{8}{3}}
\frac{h \nu^3}{c^2}
\int_{x^{({\rm in})}}^{x^{({\rm out})}}
\frac{x^{\frac{5}{3}}}{{\rm exp}\left(x\right) - 1} dx
.
\label{eqflux}
\end{equation}

\noindent{}In Eq. \ref{eqflux}, $\nu$ is the frequency,  $D$ is the distance between the star
and the Earth, $i_{\rm LOS}$ is the line-of-sight inclination with respect
to the Earth, $k_{\rm B}$ is the Boltzmann constant, $h$ is the Planck constant, 
and $x(r) \equiv h\nu/k_{\rm B}T_{\rm d}(r)$.  The discs are assumed to be passive, 
opaque (optically thick) and geometrically flat.  

The equation is degenerate with respect to three parameters: $i_{\rm LOS}$, 
and the disc temperatures at the inner and outer edges.  Fig. 5 of \cite{giretal2012}  
and Fig. 3 of \cite{xujur2014} illustrate how the degeneracy from these three parameters manifests itself in the modelling of debris discs. For an explicit example of how 10 different viewing angles can 
change the flux signature, see Fig. 1 of \cite{livetal2005}, who simulate the 
spectral energy distribution for a (so-far unrealized) au-scale WD debris disc.

{\bf DISC CHARACTERISTICS}
\ \ \ \
One such property is disc geometry. The spectral energy distributions 
generally do not indicate that WD discs are flared.  However, a couple of 
possible exceptions include GD 362 \cite{juraGD362}
and GD 56 \cite{juraGD56}.   The size distribution 
of the dust/particles/solids in the discs is unknown,
except for the presence of micron-sized grains 
\cite{reaetal2005,reaetal2009,juretalsize}.
One notable disc, which orbits WD J0959-0200, is highly variable:
\cite{xujur2014} reported a still-unexplained flux decrease
of about 35\% in under 300 days.
  
Application of Eq. \ref{eqflux}, with assumptions about $i_{\rm LOS}$, 
yields a striking result for WDs with
$t_{\rm cool}$ greater than a few Myr:  the entire disc lies within the
Roche radius of the WD.  Fits from this equation suggest that the 
discs reside in the region $r_{\rm d} \approx 0.6-1.2R_{\odot}$.
Confirmation of this approximate range arose with the discovery of 
both dusty and gaseous components in seven of these discs.
The gaseous components constrain the disc geometry.  
This distance range clearly demonstrates that (i) the discs do not extend
all the way to the WD surface (photosphere) and that (ii) the discs could not have 
formed during the MS or GB phases.  Regarding this first point, some spectral
features do suggest the presence of gas within $0.6 R_{\odot}$
(e.g. see the bottom-left panel of Fig. 3 in \cite{gaeetal2012}), but
not yet in a disc form.

\begin{figure}[!h]
\centering\includegraphics[width=3.0in]{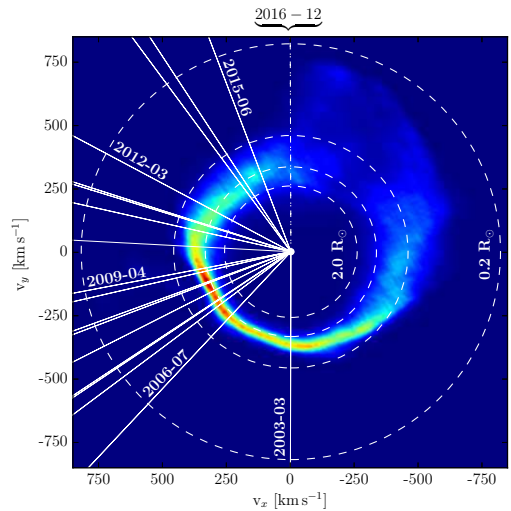}
\caption{Exact reproduction of Fig. 5 of \cite{manetal2015}. This image is a velocity space
map of the gaseous component of the debris disc orbiting the WD SDSS J1228+1040.  
The subscripts $x$ and $y$ refer
to their usual Cartesian meanings, and the WD is located at the origin. 
Observations at particular dates are indicated by solid 
white lines. The image suggests that the disc is highly non-axisymmetric and precessing on decadal timescales.}
\label{FigDisc}
\end{figure}

The first gaseous disc component found (around SDSS J122859.93+104032.9, 
also known as SDSS J1228+1040) \cite{gaeetal2006}, also exhibits 
striking morphological changes, which occur secularly and smoothly over decades 
(whereas the disc orbital period is just a few hours) \cite{manetal2015};
see Fig. \ref{FigDisc}. The figure is a velocity space intensity distribution where 
the radial white lines indicate different times from the years 2003-2016.  
Four other discs with time-resolved observations of gaseous components
are SDSS J0845+2258, SDSS J1043+0855, SDSS J1617+1620 and SDSS J0738+1835.
The first three of these -- which change shape or flux over yearly and decadal timescales
-- represent exciting dynamical objects, while the last, which exists in an 
apparently steady state (given just a handful of epochs so far), might provide an 
important and intriguing contrast.

One notable exception to all of the above WD discs is a very wide (35-150 au)
dusty structure inferred orbiting the extremely young ($t_{\rm cool} \ll 1$ Myr) WD 2226-210
\cite{suetal2007}. The interpretation of this dusty annulus representing a remnant 
exo-Kuiper belt is degenerate and is not favoured compared to a stellar origin 
\cite{claetal2014}: i.e., this annulus might represent a planetary nebula.

\subsection{Major and minor planets around WDs} \label{secOMWDplan}

\subsubsection{Orbiting WDs} \label{secOribitingWDs}

A few WDs host orbiting SBs, and they are all
exoplanetary record-breakers (as of time of writing) in at least one way.

{\bf THE FASTEST, CLOSEST AND SMALLEST SBs}
\ \ \ \
Transit photometry
of WD 1145+017 revealed signatures of one to several SBs 
(with $R_{\rm SB} < 10^3$ km) which are currently disintegrating within 
the WD disruption radius with orbital periods of 4.5-5.0 hours 
\cite{vanetal2015,croetal2015}.  \cite{gaeetal2016} have since constrained
the orbital periods of at least six SBs to within 15 seconds of 
4.4930 hours each, indicating almost exactly coplanar orbits.  
Further, within the same system, \cite{xuetal2015b} have detected
circumstellar absorption lines from likely gas streams, as well as 11
different metals in the WD atmosphere.

Because this WD is both polluted and hosts a dusty debris disc, these
minor planet(s) further confirm the interpretation that accretion onto WDs
and the presence of circumstellar discs is linked to first-generation
SB disruption (see Sec. \ref{secWDform}).  This type of discovery was 
foreshadowed by previous 
prognostications: (i) \cite{soker1999} found that for stars 
transitioning from the AGB to WD phase, their shocked winds can create 
mass ablation from surviving planets into a detectable debris tail,  
(ii) \cite{villiv2007} predicted that planets evaporating
and emitting Parker winds could be detected with spectroscopic observations, 
but was thinking of atmospheric
mass outflows at several au around GB stars.  
(iii) \cite{disetal2010} demonstrated specifically that the {\it Kepler}
space mission should be able to detect WD transits of minor planets.
Ironically, although the paper was written with the primary {\it Kepler}
mission in mind, only during the secondary mission were enough WDs observed
to achieve this discovery. 
(iv) Alternatively, \cite{spimad2012} claimed that the process of a 
stellar wind accreting onto a SB might produce a detectable coronal envelope around the SB.

{\bf THE FURTHEST AND SLOWEST EXOPLANET}
\ \ \ \  
WD 0806-661 b 
is a planetary mass ($7M_{\rm Jup}$) SB orbiting the WD at an 
approximate distance of $2500$ au \cite{luhetal2011}.  Although
some in the literature refer to the object as a brown dwarf, the mass 
is well-constrained to be in the planet
regime (see their Fig. 4).  The difference of opinions is perhaps 
partly informed by contrasting assumptions about the SB's dynamical origin 
rather than its physical properties.
The planet was discovered using direct imaging, and holds the current record
for the bound exoplanet with the widest orbit known.

{\bf THE FIRST CIRCUMBINARY EXOPLANET}
\ \ \ \
The first successfully predicted 
\cite{sigurdsson1993,thoetal1993} and confirmed \cite{sigetal2003} circumbinary exoplanet, 
PSR B1620-26AB b, orbits both a WD (with mass $\approx 0.34M_{\odot}$ and cooling
age of about 480 Myr) and a millisecond pulsar (with mass $\approx 1.35M_{\odot}$ 
and rotation period of 11 ms). The WD cooling age and pulsar rotation period 
importantly help constrain the dynamical history of the system. The planet's 
physical and orbital parameters are 
$M_{\rm SB} \sim 2.5M_{\rm Jup}$, $a \sim 23$ au, and $i \sim 40^{\circ}$ whereas the
binary orbital parameters are $a_{\rm b} \approx 0.8$ au and $e_{\rm b} \approx 0.025$.

PSR B1620-26AB b is the only known planet in a system with two 
post-MS stars, and one of the few exoplanets ever observed in a metal-poor
environment and cluster environment (the M4 globular cluster).  The 
planet name contains ``PSR'' 
because the pulsar was the first object in 
the system discovered and is the most massive object (the primary).  However, the 
planet was originally thought to orbit (and form around) the progenitor of the 
WD, and hence is more appropriately linked to that star.  Further, I do not 
classify this system as a post-CE binary 
(see Sec. \ref{secOM}\ref{secOMPCEB}) because both the system does not fit the 
definition of containing a WD and a
lower-mass MS companion, and the system is typically not included in the post-CE binary 
literature \cite{zorsch2013}.  
This combination of pulsar, WD and planet suggests a particularly fascinating 
dynamical history (see Sec. \ref{secMulti}\ref{secMulti2sNp0a}).

{\bf HINTS OF DETECTIONS}
\ \ \ \
In addition to the above observations, there are several hints of SBs orbiting WDs.
The magnetic WD GD 365 exhibits emission lines which could indicate the presence
of a rocky planet with a conducting composition \cite{lietal1998}.  Later data has been able
to rule out an SB with $M_{\rm SB} \ge 12 M_{\rm Jup}$ \cite{wicetal2010}, which
is consistent with the rocky planet hypothesis. Also, the spectral energy distribution
of PG 0010+280 may be fit with an SB with $r \approx 60$ au \cite{xuetalPG}. In order for this SB
to be hot enough for detection, it may have been re-heated; see their Sec. 3.3.2.

Further, a few tenths of percent of Milky Way WDs host brown dwarf-mass SBs 
\cite{faretal2005}.  
These companions were found to orbit at distances beyond the tidal engulfment
radius of the AGB progenitor of the WD until the notable discovery of WD 0137-349 B
\cite{maxetal2006}.  This brown dwarf has $M_{\rm SB} = 0.053M_{\odot}$. For the primary
WD star, $M_{\star} = 0.39M_{\odot}$.  The orbit is close enough --
($a \sin{i_{\rm LOS}} = 0.375R_{\odot} = 0.0017$ au) -- that WD 0137-349 B
must have survived engulfment in the GB envelope of the progenitor.
The low mass of the WD is characteristic of premature CE ejection
by a companion.

{\bf SEARCH METHODS}
\ \ \ \
For a recent short summary of the different techniques employed to 
search for planets around WDs, see Section 1 of \cite{xuetal2015}.
The discovery of SBs orbiting WD 1145+017 based on transit 
photometry \cite{vanetal2015}
highlights interest in this technique.  \cite{faeetal2011} placed 
limits ($< 10\%$) on the frequency of 
gas giants or brown dwarfs on circular orbits with orbital periods of several hours,
and mentioned that exo-moons orbiting WDs can generate 3\%
transit depths.  An important caveat to this transit method is that 
it requires follow-up with other methods, at least according to
\cite{fosetal2012}, who wrote that for WDs, 
``planet detection based on photometry alone would not be credible''.

Existing relevant formulae suppose that the SB is smaller than the star (which is not true
for giant planets or Earth-sized SBs orbiting WDs), and some formulae
make other approximations (like circularity of orbits) which would be ill-suited
for the type of transits suggested by e.g. \cite{vergae2015}.
Useful formulae from post-MS studies are provided by Eqs. 1-2 of \cite{faeetal2011},
Eq. 1 of \cite{agol2011} and Eqs. 18-19 of \cite{noretal2010}.  \cite{faeetal2011}
estimates the depth of the transit to be equal to unity if $R_{\rm SB} \ge R_{\star}$ and
instead,  to approximately equal $R_{\rm SB}^2/R_{\star}^2$ if $R_{\rm SB} < R_{\star}$.
The probability of transit and the duration of transit represent other quantities
of interest.  I display these by repackaging the fairly general expressions 
from Eqs. 9-10 of \cite{winn2014}:

\begin{equation}
{\rm probability}  = \left( \frac{R_{\star} + R_{\rm SB}}{a}  \right) 
                     \left( \frac{1 \pm e\sin{\omega}}{1 - e^2} \right)
\label{eqprobtran}
\end{equation}

\noindent{}where $\omega$ is the argument of pericentre of the orbit, the 
upper sign is for transits (SB passing in front of the WD)
and the lower sign is for occultations (or ``secondary eclipses'', when the
SB passes in back of the WD).  Both formulae assume grazing eclipses, and
that eclipses are centred around conjunctions.  Maintaining this sign convention,
I then combine Eqs. 7, 8, 14, 15 and 16 of \cite{winn2014} to obtain the 
transit/occultation duration:

\begin{eqnarray}
{\rm duration}
&=& 
2 \left( \frac{\sqrt{1-e^2}}{1 + e \sin{\omega}} \right) 
\sqrt{\frac{a^3}{G\left(M_{\star} + M_{\rm SB}\right)} }
\nonumber
\\
&\times& \sin^{-1}
\left[
\frac{R_{\star}}{a \sin{i_{\rm LOS}}} 
\left\lbrace
\left(1 \pm \frac{R_{\rm SB}}{R_{\star}}  \right)^2
-
\frac{a^2 \cos^2{i_{\rm LOS}} \left(1 - e^2\right)^2}
{R_{\star}^2 \left(1 + e \sin{\omega} \right)^2 }
\right\rbrace^{\frac{1}{2}}
\right]
\label{eqdurtra}
\end{eqnarray}

\noindent{}where $i_{\rm LOS}$ is the inclination of the orbit with respect
to the line of sight of an observer on Earth.  By convention, an edge-on 
orientation corresponds to $i_{\rm LOS} = 90^{\circ}$.

\subsubsection{Orbiting the companions of WDs}  \label{secOribitingComp}

Three stellar systems are known to harbour a planet-hosting star and
a WD:  GJ 86, $\epsilon$ Reticulum (or $\epsilon$ Ret), and HD 147513. In no
case is the WD (yet) known to be polluted nor host debris discs, planets 
or asteroids.

The star GJ 86B is an unpolluted WD \cite{faretal2013a} whose binary 
companion GJ 86A is an MS star that hosts a $M_{\rm SB} \gtrsim 4.5M_{\rm Jup}$ 
planet in a $a = 0.11$ au orbit \cite{queetal2000,mugneu2005}.  
With Hubble Space Telescope data, \cite{faretal2013a} helped constrain 
the physical and orbital parameters of the system (see their Table 4), which 
features a current binary separation
of many tens of au.  The planet GJ 86Ab survived the GB evolution of GJ 86B, 
when $a_{\rm b}$ expanded by a
factor of a few, but $e_{\rm b}$ remained fixed (see Sec. \ref{secMass}).

A similar scenario holds for the HD 27442 system.
The star $\epsilon$ Ret, or HD 27442A, hosts a $M_{\rm SB} \gtrsim 1.6M_{\rm Jup}$ 
planet in a $a = 1.27$ orbit \cite{butetal2006}. The projected separation between
$\epsilon$ Ret and its WD binary companion, HD 27442B, is approximately a 
couple hundred au \cite{faretal2011}.  The HD 147513 system is not as well constrained
\cite{desbar2007}.
However, the separation between the binary stars in that system is thought 
to be several thousand au, placing it
in the interesting ``non-adiabatic'' mass loss evolutionary regime 
(see Eq. \ref{eqpsi}).

\subsubsection{WD-comet collisions}

In the 1980s, \cite{bhatt1985}, \cite{pineault1987} and \cite{pinpoi1989} realized 
the potential for collisions between exo-comets
and their parent stars, or other stars, to produce observable signatures.
\cite{alcetal1986} more specifically suggested that comet accretion onto
WDs can constrain number of exo-Oort cloud comets in other systems.
\cite{trezyt1986} and \cite{pinpoi1989} included detailed analytics that 
may still be applicable today.  Some of this analysis was extended to binary star systems 
in \cite{pinduq1993}, with specific application to compact objects in their Sec. 4.3.
Perhaps these speculations have been realized with the mysterious X-ray
signature from IGR J17361-4441 reported in \cite{deletal2014}, 
although that potential disruption event may have just as well been 
caused by planets or asteroids rather than comets.

\subsection{Subgiant and giant star planetary systems} \label{secOMgiant}

\subsubsection{GB Planets}  \label{secOMgiantGB}

{\bf GROSS CHARACTERISTICS}
\ \ \ \ 
As of 30 Nov 2015, 79 SBs were recorded in the 
planets-around-GB-stars database \footnote{Sabine Reffert maintains this database
at www.lsw.uni-heidelberg.de/users/sreffert/giantplanets.html}
although this number may be closer to 100 \cite{witetal2015}.
About 85\% of these SBs are giant planets with $M_{\rm SB} \sim 1-13M_{\rm Jup}$,
proving that planets can survive over 
the entire MS lifetime of their parent stars.
The host stars for these SBs have not undergone enough GB evolution
to incite mass loss or radius variations which are markedly different
from their progenitor stars.  These barely evolved host stars 
are observed in their
early RGB phase, sometimes known as the ``subgiant'' phase.
Because RGB tracks on the Hertzprung-Russell diagram are so close 
to one another, RGB masses are hard to isolate; there is an 
ongoing debate over the subgiant SB-host masses 
\cite{bowetal2010,johetal2010,johetal2013,lloyd2011,lloyd2013,schwin2013}.

Regardless, the population of these SBs shows a distinct characteristic:
a paucity of planets (less than 10\%) within half of an au of 
their parent star.  In contrast, $a < 0.5$ au holds true for about 
three-quarters of 
all known exoplanets.  This difference highlights the need to better
understand the long-term evolution of planetary systems.  Further, 
a handful of GB stars have been observed to host multiple 
planets.  These systems may reveal important constraints on dynamical 
history.  For example,
the two planets orbiting $\eta$ Ceti may be trapped in a $2$:$1$
mean motion resonance (see the fourth and fifth rows of plots in Fig. 6
of \cite{trietal2014}).  A resonant configuration during the GB phase
would help confirm the stabilizing nature of (at least some types of) 
resonances throughout all of MS evolution.

Regarding the planet - giant star metallicity correlation,
\cite{maletal2013} and \cite{refetal2015} arrived at somewhat different
conclusions.  The
former concluded that planet-hosting giant stars are preferentially not 
metal rich compared to giant stars which do not host planets.
The latter, using a 
different sample, showed that there was strong evidence for a planet - 
giant star metallicity correlation.

{\bf LITHIUM POLLUTION?}
\ \ \ \
One case of particular interest is the BD+48 740 system \cite{adaetal2012},
which contains both (i) a candidate planet with an unusually high planet eccentricity
($e = 0.67$) compared to other evolved star planets, and (ii) a host star that 
is overabundant in lithium.  Taken together, these features are suggestive
of recent planet engulfment: \cite{liletal2014} wrote: 
``The first clear evidence of planet engulfment''.  
\cite{liletal2014} also helped confirm existence
of the MS planet Kepler-91b, a tidally-inspiraling extremely close-in planet 
with $r \approx 1.32R_{\star}$ and estimate that its fate might
soon (within 55 Myr) mirror that of the engulfed planet in
BD+48 740.

{\bf TRANSIT DETECTION PROSPECTS}
\ \ \ \
In principle, SBs may be detected by transits around GB stars.  The application of Eqs. 
\ref{eqprobtran}-\ref{eqdurtra} to this scenario is interesting because, as observed
by \cite{assetal2009}, despite the orders of magnitude increase in stellar radius from the MS
to GB phases, there is a corresponding increase in values of $a$ for surviving SBs.  Therefore,
the transit probability should not be markedly different.  The equation neglects however, 
just how small the transit depth becomes: $R_{\rm SB}^2/R_{\star}^2 \sim 10^{-5}$.

\subsubsection{GB debris discs}

Importantly, planets are not the only SBs that 
survive MS evolution.  \cite{bonetal2013} 
revealed the first resolved images of a debris
disc orbiting a 2.5 Gyr-old subgiant ``retired'' A 
star ($\kappa$ Coronae Borealis or $\kappa$ CrB),
although they could not distinguish between one belt from 20 to 220 au
from two rings or narrow belts at about 40 and 165 au.  This finding
demonstrates that either the structure survived for the entire main
sequence lifetime, or underwent second-generation formation.
This discovery was followed up with a survey of 35 other subgiant stars,
three of which (HR 8461, HD 208585 and HD 131496) exhibit infrared excess
thought to be debris discs \cite{bonetal2014}.
Taken together, these four disk-bearing GB stars suggest that 
large quantities of dust could survive MS evolution.

\subsection{Putative planets in post-common-envelope binaries} \label{secOMPCEB}

Some binary stars which have already experienced a CE phase are
currently composed of (i) either a WD or hot subdwarf star, plus 
(ii) a lower-mass companion.  These binaries are classified as either
``detached'' or ``semi-detached cataclysmic variables'' depending on the value
of $r_{\rm b}$. Over 55 of these binaries have just the right 
orientation to our 
line-of-sight to eclipse one another.  These systems are known
as post-CE binaries. The eclipse times in these
binaries should be predictable if there are no other bodies
in the system and if the stars are physically static objects.

In several cases, this idealized scenario has not been realized,
allowing for the exciting possibility of exoplanet detections.
\cite{zorsch2013} reviewed the potential origin of eclipse
timing variations for all known post-CE binaries (see their Table 1
and Appendix A), and emphasized
that ``extreme caution'' must be exercised when evaluating a
first-generation origin for these putative planets.  The reason
is that binary stars are complex structures and can mimic
planetary signals.  Of the 12 systems highlighted in that table
for potentially hosting planets or brown dwarfs, the existence of these SBs
remain in doubt due to
stability analyses (see Sec. \ref{secMulti}\ref{secMulti2sNp0a}).  
For SBs which are thought to exist, their
dynamical origin remains in question. Sec. 5 of \cite{zorsch2013}
discusses the possibility of second-generation formation; see also 
Sec. \ref{secFall} of this paper.

The most robust detections are the putative planets around the post-CE binary
NN Serpentis, or NN Ser. \cite{beuetal2010} and \cite{beuetal2013} found 
excellent agreement with a
two-planet fit, and \cite{beuetal2013} found resonant
solutions with true librating angles.  A recent analysis of 25 yr
worth of eclipse timing data for this system (see Fig. 2 of 
\cite{maretal2014}) strengthens the planetary hypothesis, particularly
because timings between the years of 2010-2015 matched the predicted
curve.  They do not, however, claim that these planets are confirmed,
because there still exists a degeneracy in the orbital solutions.
In fact, \cite{musetal2013} provided evidence against the first-generation nature
of these planets by effectively
backwards integrating in time to determine if the planets could have
survived on the MS.  The largest uncertainty in their study is
not with the orbital fitting but rather how the CE evolved 
and blew off in that system.

If confirmed, other reported systems which may be dynamically stable
would prove exciting.  However,
pulsation signals, which are intrinsic to the parent star, can mask 
timing variations that would otherwise indicate the presence of SBs
(e.g. \cite{heretal2015}).
\cite{chaetal2011} detected signals around the 
hot subdwarf star KIC 05807616 which could correspond to SBs
with distances of 0.0060 and 0.0076 au.  \cite{siletal2014}
instead detected timing variations around the subdwarf 
pulsator KIC 10001893 with periods of 5.3, 7.8 and 19.5 hours.

\subsection{Pulsar planets} \label{secOMpulplan}

Pulsar arrival timings in systems with a single pulsar are generally better
constrained than those in post-CE binaries because pulsars are more
reliable clocks than WD or MS stars, and radiation from a single 
source simplifies the interpretation.  Identifying the origin of
residual signals for millisecond pulsars is even easier.

These highly precise cosmic clocks, combined with a fortuitous spell
of necessary maintenance work on the Arecibo radio telescope,
provided Alexander Wolszczan with the opportunity to find
PSR B1257+12 c and d \cite{wolfra1992} (sometimes known as 
PSR B1257+12 B and C), and eventually later
PSR B1257+12 b (sometimes known as PSR B1257+12 A)\cite{wolszczan1994}. 
\cite{wolszczan2012} detailed the
history of these discoveries.  Not included in that history
is how the identities of the first two planets were almost 
prematurely leaked by the British newspaper 
{\it The Independent} in October 1991 
\cite{watts1991}.  The article referred to these planets as
``only the second and third planets to be found outside 
our own solar system'' because of the ironic assumption
that the first-ever exoplanet was the (later-retracted \cite{lynbai1992}) 
candidate PSR 1829-10 b.

The observed and derived parameters for the PSR B1257+12 system remain
amongst the best constrained of all exoplanets (see Table 1 of
\cite{wolszczan2012}), with eccentricity precision to the $10^{-4}$
level, a value for the mutual inclination of planets
c and d (which is about 6 degrees), and a derived mass for planet b of 
$0.020M_{\oplus} = 1.6M_{\rm Moon}$. However, these values are based
on the assumption that $M_{\star} = 1.4M_{\odot}$. PSR B1257+12 b has the smallest 
mass of any known extrasolar SB, given that
we do not yet have well-constrained masses for the (disintegrating) 
SBs orbiting WD 1145+017.  The other two planets are ``super-Earths'',
with masses of planets c and d of $4.3M_{\oplus}$ and $3.9M_{\oplus}$.
All three planets lie within 0.46 au
of the pulsar and travel on nearly circular ($e < 0.026$) orbits.  The orbital
period ratio of planets b and c is about 1.4759, which is close
to the $3$:$2$ mean motion commensurability. 

I have already described the one other bona-fide planet discovered 
orbiting a millisecond pulsar,
PSR B1620-26AB b, in Sec \ref{secOM}\ref{secOMWDplan}\ref{secOribitingWDs},
because that planet also orbits a WD.
For an alternate and expanded accounting of this planet as well as the
PSR B1257+12 system, see the 2008 review of neutron star research
by \cite{wolszczan2008}.  

Pulsar planets are rare.  A recent search of 151 young ($< 2$ Myr-old) 
pulsars (not ms pulsars) with the {\it Fermi} telescope yielded no planets 
for $M_{\rm SB} > 0.4M_{\oplus}$ and with orbital periods of under 1 year
\cite{keretal2015}.  The authors used this result as strong 
evidence against post-SN fallback second-generation
discs (see Sec. \ref{secFall}\ref{secFallSN}), particularly because 
theoretical models constrain these 
discs to reside within about 2 au.

\subsection{Circumpulsar asteroids and discs signatures} \label{secOMpuldisc}

Sometimes the deviations in pulse timing are not clean in the sense
that they cannot be fit with one or a few planets or moon-sized SBs.
Rather, the deviations may be consistent with other structures, such
as discs, rings, arcs or clouds.  \cite{wang2014} recently reviewed
observational results from debris disc searches around single pulsars.

The residuals for
the millisecond pulsar B1937+21 are consistent with an
asteroid disc of mass less than about $0.05M_{\oplus}$ \cite{shaetal2013},
and with constrained properties given in their Table 2.  Asteroids
can affect the timing precision of received signals from
millisecond pulsars down to the ns level.
Their Fig. 5 nicely compares their sensitivity limits for asteroids
in PSR B1937+21 with the PSR B1257+12 planets.  Unfortunately, the
asteroids interpretation is degenerate with other possibilities,
and difficult to test (see their Sec. 7).

Previously, \cite{lazetal2004} and \cite{lohetal2004} placed limits on the 
masses of dust discs in other millisecond pulsars.  However, unlike for 
PSR B1937+21, these other pulsars did not have timing residuals that were 
fit to specific asteroid belt or disc architectures \cite{shaetal2013}.
\cite{bryetal2006} searched for a dust disc in PSR B1257+12,
but were not able to exclude the presence of a Solar 
system-like asteroid belt with a mass as large as $0.01 M_{\oplus}$ and SBs with
radii up to 100 km.   The X-ray pulsar or young magnetar
4U 0142+61 might host a disc \cite{wanetal2006,wanetal2008}, although
the infrared excess in that system instead could be explained by magnetospheric
emission \cite{beltho2007}. Also, 
anomalous timing and radio emission signatures of the
pulsar PSR J0738-4042 \cite{broetal2014} cannot be explained by known stellar evolutionary
processes. Such abrupt changes can be caused by encounters with asteroids,
as argued by the authors.

\cite{corsha2008} expect asteroids to enter pulsar magnetospheres 
and create the largest detectable signals for large spin periods, 
large spin-down ages,
large magnetospheres, low surface temperatures, and low nonthermal
luminosities (see their Fig. 1).  They also posed that 
interstellar comets impacting with circumpulsar discs
may produce observable episodic behaviour.  They stated that for an au-scale 
dense disc with a high optical thickness, events may occur at the rate
of once per year.   \cite{mitrofanov1990} postulated that these events
may produce gamma ray bursts and prompt a ``re-ignition''
of the pulsar.

\section{Stellar mass ejecta} \label{secMass}

Having motivated the study of post-MS planetary systems, I now turn
to important forces in these systems.
Stellar mass loss is arguably the most important driver of all post-MS forces
because it non-negligibly affects the orbits of all SBs at 
all distances (see Fig. \ref{FigForces}).  In this vein, the 
classical mechanics-based mass-variable
two-body problem has gained renewed importance with the mounting
discoveries of post-MS planetary systems.  For decades, this
same problem has been relevant and applicable to binary stars, 
which represented the physical picture envisaged by early (pre-exoplanet era)
studies.

The mass variable two-body problem is dynamically rich and
not energy conserving.  The problem can be broken down into the following
questions: (i) How many bodies are losing mass? (ii) Is the mass loss isotropic? 
(iii) Is the mass loss instantaneous?  (iv) Does a natal kick accompany the mass loss?
(v) Are the bodies assumed to be point masses? (vi) How are one or both bodies 
dragged through the ejecta?  These questions are not mutually exclusive, complicating 
a straightforward presentation. I proceed by grouping the first four 
questions into Sec. \ref{secMass}\ref{secMassnonacc}, the fifth question into 
Sec. \ref{secMass}\ref{secMassacc}, and
the last question into Sec. \ref{secMass}\ref{secMasswind}.

\subsection{The mass-variable point-mass two-body problem} \label{secMassnonacc}

Published investigations of the problem itself date 
back to the late 19th century: \cite{razbitnaya1985} provided 
a brief historical review.  Table 1 of that paper exhaustively 
lays out the equations of motion in 22 different cases, depending
on the reference frame, relative velocities, whether one or both
of the masses is variable, and whether the mass loss is isotropic.
In all cases, both bodies are assumed to be point masses.
For most post-MS planetary science applications, when the SB mass
is considered fixed, rows \#6 (anisotropic mass loss) and 
\#7 (isotropic mass loss) of that table provide the relevant equations.
When the SB itself is shedding mass (perhaps due to atmospheric evaporation;
see Sec. \ref{secRad}\ref{secRadGB}\ref{secPlanAtm} of this paper) or when 
its accreting the mass ejected from the star, then the equations in 
rows \#1 (anisotropic mass loss) and \#2 (isotropic mass loss) should 
instead be used.

\subsubsection{Isotropic mass loss setup}

In the most general point-mass isotropic mass loss case where
both bodies are losing mass (row \#2),
the equation appears to be no different from the fixed masses case:

\begin{equation}
\frac{d^2\vec{r}}{dt^2}
=
-\frac{G \left(M_{\star} + M_{\rm SB}\right) }{r^3} \vec{r}
.
\label{eqml1}
\end{equation}

\noindent{}What is different is that both masses are functions of time.
In order to help understand the implications and provide physical
intuition for their meaning, John D. Hadjidemetriou and Tuken B. Omarov
independently, and using different approaches in different languages, derived 
in 1962-1963 the
corresponding equations of motion for isotropic mass loss 
in planetary orbital elements \cite{omarov1962,hadjidemetriou1963}.
The lack of internet and political considerations in the early 1960s prevented 
the eventual good friends from knowing about each other's work until long after.

Both Hadjidemetriou and Omarov realized that Eq. \ref{eqml1} may be expressed as

\begin{equation}
\frac{d^2\vec{r}}{dt^2}
=
-\frac{G \left(M_{\star}(t=0) + M_{\rm SB}(t=0)\right) }{r^3} \vec{r}
-\frac{1}{2\left(M_{\star}(t) + M_{\rm SB}(t)\right)}
\frac{d\left(M_{\star}(t) + M_{\rm SB}(t) \right)}{dt} \frac{d\vec{r}}{dt}
\label{eqml2}
\end{equation}

\noindent{}such that the classic orbitally static two-body problem 
is perturbed by a single mass loss term.  Isolating this term helps
quantify the motion. Equation \ref{eqml2} is derived from 
Eq. \ref{eqml1} by determining the total derivative of velocity
with respect to time.  This derivative is the sum of the partial 
derivative with $\left(M_{\star} + M_{\rm SB}\right)$ held constant, plus the
partial derivative with $r$ held constant. The results of their derivations
of the equations of motion entirely in terms of planetary 
elliptical orbital elements are as
follows.  I write them in terms of both true anomaly $f$ and eccentric anomaly $E$:

\begin{eqnarray}
\frac{da}{dt} &=& 
-\frac{a \left(1 + e^2 + 2 e \cos{f}\right)}{1 - e^2}
\frac{\dot{M}_{\star} + \dot{M}_{\rm SB}}{M_{\star} + M_{\rm SB}}
=
-a\left(\frac{1+e\cos{E}}{1 - e\cos{E}} \right)
\frac{\dot{M}_{\star} + \dot{M}_{\rm SB}}{M_{\star} + M_{\rm SB}} 
,
\label{eqml3a}
\\
\frac{de}{dt} &=& -\left(e + \cos{f} \right)
\frac{\dot{M}_{\star} + \dot{M}_{\rm SB}}{M_{\star} + M_{\rm SB}} 
\ \ \ \ \ \ \ \ \ \ \ \ \ \ \ \ \ \ 
=
-\left[ \frac{\left(1 - e^2\right) \cos{E}}{1 - e \cos{E}} \right]
\frac{\dot{M}_{\star} + \dot{M}_{\rm SB}}{M_{\star} + M_{\rm SB}}
,
\label{eqml3e}
\\
\frac{di}{dt} &=& 0
\ \ \ \ \ \ \ \ \ \ \ \ \ \ \ \ \ \
\ \ \ \ \ \ \ \ \ \ \ \ \ \ \ \ \ \ \ \ \ \
\ \ \ \ \ \ \ \ \ \ \ \ \ \ \ \ \ \ \ \
= 0
,
\label{eqml3i}
\\
\frac{d\Omega}{dt} &=& 0
\ \ \ \ \ \ \ \ \ \ \ \ \ \ \ \ \ \
\ \ \ \ \ \ \ \ \ \ \ \ \ \ \ \ \ \ \ \ \ \
\ \ \ \ \ \ \ \ \ \ \ \ \ \ \ \ \ \ \ \ 
= 0
,
\label{eqml3bigO}
\\
\frac{d\omega}{dt} &=& -\left(\frac{\sin{f}}{e}\right) \frac{\dot{M}_{\star} + \dot{M}_{\rm SB}}{M_{\star}+M_{\rm SB}} 
\ \ \ \ \ \ \ \ \ \ \ \ \ \ \ \ \ \
\ \ \ \,
=
-\left[ \frac{\sqrt{1-e^2} \sin{E}}{e \left(1 - e \cos{E} \right) } \right]
\frac{\dot{M}_{\star} + \dot{M}_{\rm SB}}{M_{\star} + M_{\rm SB}}
.
\label{eqml3smallo}
\end{eqnarray}

\noindent{}Here $\Omega$ and $\omega$ refer to the longitude of ascending node
and argument or pericentre. In order to complete the sets, the evolution of the anomalies themselves are

\begin{eqnarray}
\frac{df}{dt} &=& -\frac{d\omega}{dt} + \frac{n \left(1 + e \cos{f}\right)^2}{\left(1 - e^2\right)^{\frac{3}{2}}}
,
\label{eqml3f}
\\
\frac{dE}{dt} &=& -\frac{1}{\sqrt{1-e^2}} \frac{d\omega}{dt} + \frac{n}{1 - e \cos{E}}
.
\label{eqml3E}
\end{eqnarray}

\noindent{}In these equations, 
$n =  G^{1/2}(M_{\star} + M_{\rm SB})^{1/2}a^{-3/2}$ represents the mean motion.  
Between the mid-1960s and the 2010s, these relations have not appeared often 
in other literature. One notable example is \cite{aleetal1976}, whose 
Eqs. 3.11-3.13 can be reduced to the above relations. Equation \ref{eqml3a} 
illustrates that the semimajor axis always increases due to isotropic mass loss.
Together Eqs. \ref{eqml3a} and
\ref{eqml3e} may be combined to show that the orbital pericentre also always increases.
The argument of pericentre is not defined for circular orbits, and changes rapidly
for near-circular orbits.

As observed by \cite{adaetal2013}, a danger of using osculating Keplerian orbital 
elements is the interpretation of their non-Keplerian time evolution: for example, 
apparent oscillations in $e$ can mask the motion's actual smooth spiral.
\cite{iorio2010} further emphasised the importance of considering the time evolution
of all the orbital elements concurrently to obtain a true picture of the motion.
An alternate way of describing the motion is through a radial 
equation of motion \cite{jeans1924}.  \cite{adaetal2013} recently 
developed a dimensionless form of this equation 
and derived resulting relations for time-dependent mass-loss 
prescriptions in this framework.

\subsubsection{Anisotropic mass loss setup} \label{SecMassAniso}

Equations \ref{eqml3a}-\ref{eqml3E} assume that the mass is lost isotropically.  
If one removes this assumption, and models direction-dependent mass loss and 
differential mass ejecta speeds, then additional terms appear in the equations.  
These terms are given by Eqs. 34-38 of \cite{verhadtou2013}
\footnote{
Appropriately, after a distinguished career pioneering the dynamical field of periodic orbits, 
Hadjidemetriou returned to the topic of mass loss, and in his final paper \cite{verhadtou2013}
helped derived the anisotropic mass loss equations of motion and applied them to pre- and post-MS
planetary systems.  Confirmation of some of these equations can be found in Omarov's
work \cite{omarov1962} in his Eq. 
13. For English-language context, see also Eqs. 19-24 of \cite{omarov1964}.}
and are expressed in 
terms of mass flux relations from their Eq. 4.  These modified equations 
demonstrate that (i) the inclination and longitude of ascending node do change 
with time, and that (ii) anisotropy becomes more important the further away an SB is, 
as the anisotropic terms contain an additional factor of $\sqrt{a}$.  
\cite{verhadtou2013} proved that mass-loss variations must be asymmetric about 
the stellar equator in order to affect the motion of SBs.  They showed 
that for most post-MS applications with SBs within a few hundred au, 
isotropy is an excellent approximation.

Although motivated by atmospheric evaporation on the MS, the 
mass loss formalisms presented by \cite{iorio2012} and 
\cite{bouetal2012} provide potentially useful alternative 
approaches. \cite{iorio2012} expressed the mass loss term in
radial and tangential components, and then applied the Gaussian
equations in that coordinate system; a variation on Omarov's
approach \cite{omarov1962}. \cite{bouetal2012} instead treated radial 
mass loss from a solid angle patch of the atmosphere of one body
(different from the one-dimensional jets from \cite{verhadtou2013})
so that the mass emanates in a conical shape.  

\cite{paralc1998} modeled anisotropy of mass loss in a different manner:
through a kick, or a velocity impulse, of the star. 
Their equations of motion (in their Eq. 1, which must contain a typographic
error) include a term
which requires a-priori knowledge of the star's final (WD) mass
given a velocity impulse prescription.

\subsubsection{Isotropic non-instantaneous mass loss solutions}

\cite{veretal2011} extensively analyzed Eqs. \ref{eqml3a}-\ref{eqml3E}, 
and observed that they can be derived via either perturbation 
techniques or the vis-viva equation, combined with conservation 
of angular momentum expressed in polar coordinates. They also 
found importantly that the osculating pericentre monotonically
increases with mass loss (their Eq. 21), regardless of the mass 
loss prescription chosen. Although early studies like 
\cite{hadjidemetriou1966} used the equations to show that orbital 
eccentricity changes are triggered by high mass loss, even for an 
initially circular orbit, the mounting observations
of post-MS systems have led to a renewed interest in this problem.

{\bf PARAMETERISING SOLUTIONS}
\ \ \ \
Equations \ref{eqml3a}-\ref{eqml3E} 
do not appear to have a complete analytical solution.  They do 
however admit two well-defined solution regimes 
and a difficult-to-characterize transition regime.  In order to 
identify these solution regimes, consider the (dimensionless) 
scaled ratio of the timescales of orbital period to mass loss

\begin{equation}
\Psi 
\equiv
\frac
{\dot{M}_{\star}}
{M_{\star}n}
= 
\frac{1}{2\pi}
\left( \frac{\dot{M}_{\star}}{M_{\odot} \ {\rm yr}^{-1}} \right)
\left( \frac{M_{\star}}{M_{\odot}} \right)^{-\frac{3}{2}}
\left( \frac{a}{\rm au}  \right)^{\frac{3}{2}}
.
\label{eqpsi}
\end{equation}

\noindent{}When $\Psi \ll 1$, the system is commonly said to be in the ``adiabatic'' regime.
When $\Psi \gg 1$, the system is in a regime without a widely-used moniker.  
The transition regime lies in-between these two 
extremes, and $\Psi$ itself changes with time (Eq. 16 of \cite{veretal2011}) so that if 
mass loss continued indefinitely, the
system would enter the $\Psi \gg 1$ regime from the adiabatic regime.  The term ``adiabatic'' 
in this context does not refer to heat, but rather
conservation of eccentricity.  The reason is that when $\Psi \ll 1$, the right-hand-sides
of Eqs. \ref{eqml3e} and \ref{eqml3smallo} may be approximated as zero.  Further,  
in this limit, Eq. \ref{eqml3a} simplifies 
to: $da/dt = -(M_{\star} + M_{\rm SB})^{-1}(dM_{\star}/dt)$.

As indicated by the maximum mass loss rates in Fig. \ref{FigSpec} and Eq. \ref{eqpsi}, 
$\Psi \ll 1$ for the vast majority of known exoplanets (because $a < 100$ au and  
$M_{\star}^{({\rm MS})} \le 3M_{\odot}$ in almost every case).  Consequently, these 
exoplanets will expand,
but not stretch, their osculating orbits during post-MS mass loss by a factor approximately equal
to $(M_{\star}^{({\rm MS})}/M_{\star}^{({\rm WD})})$.  During this process, $\Psi$ remains much less
than unity.

{\bf CRITICAL ANGLES}
\ \ \ \
However, exo-Oort clouds, exo-scattered discs, and some exoplanets (like WD 0806-661 b 
from \cite{luhetal2011}) have a more complex post-MS fate.  For these systems, either $\Psi \sim 1$
or $\Psi \gg 1$.  In the latter case, special solutions of the equations of motion exist,
as outlined in Sec. 2.6.2 of \cite{veretal2011}.  Related fundamental properties of the two-body
problem derived in Sec. 2.5 of \cite{veretal2011} are the two critical true anomalies $f_{\rm crit}$

\begin{equation}
f_{\rm crit} \equiv \pi \pm \frac{1}{2} \cos{\left(\frac{7}{9}\right)}
\ \ \ \approx 160.5^{\circ} \ {\rm and} \ 199.5^{\circ}
\end{equation}

\noindent{}at which the orbital eccentricity incurs an initial decrease at the onset of mass
loss.  Consequently, these values
mark the boundaries at which some eccentric SBs for $\Psi \gg 1$ are protected from escape
(Fig. 16 of \cite{veretal2011} and Figs. 9-10 of \cite{verwya2012}).

{\bf TRANSITION TO NON-ADIABATICITY}
Based on angular momentum conservation, nonadiabatic motion does not ensure that the expansion 
of the semimajor axis proceeds faster than the adiabatic rate unless the orbit
is initially circular (see Eqs. 1-3 of \cite{musetal2013}). The transition regime can 
be wide in extent, spanning an order of magnitude in $a^{({\rm MS})}$
(see e.g. the bottom two panels in Fig. 10 of \cite{verwya2012}).
In the top panel the mass loss is strong enough to ensure $\Psi \gg 1$.
For the most
nonadiabatic of all the plots in that top panel (bottom-rightmost plot), 
the analytic values of $f_{\rm crit}$ well-predict which 
systems eventually feature escape. 

\cite{debsig2002} mentioned
an important caveat to the adiabatic transition: the mass loss rate
(from the orbit, and not the star) is dictated by the
asymptotic wind velocity and not the ejection velocity.  If, 
at Oort cloud distances, this speed is several orders of magnitude 
less than the escape speed, then the lower wind
crossing time effectively decreases the mass loss rate.  Consequently,
$\Psi$ might decrease enough for the evolution to still be adiabatic.

{\bf MASS LOSS PRESCRIPTIONS}
\ \ \ \
Up until now, I have not assumed a particular analytic form for the
mass loss prescription.  Real stars do not follow simple prescriptions,
particularly if they lose significant mass on both the RGB and AGB
and pulse violently.
Regardless, separately each branch has been fit to linear or exponential
forms.  Very early work from the 1890s \cite{mestschersky1893,mestschersky1902} assumed 
dependencies with time.  In 2009, \cite{rahetal2009} summarized 
(in their Sec. 3) different mass-loss prescriptions used throughout the literature.

More recently, Fig. 4 of \cite{voyetal2013} provided a comparison of
the eccentricity excitation due to both linear and exponential
mass loss, supposing the same total amount of mass lost.
\cite{adaetal2013} adopted a dimensionless mass loss prescription
index dependence, and used it to derive an explicit criterion for SB escape 
(see their Eq. 86).

{\bf A HAMILTONIAN FRAMEWORK}
\ \ \ \
\cite{rahetal2009} also treated the point-mass two-body 
non-instantaneous mass loss problem
in the Hamiltonian framework. In their Sec. 2, they provided 
a brief historical review of the problem with many references
not listed here.  One such reference is the classically entertaining paper
of \cite{deprit1983}, which discusses, among other topics, time-dependent
Delaunay transformations.

\subsubsection{Isotropic instantaneous mass loss solutions}

Now consider the limit $\Psi \rightarrow \infty$, which is relevant
(although technically not ever realized) for SNs.  An advantage of modeling
``instantaneous'' or ``impulsive'' mass loss is that no assumptions
about mass loss rates nor wind speeds are necessary: \cite{hadjidemetriou1966}
showed in this case how the stability of these system crucially
depends on the true anomaly at mass loss, with periastron and
apastron being the limiting cases.  Nevertheless,
a SN process often creates a velocity ``kick'' from asymmetries within the exploding
star.  The link between the mass lost and
the magnitude and direction of the kick is unclear.  Therefore, one best
treat the general case of impulsive mass lost plus an arbitrary kick velocity
equal to
$\vec{v}^{({\rm NS})} - \vec{v}^{({\rm MS})}$,
which reduces to the mass-loss only situation in the limit of a zero kick velocity.

Early treatments of this problem \cite{hills1970,michel1970} predate 
confirmed exoplanets, and were motivated by a potential pulsar planet at the time.
Both studies, in 1970, considered the effect of instantaneous mass loss on
a planet's orbit due to a SN.  Only 13 years later, however, was a comprehensive
and general treatment presented \cite{hills1983}.  

{\bf EXPLICIT EXPRESSIONS}
\ \ \ \ 
\cite{hills1983} used angular momentum and energy conservation, and the assumptions
of instantaneous mass loss from a point-mass MS star, to provide explicit
expressions for $a^{({\rm NS})}$, $e^{({\rm NS})}$, and the critical value of 
$M_{\star}^{({\rm NS})}/M_{\star}^{({\rm MS})}$ for which the SB escapes, in terms of
only MS orbital parameters (including the true anomaly $f$). The expressions 
for the semimajor axis and mass ratio
include an arbitrary kick, but the eccentricity expression does not because an 
additional degree of freedom would have to be introduced (see his Section IIIb).
By repackaging his Eq. 13, including an arbitrary kick, I obtain

\begin{eqnarray}
&a^{({\rm NS})}&
=
\left[
a^{({\rm MS})}
\frac{M_{\star}^{({\rm NS})} + M_{\rm SB}}{M_{\star}^{({\rm MS})} + M_{\rm SB}}
\right]
\nonumber
\\
&\times& \Bigg[
1 - \frac{2 \left(1 + e^{({\rm MS})} \cos{f^{({\rm MS})}} \right)}{1 - {e^{({\rm MS})}}^2}
\left( 
\frac{M_{\star}^{({\rm MS})} - M_{\star}^{({\rm NS})}}{M_{\star}^{({\rm MS})} + M_{\rm SB}}
\right)
- \frac{a^{({\rm MS})} \left({v^{({\rm NS})}}^2 - {v^{({\rm MS})}}^2 \right)  }{G \left(M_{\star}^{({\rm MS})} + M_{\rm SB} \right)} 
\Bigg]^{-1}
,
\nonumber
\\
&&
\label{eqainst}
\end{eqnarray}

\noindent{}which I have written in a form that illustrates how the adiabatic approximation
(first square brackets) is modified.  The SB will escape the system if, including
an arbitrary kick (Eq. 15 of \cite{hills1983})

\begin{equation}
M_{\star}^{({\rm MS})} - M_{\star}^{({\rm NS})} > 
\left[
\frac{M_{\star}^{({\rm MS})} + M_{\rm SB}}{2}
\right]
\left(
\frac{1 - {e^{({\rm MS})}}^2}
     {1 + e^{({\rm MS})} \cos{f^{({\rm MS})}}}
\right)
\left( 
1 - \frac{a^{({\rm MS})}\left({v^{({\rm NS})}}^2 - {v^{({\rm MS})}}^2 \right)}
{G \left(M_{\star}^{({\rm MS})} + M_{\rm SB}\right)} 
\right)
.
\label{eqminst}
\end{equation} 

\noindent{}I write Eq. \ref{eqminst} in a form which illustrates how the oft-used notion that
half of the original mass (in square brackets) must be lost for escape to occur is modified
by other parameters.  For high values of $e^{({\rm MS})}$, this modification may be 
severe (see Fig. 12 of \cite{veretal2011}). In Eqs. \ref{eqainst}-\ref{eqminst}, the
expression 
${v^{({\rm NS})}}^2 - {v^{({\rm MS})}}^2$ may be replaced by
an expression with the kick velocity, kick angle and $v^{({\rm MS})}$.
The final eccentricity, without an arbitrary kick, is the following repackaged form of their Eq. 6:

\[
{e^{({\rm NS})}}^2 = 1 - \left(1 - {e^{({\rm MS})}}^2  \right)
\ \ \ \ \ \ \ \ \ \ \ \ \ \ \ \ \ \ \ \ \ \ \ \ \ \
\ \ \ \ \ \ \ \ \ \ \ \ \ \ \ \ \ \ \ \ \ \ \ \ \ \ 
\ \ \ \ \ \ \ \ \ \ \ \ \ \ \ \ \ \ \ \ \ \ \ \ \ \
\ \ \ \ \ \ \ \ \ \ \ \ \ \ \ \ \ \ \ \ \ \ \ \ \ \ 
\]

\vspace{-0.4cm}
\begin{equation}
\times
\left[
\frac
{M_{\star}^{({\rm NS})} + M_{\rm SB}}
{M_{\star}^{({\rm MS})} + M_{\rm SB}}
\right]^{-2}
\left[
1 
- 2 \left( \frac{M_{\star}^{({\rm MS})} - M_{\star}^{({\rm NS})}}
                {M_{\star}^{({\rm MS})} + M_{\rm SB}}  \right)
    \left(  
           \frac{1 + e^{({\rm MS})} \cos{f^{({\rm MS})}}}
                {1 - {e^{({\rm MS})}}^2 }
    \right)
\right]
.
\label{eqeinst}
\end{equation}

{\bf MODERN RE-DERIVATIONS}
Later studies have attempted to reproduce special cases of
Eqs. \ref{eqainst}-\ref{eqeinst}, without kicks,
and typically under the (usually secure) assumption that
$M_{\rm SB} \ll M_{\star}$. \cite{veretal2011} provided the 
mass ratio boundedness condition in this case in their Eq.
48, except the sign of their inequality should be flipped.
Even with the above assumption, the expressions for the
mass ratio and final eccentricity in Eqs. 1-2 of
\cite{voletal2014} do not appear to exactly agree with a reduction
of the equations from \cite{hills1983}.  Other, alternative 
formulations of the orbital evolution resulting from SN kicks have 
been developed (see, e.g. Appendix A of \cite{huretal2002}).

Equations \ref{eqainst}-\ref{eqeinst} are useful because they are explicit
closed equations for which functional dependencies can quickly be read off.
However, if one wished to apply impulsive mass loss together with a kick in
an arbitrary Cartesian direction, and derive the post-kick orbital elements,
one could instead use the relations in Appendix B of \cite{stoetal2015}. 
In the most general case, where the system is in an arbitrary frame,
the equations in Appendix A of \cite{jacetal2014}, along with those in their
Sec. 2, provide a starting point for deriving explicit expressions of
post-kick elements.

\subsection{The mass-variable solid body two-body problem} \label{secMassacc}

So far, the vast majority of post-MS planetary investigations have
treated orbital changes due to stellar mass loss assuming that the SB
is a point mass.  However, as models become more detailed and 
take into account physical changes to SBs,
one may wish
to lift the point mass assumption.

\subsubsection{Orbital evolution}

If the SB is treated as a solid body, then it will increase in mass through
accretion by the stellar wind.  Consequently, 
Eq. \ref{eqml1} is no longer strictly usable.  Because mass 
is no longer ejected from the orbit in the direction of the SB,
the mass loss becomes anisotropic, leading to more complex
equations (Sec. \ref{secMass}\ref{secMassnonacc}\ref{SecMassAniso}).  
\cite{hadjidemetriou1963} derived compact expressions
in his Eq. 44 for how the orbital motion would change in this case,
but assumed that the ejecta was moving at the speed of light,
and did not account for gravitational focusing.  In any case,
incorporating solid body accretion into the orbital equations of motion 
would need to account for both extra terms in the equations of motion
and the time-dependent change in the SB mass.

\subsubsection{Physical evolution}

The level of accretion onto the SB itself might be significant.  Small enough SBs
like pebbles might accrete an amount of mass so great that the pebble
might be transformed into a boulder, or just be destroyed.  Large SBs
like planets with atmospheres could have the composition and nature of those
atmospheres permanently altered.

What is the accretion rate onto the SB?  The literature is replete with 
answers to this question.  Many answers
differ depending on whether the authors assumed, for example, that gravitational
focusing is important, the SB is on a circular orbit, and the wind speed
is much greater than $v$.  Some expressions include an explicit dependence
on $\dot{M}_{\star}$, whereas others instead show a dependence on 
$\rho_{\rm wind}$.  These two important quantities are often related
through a constant-velocity spherically symmetric (isotropic) mass 
loss assumption, and when the 
``ejected mass is stratified in concentric spherical layers of 
constant density'' \cite{hadjidemetriou1966}.  Consequently (his Eq. 2)

\begin{equation}
\rho_{\rm wind} = \frac{\dot{M}_{\star}}{4 \pi r^2 \left|v_{\rm wind}\right|}
.
\label{eqwind}
\end{equation}

\noindent{}By using Eq. \ref{eqwind}, one may compare different
expressions for $\dot{M}_{\rm SB}$ which appear in the literature.
These include Eq. 2 of \cite{dunlis1998}, Eqs. 1-2 of \cite{jura2008},
Eq. 1 of \cite{villiv2009}, Eq. 11 of \cite{pasetal2012} and 
Eq. 4 of \cite{spimad2012}.  The formulation from \cite{huretal2002} 
is expressed in terms of $e$: repackaging their Eqs. 6-9 gives

\begin{equation}
\left\langle  
\frac{dM_{\rm SB}}{dt}
\right\rangle
=
-
\frac{dM_{\star}}{dt}
\frac{3}{4 a^2 \sqrt{1-e^2}}
\left( 
\frac{G M_{\rm SB}}{v_{\rm wind}^2}
\right)^2
\left[
1 + \frac{G^2\left(M_{\star} + M_{\rm SB}\right)^2}
{a^2 v_{\rm wind}^2}
\right]^{-3/2}
.
\end{equation}

By using expressions for the accretion rate onto the SB, 
\cite{dunlis1998}, \cite{jura2008} and 
\cite{spimad2012} then estimate the total mass
accreted by the SB.  However, doing so necessitates integrating with respect
to time, and $\dot{M}_{\star}$ is a nonmonotonic function of time, initial
mass, and metallicity along the GB.  Nevertheless, \cite{spimad2012} found 
that brown dwarfs and large planets could accrete
an amount of mass that is a multiple of their MS atmospheric masses.

\subsection{Stellar wind/gas/atmospheric drag} \label{secMasswind}

Up until now, I have neglected the physical interaction between the 
ejected stellar mass and the SB.  This assumption might be good for large
SBs like planets, and/or tenuous stellar winds.  However, for typical GB stars,
the interaction between the stellar wind and smaller SBs like 
pebbles or asteroids may be significant.  I also include here the treatment 
of SBs within stellar atmospheres.  Although strictly not under the umbrella of
``stellar ejecta'', stellar atmospheres provide a similar 
medium within which drag takes place, just with a (typically) higher density
than the stellar wind.  Both the wind and atmosphere are considered
to be gaseous media.

\subsubsection{Physical evolution}

Sufficiently small SBs are in danger of catastrophically disrupting due to
ram pressure.  For spherical SBs which are hydrostatic, isothermal, 
and self-gravitating, Eq. 1 of \cite{pasetal2012} provides the 
following condition for stability against disruption:

\begin{equation}
g_{\rm SB} 
\lesssim 
\frac
{2 \pi  \rho_{\rm SB} \rho_{\rm wind} v^2} 
{R_{\rm SB} \left(\rho_{\rm SB}^2 - \rho_{\rm wind}^2 \right)}
.
\end{equation} 

\noindent{}where $g$ is surface gravity. They then specialize 
to the case $\rho_{\rm SB} \gg \rho_{\rm wind}$
and $g \approx GM_{\rm SB}/R_{\rm SB}^2$. 
These simplifications
yield the following range of masses for which disruption would occur
(their Eq. 2).  I rewrite this equation first from the prescriptive
on atmospheric infall and then for GB winds outside of the atmosphere:

\begin{eqnarray}
M_{\rm SB} 
&\lesssim&
0.02 M_{\oplus}
\left(
\frac{\rho_{\rm wind}}{10^{-4}{\rm g} \ {\rm cm}^{-3}}
\right)^{\frac{3}{2}}
\left(
\frac{\rho_{\rm SB}}{2 \ {\rm g} \ {\rm cm}^{-3}}
\right)^{-2}
\left(
\frac
{v}
{100 \ {\rm km} \ {\rm s}^{-1}}
\right)^3
\label{eqdisr1}
\\
&\sim&
100 {\rm kg}
\left(
\frac{\rho_{\rm wind}}{10^{-16}{\rm g} \ {\rm cm}^{-3}}
\right)^{\frac{3}{2}}
\left(
\frac{\rho_{\rm SB}}{2 \ {\rm g} \ {\rm cm}^{-3}}
\right)^{-2}
\left(
\frac
{v}
{10 \ {\rm km} \ {\rm s}^{-1}}
\right)^3
.
\label{eqdisr2}
\end{eqnarray} 

\noindent{}Although approximate, Eq. \ref{eqdisr1} suggests that giant
planets and brown dwarfs should survive catastrophic disruption while 
inside stellar atmospheres, and Eq. \ref{eqdisr2} suggests that asteroids, 
comets, planets and brown dwarfs will easily survive stellar winds.
The situation for boulders and pebbles requires further analysis:
\cite{jura2008} assumes that an SB ``is destroyed if it encounters
its own mass in the wind'' due to hydrodynamic friction, whereas
\cite{donetal2010} claimed that when the drag force exceeds the 
parent star gravitational force, then the SB is ``entrained'' by the wind.

\subsubsection{Orbital evolution}

{\bf CONTEXT FOR DRAG}
\ \ \ \
Surviving SBs will have their orbits altered by interactions with the surrounding
medium.  The existing post-MS planetary literature has used a variety of
names and expressions for the resulting forces.
I group the forces into two categories: one consisting of a purely gravitational 
force due to the wake generated, and one due to friction.
I refer to the former as ``gravitational drag'' and the latter as
``frictional drag'', in line with \cite{villiv2009} and \cite{viletal2014}.
\cite{rybden2001} and \cite{wilwu2005} referred to the frictional drag as ``bow shock drag''.
\cite{schsmi2008} referred to the combined effects of gravitational drag
and frictional drag as ``dynamical friction''.
In combined expressions, drag terms containing $G$ typically refer to 
gravitational drag and those without $G$ refer to frictional drag.
\cite{schsmi2008} suggested that within the atmosphere,
if  $v \approx (2-3) v_{\rm sound}$, where $v_{\rm sound}$ is the sound speed,
then both gravitational drag and frictional drag play nearly equal roles.

{\bf GRAVITATIONAL DRAG}
\ \ \ \
However, gravitational drag is negligible outside of a stellar
atmosphere.  Consequently, the seminal pre-exoplanet era gravitational drag study 
of \cite{aleetal1976} defined a type of cutoff in their Eq. 2.1.
\cite{goldstein1987} claimed that this value is ``usually defined as the
neutral point between planet and star'', whereas \cite{livio1982}
took this value as ``the mean radius of the planet's Roche lobe'';
\cite{livio1982} went on to derive analytical relations with this assumption.
\cite{aleetal1976} then used their gravitational drag expression 
and parameterized $v_{\rm wind}$ into radial and tangential components in 
order to derive evolution equations in orbital elements (their Eqs. 3.11-3.13).
These equations also contain isotropic mass loss terms and are sufficiently general
to reduce to my Eqs. \ref{eqml3a}-\ref{eqml3smallo} in the appropriate limits.

In a seminal later study, \cite{ostriker1999} used time-dependent linear
perturbation theory to derive an expression for the gravitational drag force,
which she denotes the ``dynamical drag force''.  This expression has subsequently
been used for post-MS planetary system applications by \cite{schsmi2008},
\cite{villiv2009} and \cite{viletal2014}. For Mach numbers of interest within the
atmospheres of GB stars both Eq. 5 of
\cite{villiv2009} or Eq. 4 of \cite{viletal2014} give

\begin{equation}
F_{\rm drag}^{({\rm grav})} 
= 
\frac{2 \pi G^2 M_{\rm SB}^2 \rho_{\rm wind}}{v_{\rm sound}^2}
\label{eqdragfric}
\end{equation}

\noindent{}where $F$ denotes force, and $\rho_{\rm wind}$ is the density 
of the atmosphere.  Combining \ref{eqdragfric} with
Eq. \ref{eqwind} would not be appropriate in 
this context because the
star need not be losing mass.  \cite{schsmi2008} states that a typical
value of $v_{\rm sound}$ within the stellar chromosphere is about 8 km/s.
In their study on SB engulfment into GB stars, \cite{villiv2009} 
found that $v/v_{\rm sound} > 3$ always. They adopted $v_{\rm wind} = 5$~km/s,
whereas \cite{musetal2013} adopted a wind speed of $10-15$ km/s.  In 
general $\vec{v}_{\rm wind} \propto M_{\star}$ (footnote
\#1 in \cite{schsmi2008}).

{\bf FRICTIONAL DRAG}
\ \ \ \
The frictional drag force has appeared in several different
forms in the post-MS literature:
(i) Eq. 6 of \cite{villiv2009} and Eq. (4) of \cite{viletal2014},
(ii) Eq. 3 of \cite{donetal2010}, and (iii)  Eqs. 37-39 of \cite{verGB2015}.
The last of these references considers both the Epstein and Stokes regimes of motion,
given respectively by the upper and lower relations in:

\begin{eqnarray}
F_{\rm drag}^{({\rm fric})} 
&=
\left\{ \begin{array}{ll} 
   \left(\frac{M_{\rm SB}\rho_{\rm wind}v_{\rm sound}}{\rho_{\rm SB} R_{\rm SB}}\right)   \left( \vec{v}_{\rm wind} - \vec{v} \right)
, & R_{\rm SB}\ll \zeta\\
   \left(\frac{M_{\rm SB}\rho_{\rm wind}U}{\rho_{\rm SB} R_{\rm SB}}\right)   \left( \vec{v}_{\rm wind} - \vec{v} \right) 
                                              \left| \vec{v}_{\rm wind} - \vec{v} \right|
, & R_{\rm SB} \gg \zeta
\end{array}\right .
\label{eqdragve}
\end{eqnarray}

\noindent{}where $\zeta$ is the mean 
free path length of the gas. Eq. 56 of \cite{verGB2015} provides the following numerical
constraint on the path length: $\rho_{\rm wind}\zeta \sim 10^{-8}$ kg m$^{-2}$.  The quantity
$U$ is given by

\begin{eqnarray}
U &=\left\{ \begin{array}{lll} 
   9\left[ \frac{6 R_{\rm SB}}{\zeta v_{\rm sound}} \left| \vec{v}_{\rm wind} - \vec{v} \right| \right]^{-1}
, & {\rm Re} \le 1 \\
   9\left[ \frac{6 R_{\rm SB}}{\zeta v_{\rm sound}} \left| \vec{v}_{\rm wind} - \vec{v} \right| \right]^{-0.6}
, & 1 \le {\rm Re} \le 800 \\
   0.165
, & {\rm Re} \ge 800
\end{array}\right\}
\label{Resplit}
\end{eqnarray}

\noindent{}where Re represents the Reynolds number of the flow and is given by

\begin{equation}
\text{Re} = 
\frac{6R_{\rm SB}}
{\zeta v_{\rm sound}}
\left| \vec{v}_{\rm wind} - \vec{v}  \right|
.
\label{eqRe1}
\end{equation}

\section{Star-planet tides} \label{secTides}

The vast majority of known exoplanets will experience 
violent tidal interactions with their parent stars.  Almost every
planet discovered by transit photometry is close enough to its parent
star for tides to eventually play a significant role in its evolution.
Consequently, the last 15 years have seen a surge in tidal-based
investigations around MS stars.  However, the results of this effort, 
unlike for mass loss in the last section, have proven to be controversial.  
The physics involved in tidal dissipation is sufficiently complex that basic open
questions remain unsolved: e.g. how close does a planet have to be to the 
stellar surface to be engulfed?

Because we have room to improve our understanding of tides on the MS, extrapolating
to post-MS systems might seem difficult.  Nevertheless, much recent progress
has been made on GB star-planet tides.

\subsection{Tidal theory} \label{secTidestheory}

Important questions to answer before modeling a tidally-influenced
system are:
(1) Do SBs induce tidal bulges in the star, does the star induce
tidal bulges in the SB, or both? and (2) in each case, what is 
the dominant tidal mechanism acting?  The answers depend on the
properties of both the star and the SB; for a recent review of tides between
stars and gas giant planets with zero solid matter, see \cite{ogilvie2014}. 
Depending on the level of accuracy
one seeks, \cite{efrmak2013} proved that some classical tidal theories
should be used with caution, or not at all.  However, these theories provide
appealing analytically tractable forms that have been implemented by much of the 
community.

{\bf ONLY STELLAR TIDES MATTER}
\ \ \ \ 
For GB systems, large SBs like planets or brown dwarfs within a few au
will almost certainly 
induce a tidal bulge in GB stars because these stars will grow their 
convective envelopes 
to au-scales (in fact, \cite{musvil2012} approximated the stellar radius to be equal
to the radial extent of the envelope).  Alternatively, GB stars are unlikely
to induce tidal bulges in large SBs.  Both \cite{musvil2012}
and \cite{viletal2014} consider this possibility with a constant tidal lag
model.  By adopting a generous range (many orders of magnitude) of tidal 
quality factors, they find that planetary tides are negligible.

Stellar tides matter only if the SB is large enough: at least planet-sized.
An asteroid or comet would likely induce too small of a tidal bulge in the star
to cause a significant orbital change (for quantification, see e.g. Eq. 8 
of \cite{beanes2012}, Eqs. 10-11 of \cite{nagida2011}, or 
Eqs. \ref{zahna}-\ref{zahne} below).

{\bf ORBITAL RETARDATION}
\ \ \ \ 
The dominant tidal mechanism for GB stellar tides is turbulent viscous
dissipation in the envelope. \cite{schsmi2008} alternatively describes the process as 
``retardation of the equilibrium tide by interaction with convective motions''.  
This phenomenon has been described with
standard mixing length theory in the pre-exoplanet era
by \cite{zahn1977,zahn1978}, and has been recently employed by 
\cite{villiv2009,kunetal2011,musvil2012,norspi2013,viletal2014}.
Dissipation in the convective envelope causes the SB's orbital semimajor
axis and eccentricity to change.  To leading order in eccentricity
(an unfortunately restrictive but computationally feasible assumption)
the orbital evolution is dictated through 
\cite{musvil2012,viletal2014} 

\begin{eqnarray}
\frac{da}{dt}
&=&
- \frac{a}{9 t_{\rm conv}}
\frac{M_{\star}^{({\rm env})}}{M_{\star}}
\frac{M_{\rm SB}}{M_{\star}}
\left(
1 + \frac{M_{\rm SB}}{M_{\star}}
\right)
\left(
\frac{R_{\star}}{a}
\right)^8
\nonumber
\\
&\times& 
\left[ 2p_2 + e^2\left(\frac{7}{8}p_1 - 10p_2 + \frac{441}{8}p_3 \right) \right]
,
\label{zahna}
\end{eqnarray}

\begin{eqnarray}
\frac{de}{dt}
&=&
- \frac{e}{36 t_{\rm conv}}
\frac{M_{\star}^{({\rm env})}}{M_{\star}}
\frac{M_{\rm SB}}{M_{\star}}
\left(
1 + \frac{M_{\rm SB}}{M_{\star}}
\right)
\left(
\frac{R_{\star}}{a}
\right)^8
\nonumber
\\
&\times& 
\left[\frac{5}{4}p_1 - 2p_2 + \frac{147}{4}p_3 \right]
,
\label{zahne}
\end{eqnarray}

\noindent{}where $M_{\star}^{({\rm env})}$ is the stellar 
envelope mass, $t_{\rm conv}$ is
the eddy turnover timescale within the stellar envelope

\begin{equation}
t_{\rm conv} = 
\left[
\frac
{M_{\star}^{({\rm env})} \left(R_{\star} - R_{\star}^{({\rm env})} \right)^2}
{3L_{\star}}
\right]^{\frac{1}{3}}
\end{equation}

\noindent{}and $p_1$, $p_2$ and $p_3$ are frequency components
of the tidal force (below).  In their study of planet-GB tides, 
\cite{kunetal2011} provides the same
expression for the eddy turnover timescale (their Eq. 4)
although the expression in Eq. 31 of \cite{huretal2002} is
slightly different due to the convection zone 
reference point they use.
Regarding Eq. \ref{zahna} above, that equation contains
individual frequency components, unlike Eq. 1 of
\cite{kunetal2011}, Eq. 8 of \cite{adablo2013} and Eq. 29 of
\cite{huretal2002}.

The $p$ components are functions of $a$,
$M_{\star}$, $M_{\rm SB}$ and one's choice about the nature  
of the dissipation through a functional dependence.  This last
choice, importantly, does not critically impact the tidal model
for GB stars \cite{viletal2014}.  Consequently, by adopting
the values from \cite{villiv2009}, the frequency components
are well-approximated (i.e. calibrated with observations of post-MS binaries
by \cite{verphi1995}) by

\begin{equation}
p_l \approx \frac{9}{2} {\rm min}
\left[  
1, \left(   
\frac
{4\pi^2a^3}
{l^2 G \left(M_{\star} + M_{\rm SB} \right) \left[t_{\rm conv}\right]^2 }
\right)
\right]
\label{eqfunc}
\end{equation}

\noindent{}where $l=\left\lbrace 1,2,3 \right\rbrace$. See Section 4.4 
of \cite{musvil2012} for an analysis of the goodness of the approximation 
in Eq. \ref{eqfunc}.

By using an appropriate stellar evolution model (see Sec. \ref{secCodes}) that
computes the time evolution of the stellar envelope mass and radii, one
can then integrate Eqs. \ref{zahna}-\ref{eqfunc} to determine the
SB's orbital evolution.  Note that these equations are secular, and hence cannot
resolve changes on orbital timescales.

{\bf CONTRIBUTION OF STELLAR SPIN}
\ \ \ \
Hidden in Eqs. \ref{zahna}-\ref{eqfunc} is the omission of the GB star's
spin.  A GB star's spin is likely to be negligible due to conservation
of spin angular momentum from the MS.
Consequently, neglecting stellar spin in a model is likely a good approximation.
The larger the SB, however, the more important a role any extant stellar spin might play.
\cite{norspi2013} utilized a different expression for the semimajor axis
evolution, including stellar spin, that is similar to Eq. \ref{zahna}
but assumed circular orbits and a different parameterisation for $p$.
They also included a separate equation of motion for the stellar spin.
In principle, one may derive equations for 
$\dot{a}$, $\dot{e}$ and $\dot{s}_{\star}$, where $s_{\star}$ is the stellar
spin, to arbitrarily orders in
eccentricity and spin, akin to the second-order treatment
provided by \cite{zahn1989} in his Eqs. 17-19.

{\bf TIDAL QUALITY FACTOR}
\ \ \ \
An alternative tidal approach is one that places the unknown physics of dissipation
into a single parameter (in a similar way to the $\alpha$ viscosity
parameterisation in viscous discs): the tidal quality factor 
\cite{golsot1966}.  Eq. \ref{zahna} may be expressed in terms
of the tidal quality factor (e.g. compare Eqs. 1 and 8 of \cite{kunetal2011}).
A tidal quality factor approach also allows one to generate evolution
equations in terms of arbitrary eccentricities with no series expansion 
\cite{matetal2010}. 
However, for GB stars, the formalism from \cite{zahn1977} is better constrained
than one with a tidal quality factor partly because
(i) the time and frequency dependence of the tidal quality factor
is unknown and (ii) the ``constant time lag model'', which uses the quality 
factor, is inherently inconsistent when applied to SBs containing solid material
\cite{efrmak2013}.

{\bf ANALYTICAL ENGULFMENT DISTANCES}
\ \ \ \
By employing a series of assumptions, a few investigations have derived
explicit semianalytical and analytical formulae for the minimum distance
at which an SB would be engulfed during the GB phases.
These derivations require assumptions about both tidal effects and
prescriptions for the mass loss rates, meaning that during application
one must keep in mind the accuracy sought.  Informed by numerical simulations,
Eq. 7 of \cite{kunetal2011} provides scalings which indicate that the most
important consideration for RGB engulfment is the maximum RGB stellar radius.
The more general study of \cite{adablo2013} assumed power-law dependencies 
for tidal dissipation (their Eq. 6), and a power-law
prescription for mass loss (their Eq. 4), which includes a modification for AGB
pulse mass loss (their Eq. 27).  Their final expression for the critical
engulfment distance (their Eq. 38) is a function of these power-law indices
and other coefficients.

{\bf COMPACT OBJECT TIDES}
\ \ \ \
For DA WDs and other compact objects where the convection zone is negligible,
one must seek an alternative tidal approach than that from \cite{zahn1977}.  
In compact objects, both radiative and convective regions produce tidal dissipation,
and a coupled treatment (as in Eq. 31 of \cite{livsok1984}) might be
desirable. Also, generally, if a WD and MS star have the same convective properties,
then they will appear equivalent from the point of view of tidal dynamics.

\subsection{Simulation results} \label{secTidessims}

The emphasis in GB-planet tidal investigations is on engulfment: how could
planets survive and where and when do they die.  The implications of the
answers help determine what architectures can and should exist around WDs
and help constrain GB planet discovery space.  The differences in these
answers amongst different studies \cite{villiv2009,kunetal2011,viletal2014} 
emphasise the sensitive dependence on the stellar models adopted.

\begin{figure}[!t]
\centering\includegraphics[width=3.4in]{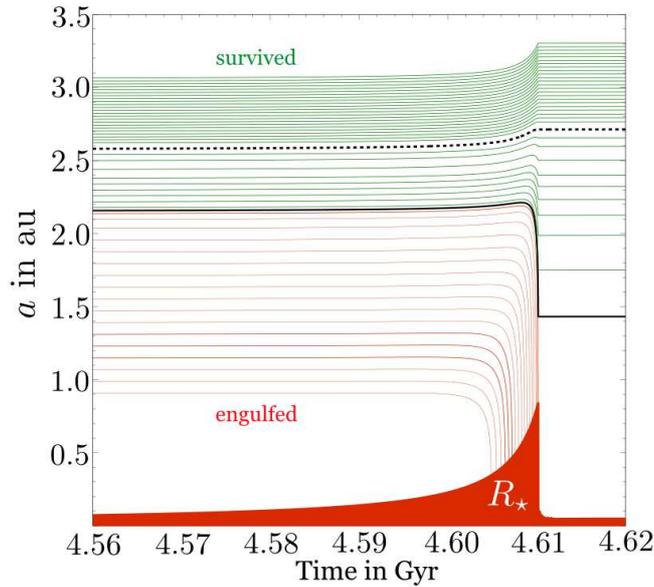}
\caption{Cosmetically enhanced version of Fig. 1 of \cite{viletal2014}.  
The prospects for survival of Jupiter-mass planets orbiting a RGB star
with $M_{\star}^{({\rm MS})} = 1.5M_{\odot}$ and evolving due to 
tides and mass loss. The
stellar surface is given by the upper curve on the solid red shape.
The red planetary tracks end in engulfment, whereas planets on green
tracks remain safe.  The solid black curve shows the closest planet
that survives, and the dotted black curve illustrates the closest planet
that is not visually affected by RGB tides on the scale of this plot.}
\label{FigRGB}
\end{figure}
\begin{figure}[!h]
\centering\includegraphics[width=5.35in]{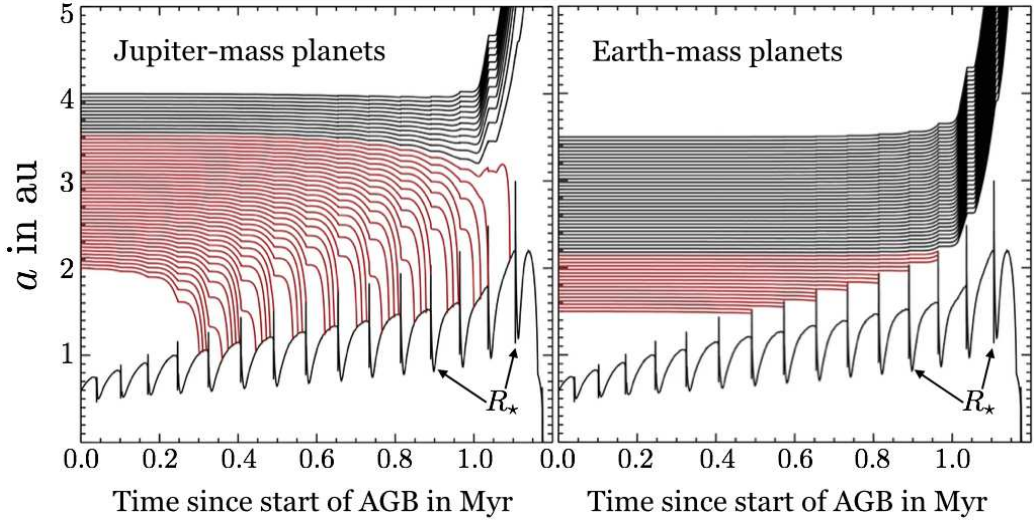}
\caption{Cosmetically enhanced version of Fig. 3 of \cite{musvil2012}.
The prospects for survival of Jupiter-mass planets (left panel) 
and Earth-mass planets (right panel) orbiting a AGB star
with $M_{\star}^{({\rm MS})} = 2.0M_{\odot}$ and evolving due
to tides and mass loss.
Unlike in Fig. \ref{FigRGB}, here the stellar surface pulses.
The left panel illustrates that surviving Jovian-mass planets must begin their
orbits at least 20 percent further away than the maximum stellar radius.
In the right panel, Earth-mass planets with starting orbits that are
within the maximum stellar radius can survive. 
}
\label{FigAGB}
\end{figure}

The process of engulfment along the RGB 
is illustrated in Fig. \ref{FigRGB}, which displays the evolution (mass loss
and tides) of a 
$M_{\star}^{({\rm MS})} = 1.5M_{\odot}$ star with Jupiter-mass planets 
that will be engulfed (red curves),
that are affected by tides but not engulfed (green curves), and survive
without being affected by tides (black curves).  Note that 
surviving planets which are ``stalled'' by GB tidal effects
harbour WD-planet distances which are smaller than that predicted
from adiabatic mass loss expansion alone.

If SBs survive the RGB phase,
however, they might still be ingested during the AGB phase.  
Fig.~\ref{FigAGB} 
illustrates the fate of Jovian-mass planets and Earth-mass planets 
around a thermally pulsing $M_{\star}^{({\rm MS})} = 2.0M_{\odot}$ AGB star.
The plots (originally from \cite{musvil2012}) demonstrate
how lower-mass planets feel weaker tidal forces than giant planets.
Consequently, the former can reside on MS orbits within the maximum
AGB radius and still survive GB evolution.  Not shown on these plots
is how eccentricity affects engulfment: eccentric giant planet orbits
can circularise more quickly than terrestrial planet orbits due to tides.
This quick circularisation acts as a protection mechanism for giant planets.  

The results of \cite{viletal2014} have an important observational link:
they show that the RGB star region void of planets (within about 0.5 au)
is too wide to have been depleted by tidal effects alone.  Therefore,
if the retired A star hosts actually have smaller masses, as argued
by \cite{lloyd2013}, the origin of the void is even a larger mystery.

\section{Stellar radiation} \label{secRad}

The importance of stellar radiation in post-MS science is highlighted
by a pre-exoplanet-era prediction \cite{steetal1990} that the Solar system's 
Kuiper belt will be eventually be evaporated from GB radiation. Further, WD and pulsar
radiation determine how volatile-rich planetary remnants behave, and may dictate
the shape and properties of surrounding debris discs.

\subsection{Giant branch radiation} \label{secRadGB}

During stellar evolution, a GB star's luminosity can reach a value that is several
orders of magnitude greater than the current Sun's (see Fig. \ref{FigSpec}). The
consequences are varied. SBs 
within hundreds
of au will feel the result, as the snowlines in those systems might extend out to 
that distance \cite{steetal1990,stoetal2015}.  Pebbles, asteroids and 
comets (Sec. \ref{secRad}\ref{secRadGB}\ref{secPebcom}) incur orbital changes more complex than those from mass loss alone,
and asteroids and comets might self-destruct through overspinning.  Within debris discs,
the size distribution and collision rates might change \cite{bonwya2010}.  
Giant planet and terrestrial planet atmospheres 
(Sec. \ref{secRad}\ref{secRadGB}\ref{secPlanAtm}) will be transformed
both physically and chemically.

\subsubsection{Effect on pebbles, asteroids and comets} \label{secPebcom}

The total radiative effect on a two-body system consisting of a star
and a solid body SB with no atmosphere yields the following expressions 
for the orbital and physical equations of motion.

{\bf ORBITAL EVOLUTION}
\ \ \ \
I obtain the orbital equation of motion by combining Eqs. 9 and 29 of \cite{verGB2015},
yielding

\begin{eqnarray}
\frac{d^2\vec{r}}{dt^2}
&=&
-\frac{G \left(M_{\star} + M_{\rm SB}\right) }{r^3} \vec{r}
+ 
\frac{A_{\rm SB}L_{\star}}{4\pi M_{\rm SB} c r^2}
\left\lbrace
\left(
1 - \frac{\vec{v}\cdot\vec{r}}{cr}
\right)
\frac{\vec{r}}{r}
-
\frac{\vec{v}}{c}
\right\rbrace
\nonumber
\\
&\times&
\Bigg[
Q_{\rm abs} \mathbb{I}
+
Q_{\rm ref} \mathbb{I}
+
\left\lbrace
\mathcal{H}\left(R_{\rm SB} - \sqrt{\frac{K_{\rm SB} P_{\rm SB}^{({\rm spin})}}{\pi \rho_{\rm SB} \mathcal{C}_{\rm SB}}} \right)
\right\rbrace
w \left(Q_{\rm abs}-Q_{\rm ref}\right) \mathbb{Y}
\Bigg]
.
\label{eqyark}
\end{eqnarray}

\noindent{}In Eq. \ref{eqyark}, $c$ is the speed of light,
$A_{\rm SB}$ is the SB's cross-sectional area,
$Q_{\rm abs}$ is the SB's absorption efficiency,
$Q_{\rm ref}$ is the SB's reflecting efficiency (or albedo),
$\mathbb{I}$ is the 3$\times$3 identity matrix, 
$\mathcal{H}$ is the Heaviside function, 
$K_{\rm SB}$ is the SB's thermal conductivity, 
$P_{\rm SB}^{({\rm spin})}$ is the SB's spin period, 
$\mathcal{C}_{\rm SB}$ is the specific thermal capacity of the SB, 
$w$ is a value between 0 and $1/4$ that indicates the extent of the temperature 
difference between the heated and non-heated sections of the SB (no difference
corresponds to $w = 0$), and
$\mathbb{Y}$ is the 3$\times$3 
Yarkovsky matrix (given through Eqs. 13-20 and 28 of \cite{verGB2015}).

The three terms in the square brackets in Eq. \ref{eqyark} 
correspond respectively to the
acceleration caused by absorbed radiation, immediately reflected radiation
and re-emission from delayed radiation.  The motion resulting from this 
last term is known as the ``Yarkovsky effect'', which has been acknowledged
as a potential driver of planetary debris \cite{zucetal2010,frehan2014}
and described further for particles in compact object debris discs 
(Appendix A of \cite{corsha2008} and Section 2.2 of \cite{rafikov2011a}), 
but not considered in more geometric detail until \cite{verGB2015} (their Sec. 2.6).  
The hallmark of the Yarkovsky effect, which is re-emission from delayed
radiation, arises from a thermal imbalance and redistribution within the SB.
Hence, if the SB is too small or spins too quickly, the Yarkovsky effect
is not triggered. When the Yarkovsky effect
vanishes, the tangential and radial components of the sum of the other two 
terms are known respectively as ``Poynting-Robertson drag'' and, confusingly,
``radiation pressure''.  The term in the first set of curly brackets is
the relativistically corrected direction of incoming radiation.

Other important features of this equation include:  (i)
The presence of the Heaviside function indicates that the Yarkovsky effect
does not ``turn on'' unless $R_{\rm SB}$ is above a critical size.  \cite{verGB2015}
found that across the entire range of realistic values of $K_{\rm SB}$, 
$P_{\rm SB}^{({\rm spin})}$ and $\mathcal{C}_{\rm SB}$, this critical value
lies between 1 cm and 10 m (their Fig. 2).  (ii) Even below this threshold,
Eq. \ref{eqyark} is applicable only when the SB is at least a few orders of magnitude
greater than the wavelength of the radiation.  \cite{verDisc2015} argued that
for WDs, this wavelength is always less than a micron.  (iii) Because, for spherical
SBs, the acceleration due to radiation is inversely proportional to $R_{\rm SB}$,
the orbits of planet-sized (thousands of km) SBs will only negligibly be affected 
by radiation.

In order to provide a rough idea of the potential importance of the Yarkovsky
effect, I compare the averaged eccentricity changes expected from the Yarkovsky effect
(Eq. 108 of \cite{verGB2015})

\begin{eqnarray}
\left\langle \frac{de}{dt} \right\rangle^{({\rm Yarkovsky})}
= 
\mathcal{O}\left(\frac{1}{c} \frac{A_{\rm SB}L_{\star}}{8\pi M_{\rm SB}na^3} \right)
&\sim&
\frac{0.08}{\rm Myr}
\left(
\frac{M_{\star}}{1M_{\odot}}
\right)^{-\frac{1}{2}}
\left(
\frac{\rho_{\rm SB}}{2 \ {\rm g} \ {\rm cm}^{-3}}
\right)^{-1}
\nonumber
\\
&\times&
\left(
\frac{R_{\rm SB}}{1 \ {\rm km}}
\right)^{-1}
\left(
\frac{a}{5 \ {\rm au}}
\right)^{-\frac{3}{2}}
\left(
\frac{L_{\star}}{10^{3}L_{\odot}}
\right)
\label{eqYark}
\end{eqnarray}

\noindent{}with those from the combined averaged effect of Poynting-Robertson drag
and radiation pressure (Eq. 109 of \cite{verGB2015})

\begin{eqnarray}
\left\langle \frac{de}{dt} \right\rangle^{({\rm PR+rp})}
= 
\mathcal{O}\left(\frac{1}{c^2} \frac{5A_{\rm SB}L_{\star}}{8\pi M_{\rm SB}a^2} \right)
&\sim&
\frac{1.8 \times 10^{-5}}{\rm Myr}
\left(
\frac{\rho_{\rm SB}}{2 \ {\rm g} \ {\rm cm}^{-3}}
\right)^{-1}
\nonumber
\\
&\times&
\left(
\frac{R_{\rm SB}}{1 \ {\rm km}}
\right)^{-1}
\left(
\frac{a}{5 \ {\rm au}}
\right)^{-2}
\left(
\frac{L_{\star}}{10^{3}L_{\odot}}
\right)
.
\label{eqPR}
\end{eqnarray}

\noindent{}The four order-of-magnitude difference in the coefficients of 
the expressions is largely attributable to the difference in powers of $(1/c)$
in Eqs. \ref{eqYark}-\ref{eqPR}.  Averaging eliminates the $(1/c)$ terms
in the expressions for Poynting-Robertson drag and radiation pressure,
but not for the Yarkovsky effect.

{\bf PHYSICAL EVOLUTION}
\ \ \ \
For the physical evolution of the SB, I display
an averaged spin equation of motion, where $s$ is spin,
(Eq. 1 of \cite{verjacgae2014} or Eq. 3 of \cite{verGB2015}):

\begin{equation}
\left\langle \frac{ds_{\rm SB}}{dt} \right\rangle
=
\frac{j}{2\pi \rho_{\rm SB} R_{\rm SB}^2}
\left(
\frac
{1}
{a^2\sqrt{1 - e^2}}
\right)
\left(
10^{17} \ {\rm kg} \ {\rm m} \ {\rm s}^{-2} \
\frac{L_{\star}}{L_{\odot}}
\right)
,
\label{eqyorp}
\end{equation}

\noindent{}an equation for radius reduction due to ablative mass loss from sublimation 
(Eq. 15 of \cite{jura2004}, Eq. 9 of \cite{jura2008} and Eq. 31 of \cite{bonwya2010}):

\begin{equation}
\frac{dR_{\rm SB}}{dt}
=
-
\frac{1.5 \times 10^{10} \text{kg} \ \text{m}^{-2} \ \text{s}^{-1}}
{\rho_{\rm SB}}
\sqrt{\frac{T_{\rm SB}^{({\rm sub})}}{T_{\rm SB}}}
\text{exp}\left( -\frac{T_{\rm SB}^{({\rm sub})}}{T_{\rm SB}} \right)
,
\label{eqsubl}
\end{equation}

\noindent{}and an equation for the average SB surface temperature,
excluding any thermal imbalance between the hemispheres of the SB
(Eq. 3a of \cite{steetal1990}):

\begin{equation}
\left\langle T_{\rm SB} \right\rangle = 
\left[
\frac{1}{\epsilon_{\rm SB} \sigma}
\left(
\frac
{\left(1 - Q_{\rm ref}\right) L_{\star}}
{16 \pi r^2}
+
\mathcal{S}_{\rm SB} \frac{dM_{\rm SB}}{dt}
\right)
\right]^{\frac{1}{4}}
.
\label{eqtemp}
\end{equation}

\noindent{}In Eqs. \ref{eqyorp}-\ref{eqtemp}, $j$
is a value between 0 and 1 which indicates the extent of the asymmetry
in the shape of the SB (no asymmetry corresponds to $j = 0$), 
$T_{\rm SB}^{({\rm sub})}$ is the composition-dependent 
sublimation temperature of the SB,
$\epsilon_{\rm SB}$ is the SB's emissivity, $\sigma$ is the Stefan-Boltzmann constant,
and $\mathcal{S}_{\rm SB}$ is the SB's specific heat of
sublimation.  See Table 1 of \cite{rafgar2012} for
a useful reference for sublimation temperatures 
$T_{\rm SB}^{({\rm sub})}$ of different 
materials, and Fig. S6 of \cite{vanetal2015} for characteristic
vapour pressures and mass loss rates for some refractory materials.  
Changes in spin due to radiation (Eq. \ref{eqyorp})
is known as the ``YORP effect''.

Five important points include (i) Eqs. \ref{eqyorp}-\ref{eqsubl} 
do not include physics arising from avalanches, micrometeoroid impacts, 
or differentiated internal layers, all potentially important sources
of modifications. (ii) Eq. \ref{eqtemp} does not include the Yarkovsky
effect, which would be affected by mass loss and any shape changes.
(iii) The spin of the SB (Eq. \ref{eqyorp}) 
is strongly tied to the value of $w$ (in Eq. \ref{eqyark}) such that a 
stationary object corresponds to $w = 1/4$.  (iv) 
\cite{faretal2013b} pointed out that if a volatile species such as water 
is contained in hydrated minerals like phyllosilicates, then the water 
will not sublimate until much higher temperatures than
$T_{\rm SB}^{({\rm sub})}$ are reached, (v) The expressions for the change
in both $R_{\rm SB}$ and $M_{\rm SB}$ in Eqs. \ref{eqsubl}-\ref{eqtemp} may
be related through an assumption of a spherical SB shape.

How does the temperature of a SB change as a result of post-MS evolution?;
Eq. \ref{eqtemp} is just an approximation.
\cite{jurxu2010} performed a more sophisticated extension of work done by
\cite{jura2008},
solving the one-dimensional heat conduction equation for asteroids which are 
assumed to be in hydrostatic equilibrium with a uniform density and no 
surface pressure.  \cite{jurxu2010} self-consistently treated orbital 
evolution and
internal water evaporation for these uniform density asteroids for two
stellar tracks: a $M_{\star}^{{\rm (MS)}} = 1M_{\odot}$ and a
$M_{\star}^{{\rm (MS)}} = 3M_{\odot}$ star, each with a metallicity of
$Z_{\star} = 0.0019$.  The thermal evolution is complex: sometimes
at the tip of the AGB, the highest temperatures were found to occur
beneath the surface.

A SB may spin so fast that it will tear itself apart. This possibility is not
remote.  In fact, nearly every asteroid in the Solar system's asteroid belt 
with $R_{\rm SB} = 100$m-10km will break apart in this manner after the Sun 
turns off of the MS \cite{verjacgae2014}.  The result is a potential debris
field which could extend to thousands of au.  The critical spin period of 
2.33 hours at which breakup occurs is well-grounded in Solar system observations 
(See Fig. 1 of \cite{seth2014}).  More generally, by assuming the SB is a strengthless
rubble pile, I can write the critical spin $s_{\rm crit}$ in terms of $\rho_{\rm SB}$ 
as (Eq. 2 of \cite{verjacgae2014})

\begin{equation}
s_{\rm crit} = 2\pi \sqrt{\frac{G\rho_{\rm SB}}{3\pi}} =  7.48 \times 10^{-4} \frac{\rm rad}{\rm s} 
\left(\frac{\rho_{\rm SB}}{2 \ {\rm g} \ {\rm cm}^{-3}} \right)^{1/2}
.
\end{equation}

\subsubsection{Effect on planetary atmospheres} \label{secPlanAtm}

GB radiation may cause SB atmospheres to evaporate, either partially
or fully.  If the SB is all atmosphere, then one can estimate the maximum
SB mass that would completely evaporate, particularly when inside
of the stellar envelope \cite{soker1998,neltau1998,wicetal2010}.
Further, Fig. 1 of \cite{goldstein1987} explicitly illustrates 
that different atomic and molecular species will escape a planetary atmosphere
at different temperatures. These temperatures vary by thousands
of degrees, and showcases how different species may be ripped away 
layer-by-layer as the tip of the AGB is approached.

{\bf PARAMETERISING ESCAPE}
\ \ \ \
If the atmospheric loss is great enough, then the atmosphere
will change structure.  Assume that the SB is all atmosphere.  
\cite{villiv2007} defined a helpful dimensionless
quantity

\begin{equation}
\Gamma \equiv \frac{GM_{\rm SB} M_{\rm H}}{k_{\rm B} R_{\rm SB} T_{\rm ex}}
\end{equation}

\noindent{}where $T_{\rm ex}$ is the temperature of the exosphere
(which may be different from the SB's effective temperature), 
$M_{\rm H}$ is the mass of a hydrogen atom
and $k_{\rm B}$ is the Boltzmann constant.  The quantity $\Gamma$
represents the ratio of the gravitational potential energy to the
particle kinetic energy, or the square of the ratio of the atmospheric
escape velocity to thermal velocity.  \cite{villiv2007} found that
the atmospheric structure is maintained for $\Gamma \gtrsim 20$.

Determining $T_{\rm ex}$ depends on how one models 
heat transfer within a SB atmosphere.  This transfer
is determined by a combination of convection and
conduction. \cite{villiv2007} neglected convection because SBs orbiting
GB stars are unlikely to be tidally synchronized.  Consequently, they solved
a conduction-dominated energy equation (their Eqs. 6 and 8)
in order to obtain an expression which can be implicitly
solved for $T_{\rm ex}$ 
(their Eq. 7) given a wide variety of thermal properties (conductivity,
thermospheric base temperature, intensity reaching atmosphere).

{\bf QUANTIFYING ESCAPE - NUMBER DENSITY FLUX}
\ \ \ \
If the atmospheric structure is maintained ($\Gamma \gtrsim 20$),
then a lower limit
to the escape rate by thermal evaporation can be expressed through 
a number flux $\Phi$ as (Eq. 9 of \cite{villiv2007})

\begin{equation}
\Phi = -\sqrt{\frac{k_{\rm B} T_{\rm ex}}{2\pi M_{\rm H}}}
       \delta 
       \left(1 + \Gamma \right)
       {\rm exp}\left(-\Gamma \right) 
\label{eqatmesc}
\end{equation}

\noindent{}where $\delta$ is the number density of the escaping constituent.
The restriction that $\Gamma \gtrsim 20$ limits the use of Eq. \ref{eqatmesc}.  
A more complete treatment requires considering
potential feedback factors, ram pressure stripping, photodissociation, immersion
into a postshocked hot bubble, and the solution of
the atmospheric structure equations; estimating the result with a single
explicit expression like their Eq. 10 still requires many assumptions
that rely on results of numerical simulations anyway.  Also, as suggested
by \cite{spimad2012}, extant pebbles and boulders could be accreted onto
the SB's atmosphere during post-MS evolution due to $N$-body interactions
(see Sec. \ref{secMulti}), altering the composition and structure of the atmosphere.

{\bf QUANTIFYING ESCAPE - MASS FLUX}
\ \ \ \
With the same functional form of Eq. \ref{eqatmesc}, 
\cite{livsok1984} provided an expression (in their Eq. 6) for 
SB atmospheric mass loss while in the midst of a red giant.
More recent treatments are provided by \cite{soker1999} and 
\cite{beasok2011}.  \cite{beasok2011} incorporated both
evaporation and ionization of the evaporated gas,
and pointed out that recombination of the ionized gas is 
not necessarily an important process for MS exosystems, 
unlike in post-MS exosystems.  By manipulating and rewriting 
their Eqs. 7 and 10, as well as assigning a negative sign
for mass loss, I obtain

\begin{equation}
\frac{dM_{\rm SB}}{dt} \approx 
-\frac{1}{2} 
\left[
\frac{2\pi GM_{\rm SB} v_{\rm out}^2}{\eta \Pi \mathcal{R}}
+
\sqrt{ 
\frac{4 \pi^2 G^2M_{\rm SB}^2 v_{\rm out}^4}{\eta^2 \Pi^2 \mathcal{R}^2}
+
\frac{8 \pi R_{\rm SB} v_{\rm out}^2 \dot{H}_{\rm EUV}}{\Pi \mathcal{R}}
}
\right]
\end{equation}

\noindent{}where $v_{\rm out}$ is the outflow speed, $\eta$ is the efficiency
factor for channeling EUV radiation to evaporation, $\Pi$ is the
average energy of ionizing photons, $\mathcal{R}$ is a recombination
rate coefficient, and $\dot{H}_{\rm EUV}$ is the total EUV power from a given
wavelength range.  Some characteristic values might be 
$v_{\rm out} = 2 \times 10^5$ m/s, $\eta = 0.2$, $\Pi = 20$ eV, 
$\mathcal{R} = 5 \times 10^{34}$ m$^3$kg$^{-2}$s$^{-1}$ and
$\dot{H}_{\rm EUV} = 4 \times 10^{23}$ J s$^{-1}$.

{\bf GIANT PLANET ATMOSPHERES - CHEMICAL CHANGES}
\ \ \ \
\cite{villiv2007} and \cite{spimad2012} are important studies 
because they speak directly to the fate of the giant planets
in our Solar system and, in particular, Jupiter.  We know that Jupiter
will survive Solar mass loss and not undergo scattering
instability, but will it evaporate completely?  Sec. 4
of \cite{spimad2012} is devoted to the chemical changes
Jupiter will undergo during the Sun's post-MS evolution,
as well as what Jupiter's spectrum would look like (their Fig. 6).
This spectrum includes some abundance alterations in carbon, methane
and water.

\subsection{Compact object radiation} \label{secRadWDNS}

\subsubsection{White dwarf radiation}

{\bf COMET SUBLIMATION}
\ \ \ \
WD radiation causes an SB to lose mass, particularly if the SB is
volatile rich.  This mass loss may arise in the form of sublimation or
outgassing. \cite{verSub2015} made a distinction between the two terms, referring
to the former as the homogeneous mass exodus of surface particles and the latter
as localized violent eruptions.  In either case, from Eqs. \ref{eqml3a}-\ref{eqml3E}, 
one can see that whatever mass is lost from the system
by the SB will negligibly change its orbit.  Instead, the largest orbital change will
come from the kick due to mass loss-induced linear momentum recoil.  \cite{verSub2015}
did not treat the kick as impulsive.  They assumed the SB was continuously losing mass
in a manner physically motivated by Solar system comets.  

By assuming that an SB has a volatile content $M_{\rm SB}^{({\rm vola})}$ which
is subject to sublimation and is made 
of a single species with molecular or atomic mass $M_{\rm vola}$, they found 
that the equations of motion become (their Eq. 6)

\begin{equation}
\frac{d^2\vec{r}}{dt^2}
=
-\frac
{G \left(M_{\star} + M_{\rm SB}\right)}{r^3}\vec{r}
-
\left\lbrace
\mathcal{H}\left(M_{\rm SB}^{({\rm vola})}\right)
\right\rbrace
\frac{D_0 v_{0} M_{\rm vola}}{M_{\rm SB} - M_{\rm SB}^{({\rm vola})}}
\left( \frac{r_0}{r} \right)^{\frac{9}{4}}
\frac{\vec{r}}{r}  
,
\end{equation}

\noindent{}where $\mathcal{H}$ is the Heaviside function and $D_0$, $v_0$
and $r_0$ are fiducial values representing, respectively, the number of 
molecules or atoms of the volatile species sublimating per time,
the mean ejection velocity of that species, and distance to the WD.

Using this formulation, \cite{verSub2015} found that for nearly parabolic
orbits ($e \gtrsim 0.998$) -- as expected of SBs which pollute WDs --  
the orbital pericentre
robustly resists changes to sublimative forces even though other orbital
parameters might easily change.  They generalised this result to arbitrary
integral power-law dependencies on distance (their Appendix A).  Therefore,
comets whose orbital pericentre are outside of the WD disruption sphere will not
self-perturb themselves into the disruption sphere unless acted on by other agents.

\cite{stoetal2015}, using a different set of assumptions than \cite{verSub2015},
approximated the location, $r_{\rm sub}$ at which a volatile-rich SB completely 
sublimates.  By generalising Eq. 12 of \cite{stoetal2015}, I obtain

\begin{equation}
r_{\rm sub} \sim
\frac
{9 L_{\star}^2}
{64 G M_{\star} R_{\rm SB}^2 \rho_{\rm SB}^2 \mathcal{T}_{\rm SB}^2}
\label{eqStoSub}
\end{equation}

\noindent{}where $\mathcal{T}_{\rm SB}$ is the SB's latent heat of transformation
(given as $3 \times 10^{10}$ erg/s) and is claimed to be
similar for ice and silicates.  If instead the comet partially
sublimates at each pericentre passage, and the amount
of mass lost per passage decreases steeply enough, then
the total mass lost can be approximated by an converging
sum as in Eq. C2 of \cite{stoetal2015}.
 
\cite{stoetal2015} also, in their Footnote \#4, estimates the terminal speed
of ice ejected to be

\begin{equation}
v_{\rm term} = \sqrt
{  \frac
{14 k_{\rm B} T_{\rm ice}}
{27 M_{\rm H}}
}
\end{equation}

\noindent{}where $k_{\rm B}$ is the Boltzmann constant, $T_{\rm ice}$ is the
sublimation temperature of ice and $M_{\rm H}$ is the mass of a hydrogen atom. 
They commented that the ejection velocity for dust is
highly model-dependent (and hence more unconstrained than ice).
They also discussed and quantified the potential distribution
of debris left over from cometary sublimation (see their Secs. 3.2-3.3).
These distributions rely on the specific orbital energy of the sublimating
debris.

{\bf DRAG ON SBs}
\ \ \ \
Besides mass loss, the pressure of WD radiation on SBs will cause
them to drift inward according to Eq. \ref{eqYark}.  Belts or
rings with nonuniform size distributions might then respond to
WD radiation in complex ways.  I discuss the implications of 
this process in the context of forming WD discs from first-generation
SBs in Sec. \ref{secWDform}.

{\bf SUBLIMATING WD DEBRIS DISCS}
\ \ \ \
WD radiation plays a crucial role in the morphology and 
evolution of WD debris discs.  This radiation sublimates dust
into gas and mingles the two.  I discuss the implications of this
sublimation in Sec. \ref{secWDaccr}.

{\bf EFFECTS ON SB ATMOSPHERES}
\ \ \ \
The relatively small size and quickly dimming nature of WDs
provide for interesting SB atmospheric dynamics.  In contrast
to the extreme MS planet Kepler-91b, which is so close to its
parent star that 70\% of the planet atmosphere is illuminated by
the host star (Fig. 10 of \cite{liletal2014}), in WD systems much
less than 50\% of a giant planet atmosphere would be illuminated
(the radius of a WD is an order of magnitude smaller than that
of a giant planet).
I summarise SB atmospheric studies from WD systems in 
Sec. \ref{secOther}\ref{secOtherClimate}
in the context of climate and habitability.

\subsubsection{Pulsar radiation}

The radiation emitted from pulsars contains highly energetic
and relativistic particles.  The consequences for SBs are
different than in MS, GB and WD planetary systems. 
Also, processes like ohmic dissipation, or induction heating
(see Appendix B of \cite{corsha2008}), might not be as prevalent
in MS, GB or WD systems.  Further, pulsar radiation evaporates 
and ionizes infalling bodies.  When remaining charges are 
accelerated to relativistic energies, they can interact with 
the magnetic field to produce observable signatures \cite{broetal2014}.

{\bf IMPLICATIONS FOR SECOND-GENERATION FORMATION}
\ \ \ \
In the context of second-generation formation, relativistic 
particles can break apart complex nuclei and prevent 
metals from forming \cite{milham2001}. Therefore, shielding 
from particle radiation may be necessary to allow planet 
formation for a high-enough luminosity. 
\cite{milham2001} considered how the luminosity of a
pulsar changes as it is being spun-up, with implications
for second-generation planet formation (see Sec. \ref{secFall}).

{\bf IMPLICATIONS FOR FORMED SBs}
\ \ \ \
The putative asteroid disc that orbits PSR B1937+21
might be composed of many asteroid-sized objects on non-crossing orbits 
that range down to a few tenths of an au \cite{shaetal2013}.  
Consequently, the Yarkovsky
effect, driven by the magnetosphere from beamed X-rays, might play a 
significant role in the evolution of this disc.
They adopted the formalism of \cite{corsha2008} and find that 
only SBs with $R_{\rm SB} \gtrsim 5$ km could survive infall due to the
Yarkovsky effect if in the nascent disc $a \gtrsim 1$ au.
This estimation provides a strong constraint on the potential
present-day size distribution in the disc.

\section{Multi-body interactions} \label{secMulti}

The forces I described in Secs. \ref{secMass}-\ref{secRad} involve the star 
and SB only: effectively, 
the two-body problem.  Now I consider the $N$-body problem, where $N \ge 3$ and 
one of the bodies is the parent star.  Because the three-body problem is generally
unsolvable, $N$-body interactions -- particularly in post-MS systems with
so many additional forces involved -- are often modelled
with numerical simulations, a prominent theme in this section.  The
actual computer codes used by the authors are not mentioned here, but rather
in a dedicated section (Sec. \ref{secCodes}) for easier reference.

A single $N$-body numerical integration of a chaotic system is not a deterministic prediction.  Fig. 1b of \cite{lasgas2009} illustrates the maximum eccentricity that the planet Mercury will achieve over the next 5 Gyr: this value can vary by nearly unity for two simulations with equivalent initial conditions except for a variation of 0.38 mm in Mercury's initial semimajor axis.  The authors refer to the search of a precise solution over this interval as ``hopeless''.

Although the long-term MS orbital evolution of planets and planetesimals has 
been studied extensively (see \cite{davetal2014} for a review),
rarely have previous investigations carried out numerical integrations
for the entire duration of the MS.  Consequently, the initial orbital 
architectures for post-MS studies are poorly understood, much less the subsequent
multi-body interactions with the complexities introduced by GB evolution.

{\bf COMMON SIMULATION FEATURES}
\ \ \ \
A common assumption is that all SBs change their orbits simultaneously 
due to stellar mass loss.  In reality, there is an ejecta-lag crossing time.  
The orbit of the closest SB will shift before the orbits of any other SBs.  
The consequences of this assumption have not yet been explored in detail, 
but are likely to be greatest, as pointed out by \cite{veretal2013}, 
(i) for widely spaced orbits, and/or 
(ii) for non-adiabatic mass loss dynamics.  According to observations 
\cite{maeetal2012}, ejecta may take on the order of 100 days to travel just 
1 au.

Also, simulations which feature mass loss do not conserve energy.
Many, however, do conserve angular momentum.  Angular momentum
conservation then provides an important check on the accuracy of simulations
and a check on the physics.
\cite{beaetal2011} demonstrated how conservation of angular 
momentum arguments alone can help exclude additional bodies 
from playing a significant dynamical role (or existing at all)
during the history of a system.

In the common case of SBs experiencing adiabatic motion due to stellar
mass loss, while the system is stable, eccentricity variations can be 
attributed primarily to their mutual interactions rather than from mass
loss.  The true anomaly and eccentric anomaly evolution is also 
negligibly affected by mass loss.  These attributes are useful for 
disentangling the effects of mutual interactions and mass loss.

\subsection{Collisions within debris discs}  \label{secMultidisc}

Given the variety of forces affecting substellar bodies in post-MS systems,
debris discs may geometrically represent ``discs'' only in a loose sense.
Nevertheless, for the purposes of this section, assume that the disc can
be represented by an annulus with inner and outer boundaries represented
by $r_d^{({\rm min})}$ and $r_d^{({\rm max})}$, where $r$ is exactly in-between those two
values.

If mutual velocities are sufficiently high for collisions to be destructive, then collisional erosion continuously replenishes smallest bodies.  Therefore, the collisional lifetimes of bodies of different sizes are key parameters.  In general, the collisional lifetime, $t_{\rm coll}$, for particles of a given size is provided by Eq. 7 of \cite{wyaetal2007} as 

\begin{equation}
t_{\rm coll} = \frac{2 <i> r^{\frac{5}{2}} \left(r_{\rm d}^{({\rm max})} - r_{\rm d}^{({\rm \min})} \right)}
               {\sqrt{G}M_{\star}^{\frac{1}{2}} \Sigma_{\rm d} \xi b}
\label{eqcoll}
\end{equation}

\noindent{}with $\Sigma_{\rm d}$ representing the cross-sectional area,
$\xi$ representing the ratio of the relative collisional velocity 
to the Keplerian velocity, and $b$ representing the fraction
of the disc's cross-sectional area in which catastrophic collisions
with these large objects may occur.  Generally, $t_{\rm coll}$ decreases
as the disc approaches the central star.  The shorter the collisional lifetime,
the more quickly collisional equilibrium is reached.  In this case,
memory of the initial mass is erased.  However, collisional equilibrium 
might not hold for all evolved discs \cite{lohetal2008}.

Various specific cases of Eq. \ref{eqcoll} have been used in
the subsequent literature.  Among the most relevant for post-MS evolution
is contained within the study by \cite{bonwya2010}.  They find that 
the disc mass evolves according to (their Eqs. 20 and 21)

\begin{equation}
M_{\rm d}^{({\rm GB})} = M_{\rm d}^{({\rm MS})}
          \left[1 + C M_{\rm d}^{({\rm MS})} \int M_{\star}^{\frac{17}{3}} dt\right]^{-1}
\label{eqMdisc}
\end{equation}

\noindent{}where

\begin{equation}
C \equiv 
\left[2.8 \times 10^{-9} \left( \frac{r_{\rm d}^{({\rm max})} - r_{\rm d}^{({\rm min})}  }{r} \right)
e^{-5/3} \text{max}\left(R_{\rm SB}\right) \mathcal{Q}^{\frac{5}{6}}{r^{({\rm MS})}}^{\frac{13}{3}}
{M_{\star}^{({\rm MS})}}^{\frac{13}{3}}     
\right]^{-1}
\end{equation}
 
\begin{figure}[!t]
\centering\includegraphics[width=4.0in]{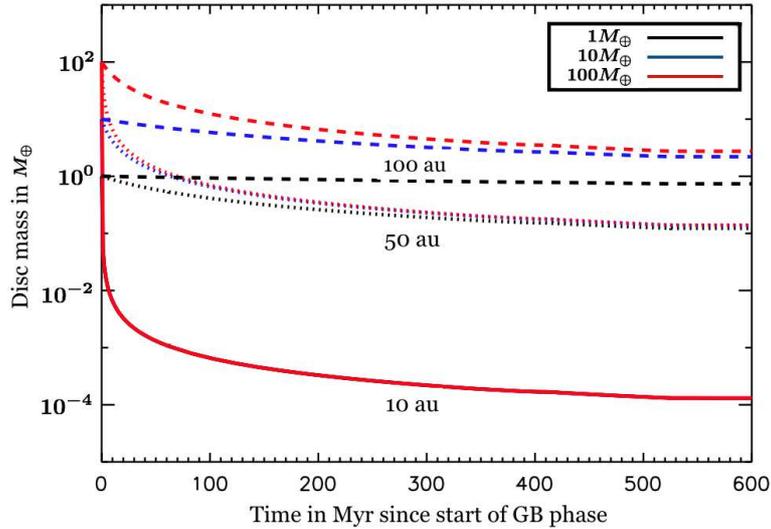}
\caption{Cosmetically enhanced version of the lowest panel of Fig. 4 of
\cite{bonwya2010}.  How GB evolution of a $M_{\star}^{({\rm MS})} = 2.9M_{\odot}$ star 
depletes a debris disc due to collisional
evolution alone.  The legend indicates the initial disc masses.
The solid lines correspond to 
$\lbrace r=10 \ {\rm au}, \ r_{\rm d}^{({\rm min})} = 7.5 \ {\rm au}, \ r_{\rm d}^{({\rm max})} = 12.5 \ {\rm au}  \rbrace$,
the dotted lines correspond to 
$\lbrace r=50 \ {\rm au}, \ r_{\rm d}^{({\rm min})} = 37.5 \ {\rm au}, \ r_{\rm d}^{({\rm max})} = 62.5 \ {\rm au}  \rbrace$,
and the dashed lines correspond to
$\lbrace r=100 \ {\rm au}, \ r_{\rm d}^{({\rm min})} = 75 \ {\rm au}, \ r_{\rm d}^{({\rm max})} = 125 \ {\rm au}  \rbrace$.
 }
\label{FigDiskDisp}
\end{figure}

\noindent{}and $\mathcal{Q}$ is the dispersal threshold for collisions.  Fig. 
\ref{FigDiskDisp} illustrates how disc mass decreases with time according
to Eq. \ref{eqMdisc}.

\subsection{One star, one planet and asteroids} \label{secmulti1s1pNa}

At least five investigations so far \cite{donetal2010,bonetal2011,
debetal2012,frehan2014,bonver2015} 
have modeled the post-MS interactions between
one planet and a collection or sometimes belt of smaller 
objects with numerical simulations.

\subsubsection{A Kuiper belt with a Neptune}

\cite{bonetal2011} modelled a $1M_{\odot}$ MS star that loses 2/3 of its mass
at a constant rate ($6.7 \times 10^{-6} M_{\odot}$/yr) over $10^5$ yr.
Because of the constant mass loss rate assumption, stellar metallicity
is not taken into account. 
They placed a Neptune-mass planet initially at 30 au in a circular
orbit, with 500 test
particles with initial semimajor axes from 30 to 47.6 au such that 
their initial eccentricities and inclinations
were less than 0.1, and 10 degrees, respectively. In effect,
this setup represents a cold exo-Kuiper belt.  Given the distances involved,
stellar mass loss is adiabatic in all their systems.

In order to develop initial conditions for the post-MS simulations, 
they integrated these systems for $10^7$ yr with a static parent star
(along the MS) to scatter away dynamically fragile members and settle
the belt. This timescale is typically three orders
of magnitude smaller than MS lifetimes, and was adopted to effectively
create a largely dynamically settled exo-Kuiper belt.

The simulations demonstrate that changing stability boundaries
from mass loss allow a fraction of the belt to be scattered inwards,
creating an inner reservoir of material after the star has become a WD.
If additional planets on circular orbits exist inward of the Neptune, 
then subsequent scattering with the test particles can be constrained 
analytically, and allow these particles to achieve WD-grazing orbits
\cite{bonwya2012}.

\subsubsection{An asteroid belt with a Jupiter}

\cite{debetal2012} simulated the orbital evolution of one planet with the
physical and orbital properties of Jupiter, along with 710 interior-lying
asteroids with orbital properties equivalent to Solar system asteroids.
The parent star contained $1M_{\odot}$ on the MS, and lost $0.46M_{\odot}$
over 1000 yr according to a prescription from \cite{schsmi2008},
corresponding to an average mass loss rate of $4.6 \times 10^{-4} M_{\odot}$/yr.
Ten simulations were run for 100 Myr, four for 200 Myr and three for 1 Gyr.
Their simulations tracked test particles until they reached $1R_{\odot}$, 
upon which they were removed.

\begin{figure}[!t]
\centering\includegraphics[width=2.7in]{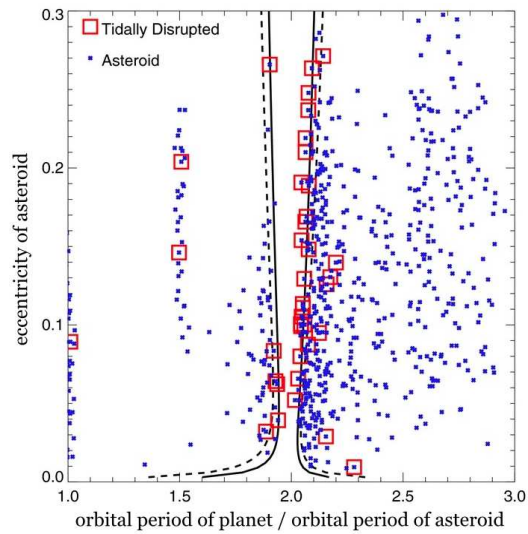}
\caption{Cosmetically enhanced version of Fig. 2 of
\cite{debetal2012}.  How the GB-induced expansion of the libration width of
the $2$:$1$ mean motion resonance between a planet and asteroids induces
the latter to be scattered towards and inside of the WD disruption radius.
The solid and dashed lines are the libration widths for, respectively,
$M_{\star}^{({\rm MS})} = 1.0M_{\odot}$ and $M_{\star}^{({\rm WD})} = 0.5M_{\odot}$.
Note that a couple asteroids near the $3$:$2$ mean motion commensurability
are also scattered towards the WD and disrupted.
}
\label{FigMMR}
\end{figure}

\cite{debetal2012} also emphasised the importance of resonant interactions between a planet
and smaller body.  They showed how the libration width of the interior $2$:$1$
mean motion resonance grows during mass loss, capturing test particles; see
Fig. \ref{FigMMR}.  The
eccentricity of these particles then increase until they potentially enter the
WD disruption distance.  Over all their simulations, about 2\% of test particles
followed this path; just one asteroid per simulation disrupted after 200 Myr.
They found that in order to match WD observations on aggregate, exo-asteroid
belts would need to be between $4 \times 10^0$ and $6 \times 10^5$ as massive as the Solar
system asteroid belt. The large range perhaps emphasizes that a single 
model has difficulty explaining the entire population of polluted WDs.

\cite{frehan2014} explored a wider region of phase space than \cite{debetal2012}
through a series of separate MS, GB and WD numerical simulations.
The MS simulations were run for 0.1 Gyr (about two orders of magnitude less
than a typical MS lifetime), the GB simulations were run for 2700 yr,
and the WD simulations were run for 1.0 Gyr.
Amongst these simulations the authors adopted planet masses of 
$0.03, 0.3, 1.0$ and $4.0 M_{\rm Jup}$ and eccentricities
of $0.02, 0.2, 0.4, 0.6$ and 0.8 (one planet per simulation).
In every case the planet's MS and resulting WD semimajor axes 
were 4 and 7.42 au, as the $1.0M_{\odot}$ MS parent star
was modelled to lose a total of $0.461 M_{\odot}$, and
at a constant ($1.7 \times 10^{-4} M_{\odot}$/yr) rate.  
The particle belt surrounded the planet in order to achieve
coverage of many first and second-order internal and external
mean motion commensurability locations.
They discussed the potential importance of the effects of wind drag and 
Yarkovsky forces (their Section 5.1), but did not include them in simulations.
These simulations importantly demonstrated that 
higher planetary eccentricities and smaller masses enhance the
accretion rate.

\subsubsection{Other configurations}

\cite{donetal2010} performed a set of simulations with a single planet
and a few planetesimals (see their Figs. 3-5).  The planet mass was a 
Jupiter-mass, and the planet was initially placed on a circular orbit at 
20 au. Planetesimals
of different sizes were placed on circular orbits at 35 au, and the $4M_{\odot}$
parent star was made to lose mass linearly at a rate of $3 \times 10^{-5} M_{\odot}$/yr 
for $10^5$ yr. Unlike the above four studies \cite{bonetal2011,
debetal2012,frehan2014,bonver2015}, wind drag (see Section \ref{secMass}\ref{secMasswind}) 
was also included 
in the \cite{donetal2010} simulations. They illustrated that in this setup, capture into 
the external $3$:$2$ mean motion resonance occurs.

\subsection{One star, multiple planets and no asteroids} \label{secMulti1sNp0a}

\subsubsection{Few-planet stability boundary changes}

{\bf TWO PLANETS}
\ \ \ \
The discovery of the post-MS PSR B1257+12 planets in 1992 
\cite{wolfra1992} provoked
a resurgence of interest in celestial mechanics and in particular
the three-body problem.  The following year, \cite{gladman1993} 
conveyed a concept from algebraic topology -- that of Hill stability
(in a Hill stable system, two planetary orbits never cross)
-- to the wider astronomical audience by expressing the relevant
criterion in orbital elements and in the context of two coplanar small
planets orbiting a star.  A decade later, \cite{debsig2002} then illustrated 
how in post-MS systems,
mass loss can convert a Hill stable system into an unstable one.
A decade after that, \cite{veretal2013} considered this transition
to Hill instability with arbitrarily eccentric and inclined orbits.
However, mounting
progress in characterising instability in the few-body problem (see, e.g.
\cite{georgakarakos2008} for a review) is not limited to Hill stability.
Several authors have since 
popularised Lidov-Kozai motions, gave attention to Lagrange instability
(a proxy for the global stability boundary), and generally 
provided an appreciation for orbital architectures previously
assumed to be unorthodox and perhaps ignorable 
(see \cite{davetal2014} for a review).

{\bf THREE PLANETS}
\ \ \ \
For systems with more than two planets, Hill stability no longer applies.  
Instead, authors have introduced the concept of a critical separation 
for instability where a single 
relation links the consecutive pairwise planet separations.  
\cite{musetal2014} derived the following estimate for the critical
separation $\Delta$ (adapted from their Eqs. 2 and 6) assuming three equal-mass planets

\begin{eqnarray}
\Delta 
&\approx& 0.7 a_{\rm inner} \left( \frac{M_{\rm SB}}{3M_{\odot}}  \right)^{\frac{1}{3}} 
\left( \frac{M_{\rm SB}}{M_{\rm Jup}} \right)^{-\frac{1}{12}}
\left( \frac{M_{\star}^{({\rm i})}}{M_{\odot}} \right)^{\frac{1}{12}}
\left( \frac{M_{\star}^{({\rm f})}}{M_{\star}^{({\rm i})}} \right)^{-\frac{1}{4}}
\nonumber
\\
&\times&
\bigg[
\log{\frac{t_{\rm f} - t_{\rm i}}{\rm Myr}} 
-
\frac{1}{2}\log{\frac{M_{\star}^{({\rm i})}}{M_{\odot}}}
+
\log{\frac{M_{\rm SB}}{M_{\rm Jup}}}
-
\frac{3}{2}\log{\frac{a}{\rm au}}
+
\log{\frac{M_{\star}^{({\rm f})}}{M_{\star}^{({\rm i})}}}
+11.0
\bigg]
\end{eqnarray}

\noindent{}where the superscripts i and f refer to initial and final states, 
and not necessarily MS and WD. This relation helps show that higher mass 
stars produce greater amounts of instability because the planet-star mass 
ratio becomes more unequal.

{\bf MORE THAN THREE PLANETS}
\ \ \ \
For more than three planets, \cite{dunlis1998} found a roughly linear
relationship between instability time and amount of stellar mass lost.
\cite{vergae2015} quantified the initial separations for which the planetary system would significantly alter its initial configuration (or become unstable, by some definitions).  They adopted the mutual separation distance formulation (see their Eq. 1) from \cite{smilis2009}, which is measured in number of ``mutual Hill radii'', denoted by $\beta$.  \cite{vergae2015} found that closely-packed giant planets with $\beta \approx 10-12$ typically unpack or become unstable during the post-MS phases.  For terrestrial planets, instead $\beta \approx 6$.

\subsubsection{GB and WD planet simulations}

{\bf TWO PLANETS}
\ \ \ \
In 2002, \cite{debsig2002} demonstrated that adiabatic mass loss changes the stability 
boundary in multi-planet systems, and can induce instability in previously stable 
systems.  They illustrated that although the ratio of semimajor axes of two 
planets remained unchanged through the GB phases, the critical Hill stability 
separation does not scale in the same way.  Consequently, during GB mass loss,
two planets can cross this boundary.  They performed numerical simulations 
which involved increasing the mass of the planets rather than decreasing the 
mass of the star.  The planet mass increase corresponded to a halving of the 
stellar mass over $10^5$ orbits of the inner planet; the simulations were 
run for up to $10^7$ of these orbits.  The authors adopted planet-star mass 
ratios of $10^{-7}$, $10^{-5}$ and $10^{-3}$.  

The complementary work of \cite{voyetal2013} a decade later 
explored how nonadiabatic mass loss
changes stability in two-planet systems. For real systems with stars that become 
WDs, this situation would arise only for two SBs which are at least several hundred 
au away from the GB star.  They found that the resulting changes in eccentricity 
enhance the prospects for instability.  One way of tracing the susceptibility of 
a two-planet system to become unstable over a Hubble time (or current age of the Universe) 
is to track the Lyapunov exponent
\cite{adaetal2013,voyetal2013}.  \cite{adaetal2013} found that although
the Lyapunov exponent varies by orders of magnitude depending on the mass loss 
timescale, both timescales are comparable.

\begin{figure}[!t]
\centering\includegraphics[width=3.0in]{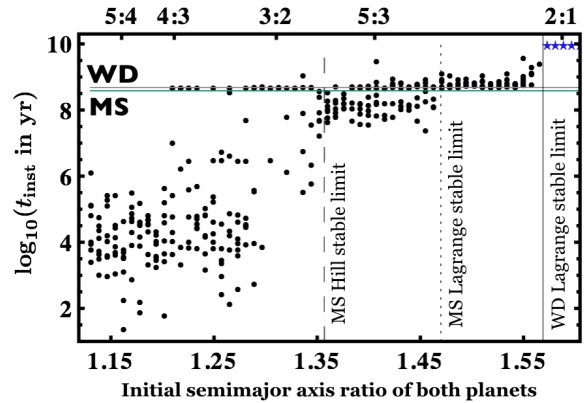}
\caption{Cosmetically enhanced version of the bottom-rightmost panel
of Fig. 9 of \cite{veretal2013}.  I show instability timescales, 
$t_{\rm inst}$, as black dots for individual
two-planet simulations with the given initial semimajor axis 
ratio and all with a parent 
star of $M_{\star}^{({\rm MS})} = 3.0M_{\odot}$.  Blue stars
indicate that all simulations sampled at that particular semimajor 
axis ratio were stable over a total of 5 Gyr of evolution. The 
two coloured horizontal lines represent the RGB
and AGB phases, and the upper axis illustrates some mean motion 
commensurabilities.  
This plot demonstrates that (i) systems which are stable on the MS 
may become unstable beyond the MS, and (ii) that Lagrange instability on the MS can manifest
itself late, right before the RGB phase.}
\label{FigHill}
\end{figure}

Motivated by instability induced by crossing the Hill stable boundary,
\cite{veretal2013} performed numerical integrations 
incorporating both planetary and stellar evolution for two planets 
across all post-formation phases of evolution.  They adopted 
$a_{\rm inner} = 10$ au, $M_{\star}^{({\rm MS})} = \left[3,4,5,6,7,8\right]M_{\odot}$, 
and ran simulations for $5$ Gyr, a long enough period to cover the MS
and GB phases 
and several Gyr on the WD phase (see their Fig. 7).  They primarily adopted 
initially coplanar planets with Jovian masses, eccentricities of 0.1, 
and initial separations that cover a range of possible 
stability outcomes (see Fig. \ref{FigHill} of this paper).

Their results demonstrated that (i) Post-MS mass loss alone can trigger instability not only during the GB phase, but well after.  In fact, because in almost every case the mass loss is adiabatic, the architecture of the system inflates without a noticeable initial change.  But that process in effect ``resets'' the system, so that subsequent dynamical instability occurs on timescales similar to those one might expect after formation on the MS; see Fig. \ref{FigHill} of this paper. (ii) The ejection of one planet perturbs the other onto a moderately to highly eccentric orbit, sometimes with an orbital pericentre of just a few au. This eccentric planet could then perturb any interior asteroids (not modelled), providing a dynamical environment for pollution. Due to conservation of angular momentum, the semimajor axis of the surviving planet is well-approximated analytically (see their Eq. 8 and Figs. 14 and 16), unlike its eccentricity, which can be determined only if the hyperbolic eccentricity and semimajor axis of the escaping planet are known.

{\bf THREE PLANETS}
\ \ \ \
In a follow-up paper that integrated systems with three Jovian planets instead of two, 
\cite{musetal2014} improved upon 
\cite{veretal2013} in several respects, including: (i) They explored multi-generational instability.
For example, with three planets, one instability can occur on the MS, with the second on the WD
(see their Fig. 5).  In about 1\% of all cases, all three planets were lost. (ii) They more accurately quantified the fraction of unstable post-MS systems which lead to direct planet-WD collisions by adopting the WD disruption radius instead of the actual WD radius, and by giving the three planets slightly noncoplanar inclinations (in line with Solar system observations).  They found that, like in the two-planet case, ejection during the WD phase is the most commonly occurring type of instability (see their Fig. 4).  These ejected planets help ensure that WD planets rarely evolve to a hot Jupiter state, and provide a potentially non-negligible contribution \cite{verray2012} to the purported free-floating planet population \cite{sumetal2011}.

\begin{figure}[!t]
\centering\includegraphics[width=3.5in]{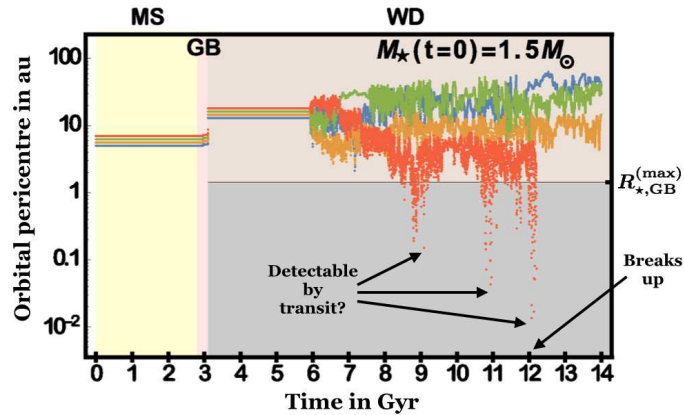}
\caption{Cosmetically enhanced version of the upper-left panel
of Fig. 1 of \cite{vergae2015}. Late unpacking of four tightly-packed terrestrial planets
throughout the MS and GB phases.  The orbit meandering which follows scattering instability
perturbs the red planet into a likely transit-detectable orbit before it enters the WD disruption
radius.}
\label{FigUnpacking}
\end{figure}

{\bf MORE THAN THREE PLANETS}
\ \ \ \
\cite{vergae2015} then extended the work of \cite{veretal2013} and \cite{musetal2014} by adopting $M_{\star}^{({\rm MS})} = 1.5-2.5M_{\odot}$, which corresponds to the primary source of progenitors of the currently observed WDs (see Fig. \ref{FigSpec}), and by considering the evolution of packed systems with four and ten planets. They also integrated these systems for the current age of the Universe, 14 Gyr, and focused on lower-mass, terrestrial planets.  As Fig. \ref{FigUnpacking} demonstrates, these packed, stable and quiescent terrestrial planets on the MS can remain so throughout the GB phase, and for several Gyr along the WD phase, before ``unpacking''. Subsequent to unpacking, the planets may meander and approach the WD closely enough for tidal circularisation to occur, increasing prospects for transit detectability by decreasing the orbital period. 
Interactions between these terrestrial planets were not strong enough to cause escape in any of their simulations, unlike for the giant planet case, representing a fundamental difference between \cite{veretal2013} and \cite{musetal2014}.  

{\bf MORE THAN THREE PLANETS PLUS MOONS}
\ \ \ \
The presence of exo-moons in WD systems can provide a potentially
important mass reservoir for heavy metal pollution. In our Solar system,
the total mass of moons exceeds that of the asteroid belt by over two
orders of magnitude.  Using the data from the simulations of \cite{vergae2015}, 
\cite{payetal2016} considered how susceptible moons are to 
planet-planet scattering which occurs during the WD phase.  They 
found that after adiabatic mass loss, the Hill radius
$r_{\rm Hill}$ of a planet increases according to (their Eq. 7):

\begin{equation}
\frac
{r_{\rm Hill}^{({\rm WD})}}
{r_{\rm Hill}^{({\rm MS})}} 
=
\left(
\frac
{M_{\star}^{({\rm MS})}}
{M_{\star}^{({\rm WD})}}
\right)^{4/3}
.
\label{HillMoon}
\end{equation}
 
\noindent{}Nevertheless, even though moons become more entrenched, and hence
stable, after post-MS mass loss, \cite{payetal2016} showed that planet-planet
scattering on the WD phase is highly destructive to moons.  They demonstrated
that moons may be regularly liberated from their parent planets. Although the
ultimate fate of the moons was not explicitly tracked, the moons may
subsequently be perturbed into WDs directly or act as agents to perturb smaller
bodies, fragments or dust into WDs.

One earlier work \cite{dunlis1998} performed a variety of multi-planet simulations with different planet-star mass scalings which mirror those measured in the Solar system.  In some cases they treated mass loss as constant, and in others used a fitted, observationally-motivated prescription.  The latter case helps demonstrate that the four outer Solar system planets are likely to survive post-MS mass loss and remain stable at least for tens of Gyr (see Sec. \ref{secFate}).

\subsubsection{Pulsar planet simulations} \label{secMulti1sNp0aPul}

The exquisite data available for PSR B1257+12 \cite{konwol2003} 
(with e.g. eccentricity accuracy at the $10^{-4}$ level)
represents an opportunity to study the dynamics of a multi-planet system 
at a standard of detail that is largely unattainable in MS planetary systems.
\cite{gozetal2005} performed an extensive analysis of the system with 1 Gyr
numerical integrations, stability maps and Lyapunov exponent analyses.  
They found (i) the system is long-term stable,
(ii) none of the planets experience (notable) secular variations in semimajor axis,
eccentricity nor inclination, (iii) the outermost two planets are locked
in a secular apsidal resonance with a libration centre of $180^{\circ}$,
and (iv) the line of nodes of the inner planet must be within 60 degrees
of the nodes of the outer planets.

\subsection{Two stars, planets and no asteroids} \label{secMulti2sNp0a}

Simulations for this physical setup cover a wide range of physical
architectures and numerical codes.  Here, I divide previous 
investigations according to whether or not the numerical codes
adopted model stellar evolution.

\subsubsection{Simulations not incorporating stellar evolution}  \label{secMulti2sNp0aWOS}

{\bf PSR B1620-26}
\ \ \ \
The fascinating PSR B1620-26 system contains a millisecond pulsar (PSR B1620-26A),
a WD (PSR B1620-26B) and a circumbinary planet (PSR B1620-26AB b) \cite{sigetal2003}.
The dynamical pathway leading to this scenario is thought to proceed along the 
following lines: 
(i) the planet forms around the MS progenitor of the
currently-observed WD,
(ii) that MS star-planet system flies by a NS and its original unknown companion
within the M4 globular cluster, (iii) this flyby induces an exchange reaction, where
the NS's companion is ejected and the NS captures the star and planet,
(iv) the planet begins orbiting both stars in a circumbinary fashion, 
(v) the MS star evolves into a GB star,
(vi) mass transfer from the GB onto the NS spins the latter up and converts it
into a millisecond pulsar, (vii) the GB star evolves into a WD.  The result is
the currently observed system.

The third part of this scenario was simulated with four-body
dynamics by \cite{sigurdsson1993}, and remains an accurate qualitative description
of the exchange despite the now-outdated orbital parameters used.  This
exchange occurs too quickly for stellar evolution to play a role.

{\bf POST-CE BINARIES}
\ \ \ \
The future dynamical stability of the putative post-CE binary planetary systems 
(Sec. \ref{secOM}\ref{secOMPCEB}) have been explored with simulations in many studies:
for HU Aquarii (HU Aqr), \cite{horetal2011} and \cite{witetal2012}; for
QS Virginis (QS Vir), \cite{horetal2013}; for NSVS 14256825, \cite{witetal2013};
for NN Serpentis (NN Ser), \cite{horetal2012}.  The overall conclusions from these
investigations is that all systems except for NN Ser undergo instability on
short ($\sim 10^3-10^4$ yr) timescales, placing doubt on the planetary interpretation
of the signals.  However, a few isolated long-term stable solutions also do 
exist for HU Aqr.
\cite{maretal2014} criticised the above studies by claiming that the initial conditions 
for their simulations are not consistent with the data, as the strong correlations 
between the orbital parameters were not taken into account.  \cite{maretal2014} 
does not necessarily believe that the qualitative stability conclusions of these 
other papers would change as a result, but thinks a re-evaluation is needed to be sure.

\subsubsection{Simulations incorporating stellar evolution}  \label{secMulti2sNp0aWS}

The intriguing dynamics of post-MS evolution in binary systems which hosts planets
has been explored with stellar evolution simulations 
by only a handful of investigations.  The first two presented here 
\cite{musetal2013,portegieszwart2013} took alternate approaches
to studying two of the post-CE binary systems from the last section: NN Ser and HU Aqu.
The third \cite{kraper2012} is an altogether different type of study that
considered how a planet can ``hop'' from one star to another, and the fourth
\cite{vertou2012} quantified where circumbinary planets are subject to escape.

{\bf NN SER}
\ \ \ \
For the purposes of investigating the stability of NN Ser,
\cite{musetal2013} took an alternate approach from \cite{horetal2012} 
by looking at its past rather than its future.
If the putative planets could not maintain stable orbits on the MS,
then the current observations are of either second-generation planets,
or of other phenomena entirely.  Through the GB phases, 
the planets' orbits increase due to adiabatic mass loss (see Sec. \ref{secMass}) 
while the stellar binary orbit shrinks because star-star tides dominate the 
evolution of their mutual orbit (see their Fig. 1). Therefore, the MS
stability of this system is more precarious than after the CE because
$a_{\rm inner}^{({\rm MS})} < a_{\rm inner}^{({\rm GB})}$, 
$a_{\rm outer}^{({\rm MS})} < a_{\rm outer}^{({\rm GB})}$
and 
$a_{\rm b}^{({\rm MS})} > a_{\rm b}^{({\rm GB})}$. 

To perform this investigation, (i) they computed what MS orbital and physical
parameters of the binary correspond with current observations by using a stellar
evolution code, (ii) analytically determined, by assuming adiabatic mass loss
and negligible other forces, the MS orbits
of the putative planets, and (iii) finally integrated them along the MS
with a planetary evolution code to assess their stability.  They found that
in nearly all ($99.74$\%) cases the planets become unstable before the end of
the MS of the higher mass star, suggesting that the currently observed 
planet candidates do not have a first-generation origin.

{\bf HU AQR}
\ \ \ \
The putative planets around the cataclysmic variable HU Aqr are less certain to 
exist because of the widespread instability found amongst the simulations
performed by \cite{horetal2011} and \cite{witetal2012}.  Nevertheless, the existence
of at least one long-term stable solution prompted \cite{portegieszwart2013} to
perform a ``reconstruction'' of the dynamical history of that system through
its CE phase.   

As a first step, \cite{portegieszwart2013} determined what physical and orbital 
binary system parameters reproduce the system seen today.  Then, he integrated 
the planets' orbits by making
assumptions about the CE, such as the CE resulting in differing mass loss rates
felt by both planets. An important outcome of that study is the ability to restrict 
CE parameters, assuming that the planets exist now in the post-CE phase.
He found that the fraction of orbital energy needed to unbind the CE during
inspiral was $45\% \pm 17\%$.

{\bf PLANET HOPPING}
\ \ \ \
\cite{kraper2012} explored the exciting possibility that
an SB can unbind from its parent star due to stellar mass loss
and then be captured by the stellar companion, performing a ``hop''.
This mechanism works because the change in the SB's orbit
is greater than the binary companion's if the expansion of both
are adiabatic (a realistic possibility for $a_{\rm b} \lesssim 10^3$ au).  
Suppose the binary companion is an MS star which remains as such
throughout the primary star's post-MS evolution. Suppose also that
the primary star becomes a WD.  Then assuming
full adiabaticity, $M_{\rm SB} \ll M_{\star}$, and $M_{\rm SB} \ll M_{\rm b}$, 
I obtain (see also Eq. 6 of \cite{perets2011})

\begin{eqnarray}
a_{\rm b}^{({\rm WD})} &=& a_{\rm b}^{({\rm MS})}
\left(
\frac
{M_{\star}^{({\rm MS})} + M_{\rm b}}
{M_{\star}^{({\rm WD})} + M_{\rm b}}
\right)
,
\\
a_{\rm SB}^{({\rm WD})} &=& a_{\rm SB}^{({\rm MS})}
\left(
\frac{M_{\star}^{({\rm MS})}}{M_{\star}^{({\rm WD})}}
\right)
.
\end{eqnarray}

\noindent{}The other orbital parameters remain the same in the
adiabatic limit (see Sec. \ref{secMass}\ref{secMassnonacc}). 
The differential orbital expansion of the binary and planetary
orbit in addition to the mass change can (i) allow a stability threshold to be crossed,
or, as explored by \cite{bonver2015}, (ii) allow the binary
orbit to be wide enough to incite future instability due
to Galactic tides. If the binary orbital expansion in non-adiabatic, 
but the planetary expansion is adiabatic, then a previously stable
system might remain so, depending on the effect of Galactic
tides.

As explained further by \cite{kraper2012} and visualised in their
Fig. 2, in order for the SB to be captured by the binary companion, 
the Jacobi constant must adhere to certain values.  Although defined for
the circular restricted three-body problem with fixed masses (including
one massless particle), the Jacobi constant still represents a reliable proxy
for capture in post-MS binaries with a single planet.

\cite{kraper2012} performed simulations with 70 different binary configurations
which include different stellar masses, binary separations in the range
$a_{\rm b}^{({\rm MS})} = 75-105$ au and a fixed $a_{\rm SB}^{({\rm MS})} = 15$ au,
$e_{\rm b}^{({\rm MS})} = 0.0$,  and $e_{\rm SB}^{({\rm MS})} = 0.0$.  They ran
simulations for the entire duration of mass loss, plus 30 Myr. Their outputs
show that about 10\% of all SBs are captured by the companion, with typically
prograde eccentric orbits satisfying $e_{\rm SB}^{({\rm MS})} > 0.4$.
Unstable systems typically featured a planet-star collision.

{\bf CIRCUMBINARY ESCAPE}
\ \ \ \ 
\cite{vertou2012} instead determined
where in phase space will a circumbinary SB's evolution become non-adiabatic
(defined as $\Psi > 1$; see Eq. \ref{eqpsi}) and consequently in danger 
of escape due to post-MS
evolution.  They illustrated this
critical SB semimajor axis in a series of contour plots, which explored dependencies
on stellar masses (both $M_{\star}^{({\rm MS})}$ and $M_{\rm b}^{({\rm MS})}$ ranging 
from $0.2M_{\odot}$ to $8M_{\odot}$),
initial binary semimajor axis ($a_{\rm b} = \left[10,50,100,500,1000,5000\right]R_{\odot}$),
initial binary eccentricity ($e_{\rm b} = \left[0.0,0.5,0.9\right]$), stellar metallicity
($Z_{\star} = Z_{\rm b} = \left[0.0001,0.0005,0.001,0.005,0.01,0.02\right]$) and CE-blowoff timescale
($[10, 10^2, 10^3, 10^4]$ yr).  The SBs themselves were not included in these
stellar evolution simulations.  

\cite{vertou2012} found that the complex physics of 
binary star evolution, which include CEs and SNs, reduces the adiabatic limit significantly 
(to several tens of au) from the single-star case (at several hundreds of au).
Therefore the post-MS contribution to the free-floating exoplanet population \cite{verray2012} 
is expected to be higher for circumbinary rather than single star systems. 
Their study also backward-engineered the adiabaticity condition to yield a potentially
useful constraint on stellar mass loss.  Suppose a SB is bound to two MS stars and 
harbours an initial
semimajor axis $a_{\rm CM}^{({\rm MS})}$ with respect to the centre of mass of the stars.  Now the primary
star becomes a WD.  During that process, the maximum value of $|\dot{M}_{\star}|$
which guarantees that the SB remains bound is (corrected from their Eq. 6, which should
contain an extra factor of $G^{1/2}$; their Eq. 4 is also missing a factor of $G^{-1/2}$)

\begin{equation}
{\rm max}(-\dot{M}_{\star}) \approx 
\sqrt{G}
\left[   
\frac
{1.4M_{\odot} + M_{\rm b}}
{ \sqrt{a_{\rm CM}^{({\rm MS})} \left(M_{\star}^{({\rm MS})} + M_{\rm b}\right)}  }
\right]^3.
\end{equation}

\noindent{}This result could be generalised to higher multiplicity cases or simplified
to the single-star case (see their Fig. 1).

\subsection{Two stars, one planet and asteroids} \label{secMulti2s1pNa}

\cite{bonver2015} considered a setup with an exo-Kuiper belt and a planet
orbiting one star of a wide stellar binary ($r_{\rm b} \sim 10^3-10^4$ au).  
After the parent star has become a WD  
the two stars experience a close encounter due to Galactic tides 
(see Sec \ref{secOther}\ref{secOtherExternal} for more details). 
This encounter perturbs the planet, which is turn perturbs the asteroids
into the WD, polluting it.  They model a 
Neptune-mass planet on a circular orbit at 30 au, and a belt of test 
particles from 30-50 au.  These simulations did not require the parent star
to evolve because it was already a WD.  The dynamics of the encounter 
between the secondary star, planet and belt reveals that the more 
intrusive the encounter, the higher the rate of inwardly scattered particles.

\subsection{Three stars only} \label{secMulti3s0p0a}

Although focused on the dynamics of hierarchical triple stellar systems
with mass loss and without SBs, both \cite{shatho2013} and \cite{micper2014} derived results 
which may be applicable to SBs in the appropriate limits.  \cite{shatho2013}
modelled linear and adiabatic mass loss from a component of the inner
binary with $N$-body integrations, and considered the resulting change
or induction of Lidov-Kozai behaviour.  \cite{micper2014} instead adopted a 
Hamiltonian formalism, derived
doubly-averaged equations including mass loss up to octupole order, and 
derived the secular equations of motion in the case when all mass loss
is adiabatic.  They also consider the general case when all three bodies lose mass.
Secular equations allow for relatively quick integrations over long timescales,
but by definition cannot resolve characteristics on orbital timescales.

\section{Formation from stellar fallback} \label{secFall}

Some mass ejected by post-MS stars might not escape the system,
and infalling material may be decelerated 
due to phenomena such as reverse shock waves \cite{chevalier1989}.
Both WDs and NSs could host remnant discs of matter through
which ``second-generation'' SBs form.  First-generation SB
formation proceeds at timescales less than 10 Myr.  If second-generation
formation is similar, then we should not expect it to occur
later than WD or NS cooling ages of 10 Myr.  Consequently, because the vast
majority of observations of remnant planetary systems (Sec. \ref{secOM}\ref{secOMWDpoll})
are snapshots in time where $t_{\rm cool} \gg 10$ Myr, we are currently not
observing activity from stellar fallback.  Further, the chemical imprint
of pollution observed within WD atmospheres is incongruous with
the chemical composition of stellar mass ejecta.

Nevertheless, the prospect of second-generation formation of SBs at very early
WD and NS cooling ages remain viable, and have been explored by
researchers for decades.  The PSR B1257+12 system was a strong
driver for this research \cite{phihan1993,podsiadlowski1993},
and recent pulsar observations \cite{shaetal2013,broetal2014} 
provide further motivation.  Incentive also arises from
putative planetary systems whose first-generation provenance
is disputed based on dynamical grounds (as in \cite{wicetal2010,musetal2013}).
In an exceptional system like NN Ser, where the WD is known to harbour
$t_{\rm cool} \sim 1$ Myr \cite{zorsch2013} , a second-generation planet 
would be the youngest known exoplanet, and help constrain the formation
process.

Second-generation formation provides intriguing possibilities for
planetary system evolution.  \cite{perets2011} highlighted some 
important concepts on this topic:
(i) post-MS evolution opens up dynamically ``previously forbidden'' 
regions of SB formation, (ii) pre-existing first-generation planets may interact
with newly-formed second-generation planets,
(iii) when each of two binary components undergoes post-MS evolution,
there exists the prospects for third-generation formation, which is
distinct from second-generation formation and (iv)  second- and third-generation 
discs may be metal-enriched in order to mirror the expected composition in 
stellar ejecta.  These systems may conspicuously stand out in very 
metal-poor environments like globular clusters (on their outskirts, 
from dynamical stability) or simply around metal-poor stars.

\subsection{Post-common envelope formation around white dwarfs} \label{secFallCE}

The amount of disc mass available for planet
formation is a fraction of the amount
of mass that remains bound after the CE.
The process for planet formation then proceeds along similar lines
to the MS case, except here the disc may be influenced by photoheating,
photoionization and radiation pressure (see Sections 4.1-4.3 of \cite{schdre2014})
in different manners.  The metal content may also be different.

\cite{beasok2014} used angular momentum arguments to suggest
that planets orbiting post-CE binaries are more likely to be first
generation than second generation.  Their Table 1 displays the minimum
angular momentum of the putative post-CE binary planetary systems,
as well as angular momentum efficiency factors for each system.
The crucial unknowns are the efficiency factors, which determine
(i) how much initial angular momentum is deposited into planets, 
and (ii) how much of the envelope ends up in the disc.
For NN Ser in particular, \cite{schdre2014} provided in 
their Table 1 some model parameters for the formation by 
gravitational instability.  \cite{voletal2014} observed that 
as second-generation planets form in systems like NN Ser, 
the formation disc may still be expanding.

In WD-AGB binary systems, the formation of a second-generation disc 
may be quantified in a similar way as in a MS-AGB binary system.
Through angular momentum arguments, \cite{perken2013} provided 
a criterion for second-generation disc formation around the 
primary star from the wind of a binary companion (their Eq. 1):

\begin{equation}
1 \lesssim 
\left( \frac{M_{\star} + M_{\rm b}}{2.5 M_{\odot}} \right)
\left( \frac{M_{\star}}{M_{\odot}} \right)^{\frac{3}{2}}
\left( \frac{R_{\star}}{R_{\odot}} \right)^{-\frac{1}{2}}
\left( \frac{a}{100 \ {\rm au}} \right)^{-\frac{3}{2}}
\left( \frac{ \left|v_{\rm b}^{({\rm wind})} - v\right| }
{10 \ {\rm km} \ {\rm s}^{-1}} \right)^{-4}
.
\end{equation}

\subsection{Post-supernova formation around neutron stars} \label{secFallSN}

{\bf GENERAL CONSIDERATIONS}
\ \ \ \
\cite{milham2001} and \cite{hansen2002}, and references therein, summarized multiple possible scenarios for the formation of a circumpulsar disc: (1) a disc forms from binary mass transfer, where the donor is disrupted or evaporated, (2) a disc forms from donor ejecta, then the donor is evaporated, (3) a disc forms from donor ejecta, and then the donor goes SN and gravitationally unbinds the disc-bearing pulsar, (4) a disc forms from ejecta from a SN-induced direct collision (where the SN kick was in the direction of the other star), (5) a disc forms from ejecta from a WD-WD direct collision, which could form a pulsar and an accompanying disc.  Second-generation pulsar planets do not necessarily need a disc if the planets either formed directly from fallback matter or the companion was evaporated just the right amount to achieve a planet mass.

\cite{milham2001}, which does not favour first-generation scenarios for the known pulsar planets, also placed restrictions on post-SN second-generation formation through stellar evolution arguments: If a pulsar was transformed into a millisecond pulsar by accreting matter, then they argue that the luminosity from the accretion would evaporate the planet. If this accretion process occurred and left a remnant second-generation disc, then the authors argued that stellar luminosity would disperse the disc too quickly too allow for formation.

In a wide-ranging dynamical study of second-generation formation and evolution, the cleverly-titled \cite{corsha2008} covered collisional migration, the Yarkovsky effect, magnetospheric currents, electromagnetic capture, heating and evaporation of SBs, and disc lifetimes.  Their main finding is that second-generation asteroids, but not planets, could easily form and create observable signatures in systems like B1931+24.  This scenario relies on (1) the SN fallback disc to be metal-rich and ``sufficiently compact and low in mass ($10^{-6}M_{\odot}$) to prevent'' massive planet formation, (2) the mass inflow into the light cylinder to be large enough to influence magnetospheric current flows on relevant timescales, and (3) the disc survival time must be long enough ($\ge 10$ Myr) to alter the electrodynamics of old pulsars.  A couple interesting ancillary findings include the fact that (i) rocks with high tensile strengths can form metal-rich gas inside the tidal disruption radius.  These second-generation rocks are not like the strengthless rubble piles seen in the Solar system because they are metal-rich and fractionated. Further, (ii) the radiative effect on second-generation asteroids near the pulsar's light cone are comparable to currently observed Solar system Yarkovsky drift rates (about $10^{-3}$ au/Myr).

{\bf PSR B1257+12}
\ \ \ \
Before delving into the possibilities for second-generation formation in PSR B1257+12,
one may ask if the planets there have a first-generation origin.  To help answer
this question, \cite{veretal2011} considered
the likely value of $\Psi$ (see Eq. \ref{eqpsi}) for the PSR B1257+12 planets during their 
SN mass loss. Even though the mass is lost from SN quickly, the small MS semimajor axis values
($ < 0.1$ au) of all three planets was too small to place them in the strongly nonadiabatic
regime.  Actually, $\Psi \sim 0.1-1.0$, and therefore the planet eccentricities
were not pumped sufficiently high to cause ejection. Therefore, in this one respect,
a first-generation scenario remains plausible.

For a second-generation origin, \cite{curhan2007} performed a dedicated modeling suite 
for the PSR B1257+12 disc.  They considered both fully viscous discs and layered 
accretion discs.  
They found that discs typically contain material out 
to 1-2 au, and are sufficiently massive to produce the  PSR B1257+12 planets.
Because the gas in these discs dissipates in under 0.1 Myr, gas giants are 
unlikely to form.  The follow-up study of
\cite{hanetal2009} considered how 
planetesimals from second-generation discs would assemble 
into a planetary architecture. 
They found that because the density of solid material must be 
sufficiently high for planet formation to commence, one can impose 
restrictions on the speed of dust sedimentation 
during disc cooling.  Their study provides further support of the 
post-SN fallback scenario (as opposed to a binary companion 
disruption scenario) for the origin of the PSR B1257+12 planets.

\subsection{Formation from tidal disruption of companions} \label{secFallTides}

{\bf PULSAR STELLAR BINARIES}
\ \ \ \
The PSR B1257+12 planets may have been formed through one of a few
different second-generation channels.  \cite{curhan2007} provided
a comparison. Instead of forming out of stellar fallback, these planets might 
have formed out of shorn-off binary star matter.  As opposed to fallback discs, 
with angular momentum values as low as $10^{42}$ J/s, disruption discs could 
harbour angular momentum values three orders of magnitude higher. Because fallback 
discs tend to be more compact than disruption discs, a fallback disc is more likely 
to reproduce the planets in the PSR B1257+12 system.  Similar to the SN fallback disc, 
in a SN disc formed from tidal disruption gas dissipates too quickly to form giant planets.

In a similar vein, \cite{shaetal2013} suggested that if the putative asteroid disc around PSR B1937+21 is a second-generation disc, then it might have formed from the ablated material of a disrupted stellar companion.  They proposed that the resulting disc was too tenuous to form planets, which is why none are detected.  Instead, the planet formation process would stop after a brief period of runaway growth, when asteroids were formed.  Formation could occur throughout a radial distance of several au, because the disc would not be truncated by the presence of planets.  This second-generation disc itself might feature many SBs on non-crossing orbits and whose evolution is described in their Sections 4.1-4.2.  They suggested that inwardly-migrating SBs will never reach the star's magnetosphere before being evaporated.

{\bf WD STELLAR BINARIES}
\ \ \ \
\cite{livetal2005} predicted the existence of second-generation debris discs, and possibly planets,
around massive WDs as a result of a previous WD-WD merger. They envisaged the merger of 
two $0.5M_{\odot}$ WDs forming a single approximately $1.0M_{\odot}$ WD with a residual disc.
Based on angular momentum arguments and a power-law surface density disc profile, they found

\begin{equation}
M_{\rm d} = 0.81 M_{\star} 
\left(  
\frac{R_{\star}}{r_{\rm d}^{({\rm max)}}}
\right)^{\frac{1}{2}}
\end{equation}

\noindent{}where the inner (minimum) disc radius is presumed to coincide with the WD surface.  They
do not model planetesimal formation, but assume that it commences when $r_{\rm d}^{({\rm max)}}$
exceeds the dust grain condensation line, which they estimated to be

\begin{equation}
r_{\rm d}^{({\rm dgline})} = 0.02 \ {\rm au} 
\left(  
\frac{R_{\star}}{6 \times 10^6 \ {\rm m}}
\right)
\left(
\frac{T_{\star}}{50000 \ {\rm K}}
\right)^2
\left(
\frac{T_{\rm d}^{({\rm dgline})}}{1600 \ {\rm K}}
\right)^{-2}
\end{equation}

\noindent{}where $T_{\rm d}^{({\rm dgline})}$ is the disc temperature at which dust grains condense.

{\bf LARGE SBs}
\ \ \ \ 
\cite{beasok2012} indicated that while the envelope of a massive planet 
that is being destroyed may form a gaseous disc, the rocky fragments could
represent small planets that then migrate within the disc; \cite{beasok2015}
made a similar suggestion.  Alternatively, \cite{guietal2011} and 
\cite{liuetal2013} intimated
that a first-generation planet may be transformed into a second-generation 
planet by being tidally disrupted.  In other words, after each close 
disruptive pericentre passage of a gas giant 
planet and an MS (or, nearly equivalently, WD) star,
the planet fundamentally changes character.  For coreless 
planets \cite{guietal2011} when free-falling ejecta accretes onto the post-encounter
planetary remnant, then its outer layers spin faster and contain more mass, changing
the atmospheric structure. Although subsequent passages may destroy the planet, other
bodies further away could perturb the disrupting planet into a stable orbit.
Cored giant planets \cite{liuetal2013} can transform into super-Earths
or Neptune-like planets through this process.

\section{White dwarf disc formation from first-generation substellar bodies} \label{secWDform}

The compact debris discs which have been observed to orbit isolated
(single) WDs at distances of ($r_{\rm d} \lesssim 1.2R_{\odot}$) cannot have 
formed during the MS or GB phases.  These discs also cannot have formed from fallback
stellar material because of their age and composition.  The canonical explanation
is that they formed from the disruption of first-generation SBs
\cite{graetal1990,jura2003,jura2008}, and
in particular, asteroids.  See \cite{farihi2016} for an observationally-based 
review of these discs.  The orbits of the progenitors are assumed to be highly eccentric
$(e \gtrsim 0.998)$ because their semimajor axes must be at the very least a few au.
Otherwise, these SBs would have been engulfed, vapourised or spun-up to fission
during the GB phases.

{\bf TIDAL DISRUPTION RADIUS}
\ \ \ \
Consequently, and emphasised by the WD 1145+017 system 
(see Sec. \ref{secOM}\ref{secOMWDplan}\ref{secOribitingWDs}), 
the process of tidal disruption plays a crucial role
in the interpretation of observations of post-MS systems.  A primary
consideration is then identifying the tidal disruption distance, $r_{\rm t}$,
which defines a disruption sphere, or Roche sphere, around the WD.
This value depends on many factors, including the SB's shape, 
spin state, composition, and orbital state, as well as if disruption
is defined as cracking, deforming or dissociating entirely.

A formulation which includes the internal strength of the SB is
(rearranged from Eq. 8 of \cite{beasok2015}): 

\begin{equation}
r_{\rm t} = 
\left[
\frac
{2GM_{\star}\rho_{\rm SB} R_{\rm SB}^2}
{\gamma_{\rm SB} + \frac{4}{3}\pi G \rho_{\rm SB}^2 R_{\rm SB}^2}
\right]^{\frac{1}{3}}
\label{eqrtgen}
\end{equation}

\noindent{}where $\gamma_{\rm SB}$ is the critical tensile strength of the SB that is disrupting.  For a strengthless SB, which is representative of many ``rubble pile'' Solar system asteroids, this equation reduces to $r_{\rm t} \approx 1.26 R_{\star} ( \rho_{\star}/\rho_{\rm SB} )^{1/3}$.  However, because of the multiple dependencies mentioned above, \cite{beasok2013} re-expressed this relation in their Eq. 1 with the following range of coefficients:  $r_{\rm t} \approx [1.3-2.9] R_{\star} ( \rho_{\star} / \rho_{\rm SB} )^{\frac{1}{3}}$.  Alternatively, for bodies with high tensile strengths, Eq. \ref{eqrtgen} reduces to Eq. 3 of \cite{corsha2008}, modulo a factor of 1.26.  Another alternative is Eq. 5 of \cite{stoetal2015}, where the internal strength dominates and the SB is characterized by solid state forces and a corresponding internal sound speed.  Then, $r_{\rm t} \propto G^{1/3} M_{\star}^{1/3} R_{\rm SB}^{2/3} v_{\rm sound}^{-2/3}$.  Typical SB speeds within the disruption radius can reach $10^3$ km/s (Eq. 7 of \cite{verleibongae2014}).

{\bf TIDAL DISRUPTION SIMULATIONS}
\ \ \ \ 
The above relations allow one to consider modelling the process of tidal disruption
with analytics and simulations.  \cite{debetal2012} performed simulations of the 
breakup of strengthless 5000-particle rubble-pile asteroids by a $0.5 M_{\odot}$ 
WD with initial orbital semimajor axes of 4.77 au and pericentres of 60, 65, 70, 75
and 80 $R_{\star}$ (so $e > 0.999$).  They modelled a single pericentre passage and analysed
the cumulative size distribution of the resulting fragments in the stream
(see their Fig. 6).  
They found that for smaller pericentres, more large fragments are generated,
and that the stream largely followed the initial orbit.

\cite{verleibongae2014} later also modelled the tidal disruption of highly-eccentric 
strengthless 5000-particle rubble-pile asteroids, but did so for over one hundred orbits.
They established that the time an SB spends within the disruption sphere
is largely independent of $a$ (their Fig. 1), enabling them to perform multi-orbit
simulations by scaling down the orbit. They found that the SB disruption eventually
creates a highly collisionless eccentric ring in the same shape as the original
orbit, in the absence of other forces.  Other forces might cause the ring to 
become collisional, as suggested by \cite{beasok2013}.

{\bf RING FORMATION ANALYTICS}
\ \ \ \ 
\cite{verleibongae2014} also derived a closed-form expression for the filling time, 
$t_{\rm fill}$, of this ring 
(their Eq. 25) supposing that the breakup occurs instantaneously at some 
$r_{\rm u} < r_{\rm t}$:

\begin{eqnarray}
\frac{t_{\rm fill}}{P_{\rm SB}}
&=&
r_{\rm u}^{\frac{3}{2}} 
\bigg[
\left\lbrace
\frac
{r_{\rm u}^2 + 2aR_{\rm SB} - r_{\rm u}R_{\rm SB}}
{r_{\rm u} - R_{\rm SB}}
\right\rbrace^{\frac{3}{2}}
\nonumber
\\
&-&
\left\lbrace
\frac
{r_{\rm u}^2 - 2a{\rm min}\left(r_{\rm crit} - r_{\rm u}, R_{\rm SB}\right)
+
r_{\rm u}{\rm min}\left(r_{\rm crit} - r_{\rm u}, R_{\rm SB}\right)  }
{r_{\rm u} + {\rm min}\left(r_{\rm crit} - r_{\rm u}, R_{\rm SB}\right) }
\right\rbrace^{\frac{3}{2}}
\bigg]^{-1}
\label{eqRingForm}
\end{eqnarray}

\noindent{}where $P_{\rm SB}$ is the orbital period of the SB and where $r_{\rm crit}$ is 
the location at which the SB's orbit would become
parabolic

\begin{equation}
r_{\rm crit} = 
\frac
{2ar_{\rm u}}
{\left(1 + \frac{M_{\rm SB}}{M_{\star}} \right) \left(2a - r_{\rm u}\right)}
\approx
\frac
{2ar_{\rm u}}
{2a-r_{\rm u}}
.
\end{equation}

\noindent{}Eq. \ref{eqRingForm} does not make any assumptions about orbital
eccentricity.

However, as indicated by Fig. \ref{FigForces}, considering gravity alone in 
a regime where
$r \lesssim 10^{-3}$ au is inadequate.  \cite{verDisc2015} modelled the long-term
effect of WD radiation on these collisionless rings.  They found that
WD radiation alone compresses and circularises orbits of SBs with
$R_{\rm SB} \approx 10^{-5}$--$10^{-1}$m to within the WD disruption radius.
See their Eqs. 20-22 for closed-form approximations of the evolution of $a$ and $e$.
Outside of this SB size range, radiative effects such as the Yarkovsky effect
and radiation scattering may dictate the motion, and have not yet been explored.
For this size range, they derive in their Eq. 23 a closed-form expression
for the shrinking time $t_{\rm shr}$, which is better expressed as

\[
\left(1 + \frac{10\left(t_{\rm shr} + t_{\rm u}\right)}{\rm Myr}  \right)^{-\frac{1}{5}}
=
\left(1 + \frac{10t_{\rm u}}{\rm Myr}  \right)
\ \ \ \ \ \ \ \ \ \ \ \ \ \ \ \ \ \ \ \ \ \ \ \ \ \ 
\ \ \ \ \ \ \ \ \ \ \ \ \ \ \ \ \ \ \ \ \ \ \ \ \ \ 
\ \ \ \ \ \ \ \ \ \ \ \ \ \ \ \ \ \ \ \ \ \ \ \ \ \ 
\ \ \ \ \ \ \ \ \ \ \ \ \ \ \ \ \ \ \ \ \ \ \ \ \ \
\]

\begin{equation}
-
2.39
\left(
\frac
{\rho_{\rm SB} R_{\rm SB} c^2 q_{\rm u}^{3/2}}
{L_{\star}(t_{\rm cool} = 0) \left[Q_{\rm abs} + Q_{\rm ref}\right] {\rm Myr}}
\right)
\left(
\sqrt{a_{\rm u}} - \sqrt{2q_{\rm u} + r_{\rm t}}
\right)
\end{equation}

\begin{figure}[!t]
\centering\includegraphics[width=2.9in]{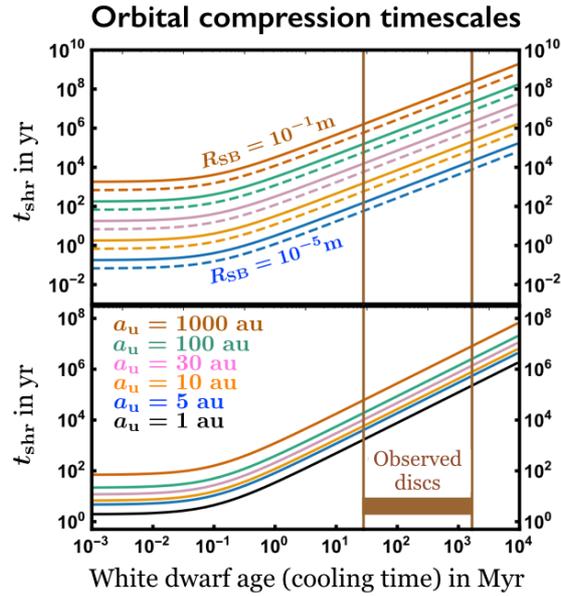}
\caption{Cosmetically enhanced version of Fig. 2 of \cite{verDisc2015}.
How particles of given radii and semimajor axis shrink due to the effects
of radiation alone, as a function of WD cooling time.}
\label{FigShrinking}
\end{figure}

\noindent{}where $c$ is the speed of light, where $Q_{\rm abs}$ and 
$Q_{\rm ref}$ are the SB's absorption
efficiency and reflecting efficiency (albedo), $t_{\rm u}$ is the cooling
age at which disruption occurred, and $a_{\rm u}$ and $q_{\rm u}$ are the
orbital semimajor axis and pericentre at disruption.  The value of
$t_{\rm shr}$ may change when additional effects, such as sublimation
are taken into account.  Fig. \ref{FigShrinking} illustrates the
shrinking time as a function of cooling age. Because the shrinking 
timescale is dependent on $R_{\rm SB}$, 
differently-sized pebbles will shrink at different rates,
fanning out the original ring.  Further, as the SBs approach the WD,
the effect of general relativity becomes more important
(see Sec. \ref{secOther}\ref{secOtherGR}).  Throughout the 
contraction, collisions may occur, which could affect subsequent shrinking times.

\cite{stoetal2015} instead adopted a constant WD luminosity
and in the high-eccentricity limit obtained a characteristic
shrinking timescale (their Eq. D6) that is proportional to 
$a_{\rm u}^{\frac{1}{2}} q_{\rm u}^{\frac{3}{2}}/L_{\star}$.  The quality of this approximation
increases as the WD gets older, when $L_{\star}$ experiences less-drastic changes.

{\bf OTHER DISRUPTION PROCESSES}
\ \ \ \ 
How the tidal disruption of an asteroid encountering a WD
compares with much better-studied tidal disruption situations in
the literature is not yet clear.  As observed by 
\cite{beasok2013}, stars which pass close to black holes
typically result in about half of their mass being ejected,
with the remaining mass residing on highly eccentric orbits.
A perhaps more relevant case are planets disrupting around
MS stars \cite{guietal2011,liuetal2013}. In a rough 
qualitative sense, this situation 
should be similar for planet destruction around WDs.

\section{White dwarf disc evolution} \label{secWDevol}

Our understanding of how WD discs evolve crucially impacts 
observational targeting strategies
and touches upon both how these discs formed (Sec. \ref{secWDform}) and how 
they accrete onto the WD (Sec. \ref{secWDaccr}).  More directly and perhaps 
just as importantly, we cannot yet explain the variability in 
the dusty components of discs such as that around WD J0959-0200 \cite{xujur2014} 
and the axisymmetric structures in the gaseous components of discs 
such as that around SDSS
J1228+1040 \cite{manetal2015}. \cite{xujur2014} speculated 
that the drop in the flux of the WD J0959-0200 disc could be attributed to a 
recent SB impact or instability within the disc.  More detailed models
may help distinguish these possibilities.

{\bf SOLID BODY EVOLUTION DURING IMPACT}
\ \ \ \
\cite{jura2003} emphasized the potential connection of WD discs
with planetary rings in the Solar system.  In line with this analogy, 
\cite{jura2008} discussed disc
evolution following additional impacts (which can generate additional dust). 
He posed that the disc evolution would be determined by size: small 
entrants would be destroyed by sputtering whereas large bodies more massive 
than the disc would imprint their trajectories on the disc.  
 
Assume that an asteroid tidally breaks up and produces a stream of 
identical grains with density $\rho_{\rm SB}$ and orbital inclination 
$i$ measured with respect to the existing WD debris disc plane. Assume 
also that the orbit of the grains passes through the disc twice per orbit, 
the orbit does not change, and that the grain material is equivalent to 
the disc material.  After many orbits, the minimum surviving grain size 
is (from Eqs. 15-17 of \cite{jura2008})

\begin{equation}
R_{\rm SB}^{({\rm min})} = 
y t \frac{\Sigma_{\rm d}}{4 \pi \rho_{\rm SB} \sin{i}}
\sqrt{\frac{G M_{\star}}{a_{\rm SB}^3}}
\end{equation}

\noindent{}where $y$ is the (dimensionless) sputtering yield, and $\Sigma_{\rm d}$
is the debris disc density.  The detectable infrared excess which results
from the dust produced in the encounters diminishes as the grain radius decreases.
\cite{jura2008} found that the excess from the encounter dissipates on a much
shorter timescale than the characteristic disc lifetime.

{\bf GAS GENERATION}
\ \ \ \ 
Now consider the incidence of gas in these discs.
When and where gas is produced represents a crucial aspect of
the disc evolution.  In principle, all dusty discs
should produce gas through sublimation as the inner rim falls towards the WD.
However, this gas has been detected in only 7 of the nearly 40 dusty 
discs so far (see Sec. \ref{secOM}\ref{secOMWDpoll}), and in those 7 discs, the gas
has a wide radial extent, overlapping with the dust.  This overlap
highlights a curious puzzle.

An alternative gas generation mechanism is collisions of solid 
particles.
\cite{jura2008} suggested that mutual velocities 
between the grains may be as high as $10^3$ km/s (perhaps in line
with Eq. 7 of \cite{verleibongae2014}), which could
generate copious amounts of gas.  However, \cite{metetal2012} 
suggested (in their Section 6.4) that this process produces
a negligible amount of gas because the initially-high
relative velocities are quickly damped through inelastic
collisions.  They pose that more gas would be generated
and persist for tens or hundreds of years immediately after
a disruption event, as with a incoming SB interacting with an
already evolving disc.

Returning to sublimation, 
the distance at which
the sublimation takes place, $r_{\rm sub}$, (often refereed to as the sublimation radius) 
is usually expressed as (Eq. 1 of \cite{rafikov2011a})

\begin{equation}
r_{\rm sub} = \frac{R_{\star}}{2}
\left(
\frac{T_{\star}}{T_{\rm SB}^{({\rm sub})}}
\right)^2
\approx 
22 R_{\star} 
\left( \frac{T_{\star}}{10^4 \ {\rm K}} \right)^2
\left( \frac{T_{\rm SB}^{({\rm sub})}}{1500 \ {\rm K}} \right)^{-2}
,
\label{eqsub}
\end{equation}

\noindent{}demonstrating that particles typically sublimate at a distance of a few tens
of WD radii. However, \cite{rafgar2012} argued that one fundamental difference between WD
debris discs and protoplanetary discs is that because the former is hydrogen poor,
values of $T_{\rm SB}^{({\rm sub})}$ should be several hundred K higher.  Also,
Eq. \ref{eqsub} does 
not include physics such as conduction or reflectance
(Eq. 1 of \cite{rafikov2011b} includes a factor incorporating particle and stellar emissivity).
Note tangentially that a direct comparison with the other sublimation distance expression
in this paper (Eq. \ref{eqStoSub}) shows significant differences: In Eq. \ref{eqsub}, the
only explicit dependence on SB properties is through its sublimation temperature.

{\bf DISC TEMPERATURE}
\ \ \ \
That equation does not actually require the presence of a 
disc.  If a disc does exist, then the temperature is not necessarily uniform throughout:
An isothermal disc would in fact not produce any detectable flux, according to 
Eq. \ref{eqflux}.

The temperature at a given location in the disc is dependent on (i) the optical thickness
at that location, (ii) the amount of shielding, (iii) heating by the WD, (iv)
heating by gas, (v) cooling by sublimating atoms, and (vi) cooling by thermal radiation
from particle surfaces.  \cite{rafgar2012} began to address these issues in their
Appendix A, and related disc temperatures in optically thin and thick parts of the disc
through (their Eqs. 1-2)

\begin{eqnarray}
T_{\rm d}^{({\rm thin})}(r_{\rm d}) 
&\approx& 
T_{\star}
\left( \frac{1}{2} \right)^{\frac{1}{2}}
\left( \frac{R_{\star}}{r_{\rm d}} \right)^{\frac{1}{2}}
,
\\
T_{\rm d}^{({\rm thick})}(r_{\rm d}) 
&\approx& 
T_{\star} 
\left( \frac{2}{3\pi} \right)^{\frac{1}{4}}
\left( \frac{R_{\star}}{r_{\rm d}} \right)^{\frac{3}{4}}
.
\end{eqnarray}

{\bf DISC VISCOSITY}
\ \ \ \ 
One key aspect of WD discs is the viscous timescale of their gaseous components

\begin{equation}
t_{\rm d,\ gas \ only}^{({\rm visc})} \sim \frac{r_{\rm d}^2}{\upsilon_{\rm d}}
\end{equation}

\noindent{}where $\upsilon_{\rm d}$ is the viscosity of the gaseous component, 
and is assumed to be constant, and $r_{\rm d}$ represents
a location in the disc.
Unfortunately, the viscous timescale parameter is unconstrained.  In fact,
the $t_{\rm d}^{({\rm visc})}$ values reported 
by \cite{beasok2013} and \cite{metetal2012} 
differ by six orders of magnitude (0.75 day to 2000 yr).
\cite{beasok2013} attributes this discrepancy 
to different assumptions about when the disc was
formed, and the resulting different values of
sound speed, viscosity parameter and scale height.

Gas will exist inside of any dusty disk where 
$r_{\rm d}^{({\rm out})} > r_{\rm sub} > r_{\rm d}^{({\rm in})}$, 
with $r_{\rm d}^{({\rm in})}$ and $r_{\rm d}^{({\rm out})}$ representing
the boundaries of the disc. Gas will also exist outside of the disc,
at $r < r_{\rm d}^{({\rm in})}$, as the gas accretes onto the WD. Gas is thought to accrete 
onto the WD in order to maintain angular momentum conservation with 
the gas's outward viscous spreading. For the seven discs
with both dusty and gaseous components, the latter is no more
massive than the former. This statement arises from estimates
of the dust mass \cite{jura2003,juraGD56,jura2008,reaetal2009}
and of the gas mass \cite{kinnear2011}.  For a disc like SDSS J1228+1040,
the dust mass may be about $10^{19}$g-$10^{24}$g whereas the gas mass is
about $10^{19}$g \cite{kinnear2011,manetal2015}.  The presence of
dust and gas, even if not in the same proportion, showcase the necessity
of considering coupled evolution models.  
As observed by \cite{metetal2012}, just how 
gas and dust could exist in the same location in different phases
is an outstanding question.

Nevertheless, considering gas only can provide initial helpful insight.
By assuming an isothermal gas-only disc is turbulent with a viscosity equal
to $\alpha v_{\rm sound}^2r_{\rm d}^{3/2}/\sqrt{GM_{\star}}$ (according to
the typical $\alpha$ parameterization with the sound speed
$v_{\rm sound}$), this disc's viscous timescale can be expressed as
(Eq. 2 of \cite{metetal2012})

\begin{equation}
t_{\rm d, \ gas \ only}^{({\rm visc})} 
\approx
2 \ {\rm yr} \times \alpha^{-1} 
\left( \frac{T_{\rm d}^{({\rm gas})}}{5000 \ {\rm K}} \right)^{-1}
\left( \frac{\mu}{28M_{\rm H}} \right)
\left( \frac{M_{\star}}{0.6M_{\odot}} \right)^{\frac{1}{2}}
\left( \frac{r_{\rm sub}}{0.2R_{\odot}} \right)^{\frac{1}{2}}
\label{eqvisc}
\end{equation}

\noindent{}where $M_{\rm H}$ is the mass of hydrogen and $\mu$ is mean 
molecular weight (equal to $28M_{\rm H}$ for
silicon). The largest source of uncertainty in Eq. \ref{eqvisc} is the unknown $\alpha$.
For an alternative formulation of the viscosity, as expressed by a power law
and with sources and sinks of mass, see Appendix B of \cite{metetal2012}.
Note also that if Eq. \ref{eqvisc} is applied to a disc containing dust
(as all WD discs do), then $T_{\rm d}^{({\rm gas})}$ would not be equal to 
$T_{\rm d}^{({\rm dust})}$ due to their different heating and cooling sources and sinks.

As a first approximation, equations like Eq. \ref{eqvisc} help provide scalings
for observed discs like SDSS J1228+1040 \cite{manetal2015}, and can help
assess the steady-state accretion assumption (see Eq. \ref{eqSteadyState}) for individual
WDs.  However, one should keep in mind that angular momentum transfer in a gaseous disc 
is more efficient than in a dusty counterpart, leading to a shorter lifetime.

{\bf COUPLED DUST-GAS EVOLUTION}
\ \ \ \ The coupling between dust and gas in WD debris discs has been studied by
\cite{bocraf2011,rafikov2011a,rafikov2011b,rafgar2012,metetal2012}.
\cite{rafikov2011b} suggested that the coupling leads to ``runaway'' accretion
due to positive feedback from newly formed gas.  Forces between concomitant
gas and dust rotating at different rates enhances solid body angular momentum
loss, causing this runaway. \cite{rafikov2011a} supported these results,
in part, by evaluating and discounting the prospect for the Yarkovsky force 
(see Sec. \ref{secRad}\ref{secRadGB}\ref{secPebcom} of this paper) opposing 
the inward Poynting-Robertson drag, even despite the unknown size distribution 
in the WD debris disc.  \cite{bocraf2011} performed a global disc analysis.
They found a  ``universal appearance'' of a sharp outer edge in optically thick discs.  
But this edge moves inward, and so can easily turn into a narrow ring.
\cite{rafgar2012} explored the effect of shielding and vapor pressure.

\cite{metetal2012} performed the most recent global time-dependent simulations of WD
discs with dust and gas, and included (i) aerodynamic coupling in two limiting cases
of optical depth (their Secs. 2.1.1 and 2.1.2), and (ii) mass exchange
(their Sec. 3).  In the optically thin case, where solid particles 
individually interact with the gas, they provide expressions
for the particle Reynolds number (their Eq. 13)

\begin{equation}
{\rm Re} = \frac{2R_{\rm SB}v_{\phi}^{({\rm rel})}}{\upsilon_{\rm m}}
\label{eqReMet}
\end{equation}

\noindent{}and frictional drag force (their Eq. 14)

\begin{eqnarray}
F_{\rm drag}^{({\rm fric})}
&=
\left\{ \begin{array}{ll} 
 \frac{4\pi}{3} R_{\rm SB}^2 \rho_{\rm gas} v_{\rm sound} v_{\phi}^{({\rm rel})}
, & R_{\rm SB}\lesssim \zeta\\
 6\pi R_{\rm SB} \rho_{\rm gas} \upsilon_{\rm m} v_{\phi}^{({\rm rel})}
, & R_{\rm SB} \gtrsim \zeta
\end{array}\right .
\label{eqdragme}
\end{eqnarray}

\noindent{}In Eqs. \ref{eqReMet}-\ref{eqdragme}, $v_{\rm sound}$ is 
the sound speed, the value $\upsilon_{\rm m}$ is the molecular shear 
viscosity (not turbulent viscosity), 
$\rho_{\rm gas}$ is the midplane gas density and $v_{\phi}^{({\rm rel})}$
is the azimuthal speed difference between the particles and gas.
Comparing these equations with Eqs. \ref{eqdragve} and
\ref{eqRe1} illustrates that, besides numerical factors of order 
unity, the difference
comes here from the inclusion of molecular shear viscosity.
Also, as in Eq. \ref{eqdragve}, the upper and lower pieces of Eq. \ref{eqdragme}
correspond to the Epstein and Stokes regimes, respectively.

\begin{figure}[!t]
\centering\includegraphics[width=2.8in]{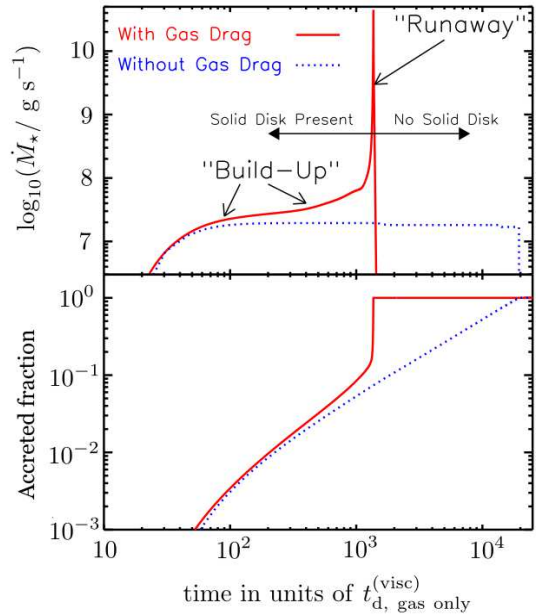}
\caption{Cosmetically enhanced version of Fig. 7 of \cite{metetal2012}.
How a coupled gas and dust WD debris disc becomes a ``runaway'' disc,
which features a burst of accretion at the inner rim.  The difference
in the dotted blue and solid red lines illustrates the importance of including
coupling in WD debris disc models.}
\label{FigRunaway}
\end{figure}

\cite{metetal2012} ultimately provide support for the runaway theory,
characterising the process as a ``buildup'' and then ``runaway'' phase
as an optically thick disturbance migrates inwards and then provides 
positive feedback at the inner rim.  See Fig. \ref{FigRunaway}.  
Their finding that eccentricity in
gas motions as small as $10^{-4}$ are highly efficient at driving runaway
accretion enables them to make an important prediction: the 
asymmetric line profiles in observed gaseous disc components indicate 
non-axisymmetric surface brightness rather than eccentric gas motions.

{\bf SECULAR EVOLUTION}
Observations indicate \cite{manetal2015} that at least one disc (SDSS J1228+1040;
see Fig. \ref{FigDisc} of this paper) 
has secular modes which act on timescales (tens of years) which are five orders
of magnitude greater than orbital timescales (hours).  The authors discuss 
several possible origins for the disc evolution: (i) an external perturber,
(ii) the youth of the disc, (iii) self-generated perturbations.
An almost certainly important consideration in all three scenarios is precession 
due to general relativity: the disc will precess over one complete orbit due to 
general relativity (see Eq. \ref{eqGRdisc}) in about 27 years, which matches with the 
secular period (the precession period ranges between 1.5 yr and 134 yr for small 
eccentricities and $a = 0.2R_{\odot}-1.2R_{\odot}$).  The authors excluded the 
first scenario based on $N$-body simulations of an external perturbing body, 
which changes the disc too quickly (only during pericentre passages).  
Either or both of the other scenarios might work, but require further modeling with 
radiative and collisional effects.  In the second scenario, the disc just formed 
from a tidal disruption and is in the process of settling and circularising.
Modelling the third scenario is difficult because the mass ratio of gas to dust in 
the disc is poorly constrained, ranging from $10^{-5}$ to unity.

\section{Accretion onto white dwarfs} \label{secWDaccr}

WDs with photospheric signatures of heavy metals are considered
to be polluted because they accrete these metals from an external origin.
Understanding the process of pollution is then vital to achieve links
with observables.

{\bf SETTLING PHASES}
\ \ \ \
Assume that a particular chemical element is accreting onto
the WD at a constant rate $\dot{M}_{\rm ele}$ with
a constant characteristic diffusion, or settling timescale 
$t_{\rm set}^{({\rm elm)}}$ through the bottom of the convection zone.
The mass of the convection zone is $M_{\rm conv}$, which is calculated
from a stellar model.
These assumptions allow one to obtain the following explicit 
formula (Eq. 5 of \cite{koester2009}) for
the time evolution of the mass fraction of that chemical
element, $X_{\rm elm}(t)$,
assuming that the element starts accreting at $t=t_{\rm start}$. 

\begin{equation}
X_{\rm elm}(t) = 
X_{\rm elm}(t_{\rm start}) \cdot {\rm exp}\left( -\frac{t-t_{\rm start}}{t_{\rm set}^{({\rm elm)}}} \right)
+
t_{\rm set}^{({\rm elm)}}
\frac{\dot{M}_{\rm ele}}{M_{\rm conv}}
\left[
1
-
{\rm exp}
\left( -\frac{t-t_{\rm start}}{t_{\rm set}^{({\rm elm)}}} \right)
\right]
\label{eq3states}
\end{equation}

Eq. (\ref{eq3states})
describes three important regimes, as explained by both \cite{koester2009} and
\cite{koester2013}: (i) onset:~$t-t_{\rm start}\ll~t_{\rm set}^{({\rm elm)}}$, 
(ii) steady state: $t - t_{\rm start} \gg t_{\rm set}^{({\rm elm)}}$,
and (iii) trailing:  $t - t_{\rm start} \gg t_{\rm set}^{({\rm elm)}}$ plus 
$\dot{M}_{\rm ele} \rightarrow 0$.
Distinguishing between these modes for observations of accretion
in helium-rich WDs (``DB WDs'') is unfortunately not yet possible, because of the
long sinking times (up to Myrs). This degeneracy 
does not arise for
hydrogen-rich WDs (``DA WDs'') which are older than about 300 Myr
(see Fig. 1 of \cite{wyaetal2014})
because their convection zones are too small and 
the element diffusion times are
too fast (typically, days to weeks). The 
typically-used steady state assumption 
importantly allows one to obtain abundance ratios between chemical 
element pairs (labeled \#1 and \#2 below) because in this regime, 
both exponential terms in Eq. \ref{eq3states} 
vanish while accretion is ongoing.  Consequently, 

\begin{equation}
\frac{X_{\rm elm\#1}}{X_{\rm elm\#2}} = 
\frac
{t_{\rm set}^{({\rm elm\#1)}} \dot{M}_{\rm elm\#1}}
{t_{\rm set}^{({\rm elm\#2)}} \dot{M}_{\rm elm\#2}}
\label{eqSteadyState}
\end{equation}

\noindent{}where the quantities on the right-hand-side are known.  Therefore,
abundance ratios of different species are quantities which are readily 
reported in WD pollution literature.

Note importantly that Eq. \ref{eq3states} does not necessarily
require the existence of a debris disc; rather just a constant stream of
pollution from an external source (perhaps from a stream of tidally-disrupted matter).
\cite{juretal2009} (and also \cite{jurxu2010}) considered the case
when a debris disc exists, was formed by a single progenitor, and decays in 
an exponential manner, with
a characteristic time of $t_{\rm d}$. In this case, they derived an
expression for the total mass of a particular element that existed in
the progenitor, denoted by $M_{\rm SB}^{({\rm elm})}$.  By also assuming
$t_{\rm d} > t_{\rm set}^{({\rm elm)}}$ and
that the currently-measured/modeled convection zone mass 
of that element only is $M_{\rm conv}^{({\rm elm})}$,
they derive (from Eq. 8 of \cite{juretal2009} or from Eqs. 7-8 of \cite{jurxu2010}),

\begin{equation}
M_{\rm SB}^{({\rm elm})}
=
M_{\rm conv}^{({\rm elm})}
\left( 
\frac{t_{\rm d}}{t_{\rm set}^{({\rm elm)}}} - 1
\right)
\left\lbrace
{\rm exp}
\left(-
\frac{t-t_{\rm start}}{t_{\rm d}}
\right)
-
{\rm exp}
\left(-
\frac{t-t_{\rm start}}{t_{\rm set}^{({\rm elm)}}}
\right)
\right\rbrace^{-1}.
\end{equation}

{\bf THERMOHALINE CONVECTION}
\ \ \ \
The above analysis is for gravitational settling alone. \cite{deaetal2013} warned 
that neglecting thermohaline (or fingering) convection could
induce steady states before they actually occur in DAZ (hydrogen and metal-enriched) 
WDs.  Thermohaline convection
occurs when upper layers of a WD atmosphere are heavier than lower layers,
which could induce localized mixing. \cite{wyaetal2014} helpfully used the analogy of salt
fingers in Earth's oceans.  However, \cite{xuetal2014} argued against the 
conclusions of \cite{deaetal2013}, claiming that (1) their thermohaline
prescription asymptotically approaches infinity at the bottom of the convection
zone, (2) one of their models does not contain any atmospheric convection and 
therefore cannot 
be representative of canonical polluted WDs like G 29-38, and (3) they do not 
consider the fuzziness of the convection zone boundary, and the consequences for mixing.
\cite{koesterWD2014} presented more detailed arguments.

{\bf CONVECTIVE OVERSHOOTING}
\ \ \ \
Because convection zones in WD atmospheres are usually defined 
as the boundaries where instabilities to convection occur
(the Schwarzschild criterion), these zones do not take into account 
real convection cells, or bubbles, that traverse or are said 
to ``overshoot'' the boundaries. \cite{freetal1996} pioneered the 
modelling of this process in WDs.  Recent progress (e.g. Sec. 5.2 of 
\cite{treetal2015}) suggests that overshooting may
represent an important consideration for polluted WDs, although 
further modelling is necessary.

{\bf ACCRETOR SIZE DISTRIBUTIONS}
\ \ \ \
Observationally-derived accretion rates can help us distinguish the size
distribution of the accreting SBs.  The metal content of the SB combined
with the type of WD (DA or non-DA) determines the sinking time through the
convection zone.  If SBs (from a disc or not) of a certain size
rain onto a particular WD frequently enough, then their accretion
signature is always present in the atmosphere.  This accretion is said
to be ``continuous'' accretion.  Alternatively, if larger SBs infrequently
penetrate the atmosphere, then they ``stochastically'' accrete onto the WD.

\cite{wyaetal2014} helped distinguish size distributions that would
produce continuous versus stochastic accretion.
They found that the critical value of the SB radius which divides
these two regimes is between about $1.7$ km and $170$ km.  The critical
radius is larger for non-DA WDs because of their larger convective zones.
\cite{wyaetal2014} also considered, section-by-section, the quality of observational
matches from accretion distributions based on (i) mono-mass, mono-rate accretion (their
Sections 4.1 and 4.3.1), and (ii) mono-mass, multi-rate accretion (their Sections 4.2 and 4.3.2).
In all these cases, they effectively debunk the idea that pollution arises
from any type of mono-mass distribution.

A more complex multi-mass distribution is required, and might have to incorporate 
the potential presence of a WD disc.  The accretion rate might depend on the
disc lifetime.  Regardless, in the course of their study, \cite{wyaetal2014}
used an analytical model that might be helpful for future investigations.
Their stochastic mono-mass accretion model is equivalent to that of shot noise in electrical
systems (see their Appendix A), providing for an analytical foundation (particularly 
their Eqs. 9 and 13) which is an excellent
match to Monte Carlo simulations (see their Fig. 4).

{\bf IMPACTOR PHYSICS}
\ \ \ \
Stochastic accretion will lead to direct impacts onto the WD by SBs
larger than pebbles.  Boulders, asteroids, comets and planets may directly
hit the WD, bypassing any disc phase.  The physics of impact
in the context of NSs and WDs has been studied in the 1980s \cite{colpet1981,alcetal1986},
and more recently for the Sun by \cite{broetal2011,broetal2015} but
requires more current WD-specific attention.
\cite{broetal2011} distinguished three possible outcomes, which are not
mutually exclusive: explosion, ablation and sublimation.  \cite{alcetal1986}
instead characterised the impact process through explosion, spreading
and sedimentation phases.

{\bf DISC ACCRETION}
\ \ \ \
The alternative to stochastic accretion is continuous accretion -- from a disc
(Secs. \ref{secWDform}-\ref{secWDevol}).
The observed discs, however, do not extend to the WD surface (or photosphere). 
The inner rims of dusty disc components reside tens of WD radii
outside of the WD surface.  The gap between the WD and the 
inner rim is then assumed to be filled with sublimated metallic gas originating
from the inner rim.  This gas is accreted onto the WD though viscous torques
\cite{rafikov2011a}.  Hence, in effect, the mass flux through the inner rim
acts as a proxy for the actual accretion rate onto the WD surface.
Estimating this mass flux is particularly important because the accretion rate
onto WDs is a quantity derived from observations (by combining sinking times
with convection zone metal content).

Computations of this mass flux necessarily rely on modeling the coupling
between dust and gas, which is not trivial, as already explained in Sec. \ref{secWDevol}
and illustrated explicitly in Eq. 4 of \cite{rafikov2011b}.
One of the dominant forces widely thought to influence this mass flux 
is Poynting-Robertson drag, for which, when acting in isolation,
several estimates have been given (Eq. A8 of \cite{faretal2010PR}, Eq. 9 of
\cite{rafikov2011a}, Eq. 5 of \cite{rafikov2011b}, Eq. 43 of \cite{metetal2012}).  
Regarding other potential contributions to this mass flux, \cite{rafikov2011a} 
argues against the Yarkovsky effect being significant (in his Section 2.2).

{\bf SB ACCRETION FREQUENCY}
\ \ \ \
The frequency and size distribution of accreting SBs depend on the
architecture of the planetary system and what instabilities it undergoes
throughout its lifetime.  Despite the vast number of free parameters in
this problem, several investigations have related WD accretion frequencies
to particular setups.  I have already discussed studies that have modelled
architectures involving planets and asteroids in Sec. \ref{secMulti}.  Here I present
three studies \cite{alcetal1986,vershagan2014,stoetal2015} dedicated to 
determining accretion rates onto WDs from
exo-Oort cloud comets.  These studies do not actually involve $N$-body interactions,
as the comets are assumed to be point masses and
instability is caused by other means (primarily stellar mass loss, Galactic
tides and stellar flybys).

Both \cite{alcetal1986} and \cite{vershagan2014}, written almost three decades
apart and through different means, came to the same conclusion:  the 
cometary accretion rate onto WDs is about one per $10^4$ yr.
Both apply isotropic GB mass loss with no natal kick, but the former
uses a ramp function prescription (their Eq. 13) and the latter
adopts prescriptions from a stellar evolution code.
The former does not incorporate Galactic tides and stellar flybys,
but the latter does.  These variations are perhaps offset by
the different assumptions that the studies adopted for stellar masses and MS Oort cloud 
architectures.
One other perhaps important difference is that \cite{vershagan2014}
adopts an escape ellipsoid surrounding the primary such that a comet
is assumed to escape even if it enters the ellipsoid without an osculating
parabolic or hyperbolic orbit.

In contrast with these two studies, \cite{stoetal2015} incorporated
a natal kick that accompanies GB mass loss, and assumed that the
mass loss is instantaneous.  They modelled the stellar mass 
to change from $1.2M_{\odot}$ to $0.6M_{\odot}$,
and the star to be kicked with velocities ranging from $0.05$ km/s to $4.00$ km/s.
They considered the effect on exo-Oort
clouds with $a^{({\rm MS})} = [1\times 10^3, 5\times 10^4]$ au.  This 
range is particularly interesting because it lies at the limit between
adiabaticity and non-adiabaticity in the non-instantaneous mass
ejecta case \cite{veretal2011,verwya2012}.  One can observe the straddling of 
this boundary in Fig. 3 of \cite{stoetal2015}.  They ultimately find
that this combination of
instantaneous mass loss and kick succeeds in producing comets
on near-parabolic orbits so that at least some will enter the
ice line and sublimate.

None of these three studies modelled interstellar comets.  
\cite{vershagan2014} justify this neglect (see their Section 3) by
(i) estimating that a typical exo-Oort cloud should be 
a few orders of magnitude more dense than the interstellar medium, 
(ii) estimating that Oort cloud
comet velocities are likely to be less than an order of magnitude smaller 
than those of interstellar comets, and (iii) pointing out that no 
Sun-grazing comet has ever been observed to harbour a hyperbolic orbit.

{\bf SPECTRAL SIMULATIONS}
\ \ \ \
An exciting alternative, 
complementary method to measure the chemical composition of planetary remnants 
through accreted material is to derive the chemical composition 
of the observed 
dusty \cite{reaetal2005,reaetal2009} and gaseous \cite{haretal2011,haretal2014} 
components of WD discs (as opposed to the atmospheric
pollution) by simulating their spectra.  
However, successful models for the gaseous discs require knowing the origin of 
the heating, particularly for the 
Ca II emission lines.  The accretion rate derived by \cite{haretal2011} 
($10^{17}-10^{18}$ g/s) is too high by several orders of magnitude to match observations.
For dusty discs, Fig. 14 of \cite{xuetal2014} compared abundances derived from
infrared spectra and atmospheric analyses; JWST should provide much better constraints.

\section{Other dynamics}  \label{secOther}

\subsection{General relativity}  \label{secOtherGR}

The sub-au scales of WD debris discs, and the multi-au scales of the orbits
of disc progenitors, suggest that general relativity (GR) may play a significant
role in the secular evolution of remnant planetary systems.  GR 
does alter $a$, $e$ and $\omega$ on orbital timescales \cite{veras2014}, 
which can be important for volatile-rich SBs with orbital pericentres
close to the disruption radius \cite{verSub2015}.  On secular timescales,
$a$ and $e$ do not change, but $\omega$ famously does.  GR will cause 
precession of the argument of pericentre over one full orbit on the
timescale

\begin{eqnarray}
t_{\omega}^{({\rm GR})} &\approx& 107 \ {\rm yr} 
\left( \frac{M_{\star}}{0.60M_{\odot}} \right)^{-\frac{3}{2}}
\left( \frac{a}{1.00R_{\odot}}  \right)^{\frac{5}{2}}
\left(1 - e^2 \right)
\label{eqGRdisc}
\\
&\approx& 0.15 \ {\rm Myr}
\left( \frac{M_{\star}}{0.60M_{\odot}} \right)^{-\frac{3}{2}}
\left( \frac{a}{1.00 {\rm au}}  \right)^{\frac{5}{2}}
\left[
\frac{1 - e^2}{1 - 0.999^2}
\right]
.
\label{eqGRout}
\end{eqnarray}

\noindent{}Eqs. \ref{eqGRdisc} and \ref{eqGRout} may be useful, respectively,
for WD debris discs and their potentially highly-eccentric progenitors.

\subsection{Magnetism} \label{secOtherMagnetism}

{\bf UNIPOLAR INDUCTORS}
\ \ \ \ 
Magnetic WDs may provide newfound opportunities to detect post-MS planets.
Both \cite{wilwu2004} and \cite{wilwu2005} demonstrated how terrestrial
planets orbiting WDs with a period of less than about 30 hours may
be detectable as electron-cyclotron maser sources.  Both papers
invoke the unipolar inductor model for WD-planet pairs \cite{lietal1998},
where a circuit is formed between the conducting planet (or planet core) --
which would be inside of the WD magnetosphere -- 
and the poles of the WD.  Consequently, the WD atmosphere is heated.  
This scenario is visualized in Figs. 1-2 of \cite{lietal1998}, Fig. 1 of \cite{wilwu2004}
and Fig. 1 of \cite{wilwu2004}.

Within this framework, \cite{wilwu2004} and \cite{wilwu2005} made 
an analogy with the detectable 
radio emissions from Jupiter due to its magnetic interaction with 
Galilean moons.  For WD planetary systems, Jupiter would be replaced with 
the magnetic WD.  The discovery space of WDs by this method is presented
in Fig. 2 of \cite{wilwu2005}, and is a function of WD magnetic moment
and SB orbital period.

One consequence of the unipolar inductor model is magnetically-induced
drag on the planet (or large enough SB to be a conductor).  This interaction
can cause an inward migration, with an inward drift speed of 
(Eq. 7 of \cite{lietal1998})

\begin{equation}
v^{({\rm drift})} = 
\frac{4 W r^2}{GM_{\star}M_{\rm SB}} 
\end{equation}

\noindent{}where $W$ is the heating rate associated with the current.
\cite{lietal1998} claimed that the drift is negligible for $r \gtrsim 200R_{\star}$.
Within this distance, the SB might already be undergoing tidal breakup
(see Sec. \ref{secWDform}).  Whether there exist an planet-mass SB 
which represents this type of unipolar conductor orbiting the magnetic WD
named GD 356 is an open question.
If so, \cite{wicetal2010} claimed that the planet is then likely to 
have a second-generation origin, in particular from the remnants of a WD-WD merger.

{\bf ACCRETION ONTO COMPACT OBJECTS}
\ \ \ \
Some metal-rich WDs are magnetic \cite{holetal2015}.  
None of these stars have observed discs.  Therefore,
the pollution could arise from stochastic accretion of 
neutral material that is not affected by magnetism.
\cite{holetal2015} asserted that rocky debris would be 
largely unaffected by even a large magnetic field, 
although \cite{metetal2012} claimed that a sufficiently
strong field could affect solid body accretion by inductive coupling.

Alternatively, impactors might sublimate and produce
charged gas.  In order for this gas to be accreted,
it must orbit faster than the magnetosphere at the Alfv\'{e}n
radius \cite{corsha2008}.  Further, this gas would
be accreted along magnetic field lines.
The Alfv\'{e}n radius, $r_{\rm Alf}$, for a typical accreting WD is 
(Eq. 63 of \cite{metetal2012})

\begin{eqnarray}
r_{\rm Alf} &\approx& 
\left( 
\frac
{3B_{\star}^2R_{\star}^6}
{2 \dot{M}_{\star} \sqrt{G M_{\star}} }
  \right)^{\frac{2}{7}} 
\\
&\approx&
1.2 R_{\odot} 
\left( \frac{B_{\star}}{{\rm kG}} \right)^{\frac{4}{7}}
\left( \frac{\dot{M}_{\star}}{10^8 {\rm g \, s}^{-1}} \right)^{-\frac{2}{7}}
\left( \frac{R_{\star}}{0.01R_{\odot}} \right)^{\frac{12}{7}}
\left( \frac{M_{\star}}{0.6 M_{\odot}} \right)^{-\frac{1}{7}}
,
\end{eqnarray}

\noindent{}where $B_{\star}$ is the star's magnetic field.

{\bf PLANETARY ATMOSPHERES}
\ \ \ \
A planet's or large moon's magnetic field could act as a form of protection from
erosion, which might otherwise proceed rapidly.
\cite{beasok2012} posed that magnetic fields can suppress evaporation
from an SB's atmosphere due to post-MS stellar radiation.  They
claimed that this suppression will occur when the ram pressure of the
outflowing gas is less than the magnetic pressure. This relation
leads to the following condition for the evaporation to be suppressed
(their Eq. 4)

\begin{eqnarray}
B_{\rm SB}^{({\rm crit})} &\gtrsim& 
\frac{2\sqrt{\dot{M}_{\rm SB} v_{\rm wind}}}{R_{\rm SB}}
\\
&\approx&
5.5 \times 10^{-3} {\rm kG} \left( \frac{\dot{M}_{\rm SB}}{10^{10} \ {\rm kg \ s}^{-1} } \right)^{\frac{1}{2}}
            \left( \frac{R_{\rm SB}}{0.8 R_{\oplus}} \right)^{-1}
            \left( \frac{v_{\rm wind}}{2 \times 10^3 \ {\rm m \  s}^{-1}} \right)^{-1}
,
\end{eqnarray}

\noindent{}where $B_{\rm SB}$ is the SB's magnetic field.

{\bf OTHER EFFECTS}
\ \ \ \
Three other important points about magnetism are:
(i) Currents in the atmospheres of SB may be driven by electric
fields which are induced by stellar winds crossing the magnetic
fields \cite{spimad2012}, (ii) For CE evolution, as a SB companion 
spirals in, it could spin up the envelope.  Consequently, more 
differential rotation could enhance magnetic fields, which could 
represent a driver for the mass loss \cite{norbla2006},
(iii) \cite{norspi2013} speculated that if the stellar wind is 
magnetically coupled to the source, then this coupling
might affect both the mass loss rate and the star's moment of inertia.
The latter could affect tidal interactions with SBs which are brown dwarfs or
massive planets (see Sec. \ref{secTides}).

\subsection{External influences} \label{secOtherExternal}

Planetary systems are not isolated entities.  They are embedded within
a galaxy and are subject to flybys and global tidal forces.  These
``external'' forces may play a crucial role in exosystem evolution.
The larger the value of $r$, the greater chance the chance that
an SB's evolution will be affected by these external forces.
Post-MS mass loss will increase the value of $r$ by at least a few,
and possibly by orders of magnitude.  Therefore, an MS system
not influenced by tides or flybys might become so after the parent
star has become a WD or NS.  

{\bf GALACTIC TIDES}
\ \ \ \ 
\cite{verevawyatou2014} quantified external
influences in the midst of post-MS mass loss.  They found that
AGB mass loss occurs on a short-enough timescale to be decoupled
from the influence of Galactic tides (their Fig. 1) but not necessarily
stellar flybys.  This result holds everywhere in the Milky Way galaxy.
During the WD phase, however, Galactic tides can play
a significant role \cite{bonver2015}. Further, Galactic tides has its 
own ``adiabatic'' regime, where adiabatic refers to conservation of semimajor axis,
and the adiabatic/non-adiabatic regime is much further away (at $\sim 10^5$ au).
Consequently, any SB near this boundary is already in the mass loss-based non-adiabatic
regime.  Their Fig. 5 presents a flowchart of evolutionary possibilities,
and their Table 1 provides equations for the adiabatic and non-adiabatic SB evolution
due to Galactic tides. 

{\bf HILL ELLIPSOID}
\ \ \ \ 
Galactic tides carve a region around the primary within which SBs orbit the primary
and not the centre of the Galaxy.  This region, which is not a sphere, is known as 
the Hill ellipsoid, with dimensions provided by Eqs. 21-22 of \cite{verevawyatou2014}.
The Hill ellipsoid lies close to the adiabatic/non-adiabatic Galactic tidal boundary,
and can help determine, for example, which Oort cloud comets escape before, during, and
after GB evolution \cite{vershagan2014}.  For $M_{\star}^{({\rm MS})} = 2M_{\odot}$, 
about three-quarters of the volume of the entire MS Hill ellipsoid includes the semimajor axis range
for which post-MS mass loss will cause a SB to escape.

{\bf STELLAR FLYBYS}
\ \ \ \ 
Although stellar flybys are unpredictable, the stellar space density in the Milky
Way is high enough to expect a close encounter at a few hundred au over a typical
Solar-like MS lifetime.  The closer to the centre of Galaxy the star-SB system resides,
the more susceptible it is to flybys.  Fig. 3 of \cite{verevawyatou2014}
quantifies the expected close encounter distance, and many of the 
above notions in this section; note that most known exoplanets reside at 
approximately 8 kpc from the Galactic centre.

{\bf WIDE STELLAR BINARIES}
\ \ \ \ 
Galactic tides trigger close encounters between
wide binary systems for the first time many Gyr into the WD phase
of the initially more massive star.
This encounter in turn triggers instabilities in extant, and perhaps
formerly quiescent, planetary systems around one or both stars \cite{bonver2015}.
This mechanism provides an avenue to pollute WDs which are many Gyr
old.  Importantly, the mass loss experienced in the post-MS evolution
of the more massive star is adiabatic with respect to its daughter
planetary system, but nonadiabatic with respect to the stellar binary companion.
Consequently, the final super-adiabatic binary separation 
will be affected by Galactic tides and create eccentricity changes.

\subsection{Climate and habitability} \label{secOtherClimate}

Can habitable climates exist on SBs orbiting compact objects?
At least a few studies \cite{agol2011,fosetal2012,barhel2013,norspi2013} 
have explored this possibility in some detail.  In just about
every case, the SB is assumed to have tidally circularized
around the compact object.
\cite{agol2011} claimed that planets in the WD continuously 
habitable zones with orbital periods
of $\approx 4-32$ hours will be circularized and tidally locked
in $\approx 10-1000$ yr.

\cite{fosetal2012} used a modified radiative transfer model to 
demonstrate that photosynthetic processes can be
self-sustaining on planets in a continuously habitable zone around a 
nonmagnetic WD.  WDs provide stable luminosity sources with ultraviolet 
radiation doses which may be less damaging than those of MS stars. 
Although the WD luminosity continuously decreases, \cite{fosetal2012} claimed
that while a WD cools from 6000K to 4000K, a planet which is about 0.01
au distant can remain in the habitable zone for about 8 Gyr, sufficiently
long to allow complex life to develop.

Alternatively, \cite{agol2011} performed computations of one-dimensional 
radiative-convective
atmosphere models with water loss at the inner edge and a maximum carbon dioxide 
greenhouse at the outer edge.  \cite{barhel2013} adopted multiple tidal
models, and found that a WD's strong extreme ultraviolet emission is a
significant barrier to sustain life.  
\cite{norspi2013} estimated that if a planet 
is perturbed close enough to a WD to be tidally circularised 
(as in \cite{vergae2015}), then the
amount of orbital energy dissipated as heat, $I$, would be (their Eq. 6)

\begin{equation}
I \sim -4 \times 10^{35} {\rm J} 
\left( \frac{M_{\star}}{0.6M_{\odot}} \right)
\left( \frac{M_{\rm SB}}{M_{\oplus}} \right)
\left( \frac{a}{R_{\odot}} \right)^{-1}
,
\end{equation}

\noindent{}enough to scupper the prospects for habitability.  \cite{agol2011} 
also considered the prospects of other SBs delivering volatiles 
to a planet after it entered the WD
habitable zone (a realistic possibility: see e.g. 
\cite{alcetal1986,vershagan2014,stoetal2015}), and \cite{norspi2013} 
estimated that the 
impacts would impart orders of magnitude more of specific energy to those
planets than the impacts which wiped out the dinosaurs.

Finally, \cite{perets2011} mentioned that second-generation formation can 
provide new developmental pathways for habitable planets unavailable to
first-generation planets, 
and \cite{spimad2012} speculated 
that a giant planet's atmosphere could be carbon-enriched 
during the GB phases by accreting a stellar wind from a star
that has undergone third dredge-up.  Similarly, the chemical imprint
of accretion in a terrestrial planet's atmosphere during the GB
phase might affect
habitability during the WD phase.

\section{The fate of the Solar system} \label{secFate}

The post-MS fate of the Solar system is of intrinsic human interest; \cite{sacetal1993}
``could not resist exploring the Sun's future''.  However, the conclusions of the 
studies that investigated this future do not agree, highlighting the need 
for additional work in this
area.  The Solar system consists of four terrestrial planets, four giant planets,
many asteroids, comets and moons, and many more smaller SBs.  The gravitational
influence of the Sun extends to $1\times 10^{5}-3 \times 10^{5}$ au (Eqs. 21-22 of
\cite{verevawyatou2014}), with objects
likely existing out to that edge through the Oort cloud.  In short, 
the fate of the Solar system is complex.

Tackling this problem requires all relevant forces to be brought to bear
(see Fig. \ref{FigForces}).  A planet will simultaneously expand its orbit,
accrete stellar ejecta, evaporate from stellar radiation, and be tidally
influenced from Solar tides.  Each planet will experience these effects
to different extents, but are close enough to the Sun ($a \lesssim 30$ au) such
that external influences are negligible, and are large enough
($R_{\rm SB} > 10^3$ km) to not be bothered by stellar wind drag, the Yarkovsky
or YORP effects.

{\bf REMAINDER OF MS}
\ \ \ \
Barring a close stellar flyby, the current gross
architecture of the Solar system should be maintained throughout
the remainder of the MS.  
Although the outer four planets will remain stable until the end of the
MS, there is a one to few percent chance that the inner four planets will
become collisional and/or unhinged \cite{lasgas2009,zeebe2015}.  The
asteroid and Kuiper belt will likely continue to collisionally deplete.

{\bf MERCURY AND VENUS}
\ \ \ \
All studies agree that Mercury will be engulfed.  The Sun's surface will expand
to such a great extent (to about 1 au) that even Mercury's expanded orbit 
(at about 0.8 au) would lie within the Sun's GB surface. Venus will almost certainly
suffer a similar fate, despite the findings of 
\cite{sacetal1993}, because of Sun-planet tides \cite{rybden2001,schsmi2008}.  Because
\cite{sacetal1993} did not incorporate tidal effects in their analysis, Venus' expanded
orbit (at about 1.5 au) does escape the Sun's reach in that (physically unrealistic) 
scenario. 

{\bf EARTH}
\ \ \ \
The Earth's fate is sensitive to both the Solar model
and tidal model adopted.  \cite{schsmi2008}, using the updated Reimers mass loss prescription
from Eq. \ref{eqtruere} and the tidal formalism of \cite{zahn1977,zahn1989}, 
found that the Earth will be engulfed unless all tweakable parameters within the model
were at one edge of their uncertainty ranges.  They predict that the Mercury, Venus and
Earth will all be engulfed about 4.3, 1.5 and 0.5 Myr before the Sun's RGB phase ends, 
which will occur about 7.59 Gyr from now.  If Earth does survive, it won't be habitable.
The post-MS state of its atmosphere and surface has not though been modelled in detail.  
\cite{schsmi2008} does speculate that the Solar system's habitable zone during the RGB tip
would extend from about 49 to 71 au, beyond the current location of Neptune.

{\bf OUTER 5 PLANETS AND ASTEROID BELT}
\ \ \ \
Mars and the giant planets will survive, but the asteroid belt will likely not.  
Dynamically, the giant planets will not undergo instability \cite{dunlis1998}
and experience a quiescent existence around the Solar WD. Mars should also 
survive dynamically because the semimajor axis ratio of Jupiter 
and Mars is greater than 3, and that
ratio will be maintained during post-MS evolution due to adiabatic mass loss.
Physically, how these five planets will fare has only started to be explored.  
\cite{villiv2007}
warned that planets in Jupiter's location might undergo significant evaporation.
\cite{spimad2012} specifically considered the fate of Jupiter's atmosphere: how
it is chemically and thermally altered by Solar wind accretion and the Sun's GB radiation.
SBs in the asteroid belt will either be 
evaporated or spun up to fission \cite{verjacgae2014} given their close
proximity to the Sun.

{\bf MOONS}
\ \ \ \
The moons of the Solar system have diverse physical characteristics
and orbit their parent planets out to about half of one Hill radius.  These
orbits will become more stable after the Sun has become a WD because of
the effect of mass loss on Hill radius (see Eq. \ref{HillMoon}) and because
the remaining planets are not expected to undergo potentially-disruptive
scattering \cite{payetal2016}.  The consequences of GB radiation on the ice
crusts and oceans of moons like Europa and Enceladus remain unexplored.

{\bf KUIPER BELT AND SCATTERED DISC}
\ \ \ \
How the Kuiper belt and scattered disc will be altered by post-MS Solar evolution remains 
unclear.  The re-distribution of Kuiper belt objects due to the Yarkovsky effect, even if
slight, may affect the edges of the belt and the resonant characteristics of many of 
these objects with Neptune.  Consequently, the extent of the depletion 
\cite{bonetal2011,debetal2012,frehan2014} might change from current levels.
An expanded scattered disc would be more susceptible to dynamical reshuffling from stellar
flybys, particularly over several Gyr during the Solar WD phase (Fig. 3 of 
\cite{verevawyatou2014}).  The critical MS distance at which scattered disc objects
might escape during post-MS phases is between $10^3$ and $10^4$ au \cite{verwya2012}.

{\bf OORT CLOUD}
\ \ \ \
The Oort cloud will be dynamically excited by the combination of stellar mass loss,
orbital expansion, Galactic tides and stellar flybys \cite{verevawyatou2014}.  
The latter two will change from their present day prescriptions because the
Milky Way and the Andromeda galaxies will collide before the Sun turns off of 
the MS \cite{coxloe2008}.  \cite{alcetal1986} and \cite{vershagan2014} disagree 
on the extent to which Oort clouds 
would be depleted during the GB phase, but interestingly are in rough agreement about
the number of these comets reaching the inner Solar system (about one in every
$10^4$ yr).  The distribution of ejected 
comets would not necessarily scale with semimajor axis \cite{verwya2012} and the 
subsequent interactions with surviving planets has not been considered in detail.

\section{Numerical codes}  \label{secCodes}

Post-MS planetary science has benefited with improvement
in numerical tools since the discovery of the PSR 1257+12 planets.  The two
most widely-used classes of tools for this science are stellar evolution codes
and $N$-body codes.  In Table \ref{TableCodes}, I list codes that have been
used in references from this paper.  Another class of codes not listed 
are model atmosphere codes, like the IRAP and La Plata codes used 
by \cite{deaetal2013}.

\begin{table}[!h]
\caption{Some numerical codes used by cited investigations.  
``Type'' refers to either a stellar evolution code,
or an $N$-body dynamics code, or both.}
\label{TableCodes}
\begin{tabular}{ccccc}
\hline
Name & Type & Ref & Used by \\
\hline
\textsc{AMUSE}          & Both      & \cite{peletal2013}  & \cite{portegieszwart2013}  \\ 
\textsc{BSE}            & Stellar   & \cite{huretal2002}  & \cite{vertou2012,musetal2013}  \\
\textsc{MESA}           & Stellar   & \cite{paxetal2011,paxetal2015} & \cite{kunetal2011,spimad2012,pasetal2012}  \\
\textsc{SSE}            & Stellar   & \cite{huretal2000}  & \cite{bonwya2010,kraper2012,veretal2013,musetal2014,vergae2015}  \\
\textsc{STAREVOL}       & Stellar   & \cite{siess2006}    & \cite{viletal2014}  \\
\textsc{STARS}          & Stellar   & \cite{staetal2004}  & \cite{schsmi2008,wicetal2010}  \\
\textsc{Hermite}        & $N$-body  & \cite{hutetal1995}  & \cite{kraper2012}   \\
\textsc{Mercury} Bulirsch-Stoer  & $N$-body & \cite{chambers1999} & \cite{veretal2011,debetal2012,rayarm2013,veretal2013,frehan2014,musetal2014,vershagan2014,vergae2015}  \\
\textsc{Mercury} Hybrid & $N$-body & \cite{chambers1999} & \cite{horetal2011,beuetal2012,horetal2012,horetal2013,frehan2014}  \\
\textsc{Mercury} Radau  & $N$-body & \cite{chambers1999} & \cite{bonetal2011,bonver2015}  \\
\textsc{PKDGRAV}        & $N$-body & \cite{ricetal2000}  & \cite{debetal2012,verleibongae2014}   \\
\hline
\end{tabular}
\end{table}

Full-lifetime integrations over the entire MS, GB and much of the WD or NS phases
remain computationally challenging for several reasons: (i) the simulations often 
require a method to combine an $N$-body code with a stellar evolution code,
(ii) this combination should ensure that errors converge as timesteps decrease
(consider Fig. 1 of \cite{veretal2013}),
(iii) the additional relevant forces from Fig. \ref{FigForces} need to be included,
and (iv) the bodies (stellar and SBs) must be on sufficiently wide orbits,
otherwise the simulations will take months or years of real time to finish.
Fortunately the transition from MS to WD expands the orbits of surviving SBs 
and decreases the parent mass, allowing for quicker integrations by an 
order of magnitude (see Fig. 8 of \cite{veretal2013}).  Consequently, the shorter
the MS lifetime, and the larger the value of $M_{\star}^{({\rm MS})}$, the more
quickly the simulations will finish.  Although energy is not a conserved 
quantity, angular momentum typically is, and represents an important benchmark 
for accuracy.  Strong close encounters
between SBs increase angular momentum errors (footnote \#7 of \cite{vergae2015}).

Two recent $N$-body codes which may prove useful for future studies are
the \textsc{GENGA} code \cite{grista2014} and \textsc{IAS15} code \cite{reispi2015}.
\textsc{GENGA} utilizes graphical processing units (GPUs) to speed up computations.
They claim their code runs up 
to 30 times faster than \textsc{Mercury} and 8 times faster than \textsc{PKDGRAV2}, and 
can handle up to 2048 massive bodies or $10^6$ test particles.
IAS15 is a 15th-order $N$-body integrator that can handle close encounters,
high-eccentricity orbits and non-conservative forces, with systematic errors 
below machine precision.  Section 1 of \cite{reispi2015} usefully explains the 
concept of symplecticity, and how symplectic integrators encounter difficulties 
when modeling binary systems or incorporating non-conservative forces.

\section{Future directions}  \label{secIdeas}

The fast-growing field of post-MS planetary science 
demands that theorists and observers work in concert to achieve shared goals,
but should not obscure the importance of purely theoretical pursuits for 
such a dynamically rich and unexplored topic.  In this section, I present 
ideas for future directions for both groups.

\subsection{Pressing observations}  \label{secIdeasPres}

\subsubsection{Continued monitoring of known objects}  \label{secIdeasPresMon}

\ \ \ \ {\bf WD 1145+017}
\ \ \ \
The one to several disintegrating SBs orbiting the polluted and disc-bearing WD 1145+017 
\cite{vanetal2015}
requires vigilant monitoring (i) first to constrain the number of SBs and their 
masses, (ii) to detect the onset of future breakup events from the large SBs, as 
they reside within the disruption radius, (iii) to determine how the surrounding disc 
varies due to direct injection of material, and (iv) to detect variability in the 
accretion rates onto the WD.  A chemical analogy may also be made with the
planets breaking up around MS stars.  \cite{bocetal2015} provided recent and 
strong evidence for a low-mass rocky planet disrupting around the MS star 
KIC 12557548 B, following up on the suggestion from \cite{rapetal2012}.
Two other potentially disintegrating planets (KOI-2700b; \cite{rapetal2014},   
and K2-22b; \cite{sanetal2015}) raise the possibility even more
that analysis of additional light curves from these systems
could help measure grain composition of the broken-up debris.

{\bf WD J0959-0200}
\ \ \ \   
Continued monitoring of the variable disc around WD J0959-0200 \cite{xujur2014} is important.  Given that the flux dropped almost 35\% within a 300-day period, there is a possibility that (i) instability in the disc will create an observable flare, as predicted by \cite{beasok2013}, and (ii) photospheric abundances can be measured during the flux change, which will allow us to observe variability in accretion rates.  Detecting variability in accretion onto DA WDs (like WD J0959-0200) may crucially
constrain sinking timescales and disc lifetimes. Further, the WD J0959-0200 contains a gaseous component, and monitoring the gas and dust simultaneously might help us understand the interplay between dust, gas and variability.

{\bf SDSS J1617+1620}
\ \ \ \
The variability of the gaseous component of the WD disc orbiting SDSS J1617+1620 since the year 2006 \cite{wiletal2014} generates the exciting possibilities that (i) potential future flare-ups due to repeated impacts of leftover debris from a tidal disruption event could be observed, and (ii) we can see the inner rim of the disc moving outward through changes in the width of Ca II lines.

{\bf SDSS J1228+1040}
\ \ \ \
Following the evolution of the axisymmetric and eccentric WD gaseous and dusty disc shown in Fig. \ref{FigDisc} is particularly important to constrain WD disc evolution theory.  \cite{manetal2015} provided a clear prediction: In December 2016, the WD debris disc in SDSS J1228+1040 should reach halfway through its precession cycle.

{\bf NN SER}
\ \ \ \
Additional observations of NN Ser, particularly over the next couple years, will significantly help pin down the constraints for the number and properties of planets in this post-CE binary.  The eclipse timing trajectory features a crest in the years 2019-2020 (Fig. 9 of \cite{maretal2014}) which will help confirm or refute the nature of the putative planets.

\subsubsection{Monitoring new objects}  \label{secIdeasPresNew}

\ \ \ \ {\bf GB PLANETS}
\ \ \ \
One of the most pressing theoretical concerns is the uncertainty about tidal dissipation
mechanisms and their relation to planetary rheology.  Planet-bearing GB systems can help 
us constrain some of the relevant physics.  The tidally-based RGB study of \cite{viletal2014} 
illustrated that the region void of planets is too wide to have been 
depleted by tidal effects alone: providing us with a mystery.  We need observations
of greater numbers of GB planets in order to obtain a more detailed parameter
distribution of these objects, and in particular to identify a cluster at
the tidal disruption edge \cite{norspi2013}.  Separately, a greater sample of 
GB planets will help us resolve the difference in the planet-metallicity correlation 
in MS systems vs. GB systems \cite{maletal2013,refetal2015}. 

{\bf MASSIVE POLLUTED WDs}
\ \ \ \
How massive can host stars of exoplanets be?  This fundamental question
informs planetary formation and evolution, and current technology dictates
that the answer is best obtained through observing high-mass WDs for signatures
of pollution.  For example, the polluted WD SDSS J1228+1040, which also happens
to be the first WD discovered with a gaseous disc component, had a likely
progenitor MS mass of $M_{\star}^{({\rm MS})} \sim 4M_{\odot}$.  No MS planet
has yet been discovered around such a massive star.

{\bf MAGNETIC POLLUTED WDs}
\ \ \ \
Planetary remnants, through split metal lines in the spectrum, help us understand 
magnetism in WDs, particularly in the oldest and coolest WDs.  Therefore, we need to 
identify additional DZH stars \cite{holetal2015}, and further build up enough of 
a sample to distinguish the incidence of magnetic hydrogen-rich and magnetic 
helium-rich WDs.  Further, a planetary core heating up a DH WD might have optically 
detectable H$\alpha$ emission \cite{lietal1998}.

{\bf BINARY SYSTEMS WITH WDs}
\ \ \ \
\cite{kraper2012} explicitly suggested from their theoretical results that future 
surveys might find planets orbiting WDs in WD-MS binaries within the progenitor 
AGB radius.  Further, \cite{perets2011} claimed that WD-MS and WD-WD systems 
with separations of tens or hundreds of au should be prime targets.  He
also provided specific examples of candidate second-generation
exoplanetary systems. The likelihood of a second-generation disc formation scenario
from binary stellar winds may be assessed from the frequency of wide WD 
companions to WD debris discs \cite{perken2013}. 

{\bf JAMES WEBB SPACE TELESCOPE (JWST)}
\ \ \ \
\cite{stoetal2015} specifically studied the capacity of JWST to discover new WD 
debris discs, and found the prospects excellent (see their Eq. 35). 
JWST observations of young WD debris discs could 
help distinguish an origin from exo-Oort cloud disruption versus from a 
remnant exo-Kuiper belt.
The former would produce a different and detectable brightness profile
(see Sec. 4.4 and Figs. 7-9 of \cite{stoetal2015}). JWST, along with SPICA, 
should also find older, fainter discs \cite{beretal2014}.  Further,
the European Mid-InfraRed Instrument on-board JWST will 
identify specific minerals in the debris orbiting WDs (currently G 29-38
is the only WD with a high-quality infrared spectrum; see Fig. 4 of 
\cite{reaetal2009}), and this data can be combined 
with the known atomic abundances measured with HST. 
For pulsar planetary systems, \cite{shaetal2013} suggested that tens to hundreds of 
hours of JWST time could detect the putative asteroid disc around B1937+21.  
Identifying the disc is particularly 
important partly because they suggest that second-generation circumpulsar planetesimal 
formation may be common; JWST observations will provide the timing 
precision ``noise floor'', which in turn 
informs us about the stability of these stars.  Finally, \cite{wang2014} claimed that the 
ability of a pulsar to heat up circumstellar material is still unclear, and can 
be resolved with JWST along with WISE and the TMT.

{\bf OTHER TELESCOPES}
\ \ \ \
ALMA, due to its high spatial resolution, can directly resolve exo-Kuiper 
belts orbiting WDs \cite{stoetal2015}.  \cite{distefano2011} touted the 
potential of gravitational lensing to discover WD planetary systems, 
particularly with Pan-STARRS and LSST.  The putative subdwarf 
planets inferred from pulsations may be confirmed or refuted with
the PEPSI at LBT \cite{siletal2014}. Gaia, LSST and PLATO will 
repeatedly (tens to thousands of times) survey $10^5 - 10^6$ WDs,
and by the year 2020, the combined efforts of the ESA missions Gaia 
and Euclid will increase the number of known WD debris discs by
at least an order of magnitude.
PLATO will find planets orbiting WDs.  Gaia, LSST and WFIRST will help 
detect free-floaters, which may help constrain the fraction of ejected
planets due to mass loss \cite{verevawyatou2014}, particularly if high
mass stars are targeted.  Gaia will also provide better constraints on
the wide binary population, potentially allowing for a test of the pollution
mechanism suggested by \cite{bonver2015}.  The SKA will also aid post-MS planetary
science through the radio portion of the electromagnetic spectrum: the 
array will improve the chances of detection of radio emissions from 
WD-terrestrial planet systems \cite{wilwu2005} and help confirm the
existence of circumpulsar discs \cite{corsha2008}.

{\bf OTHER CONSIDERATIONS}
\ \ \ \
The drastic increase then decline in luminosity from a
tidal disruption event could create detectable outburst events.  
\cite{beasok2013} made a comparison with dwarf novae (see their Table 1), 
which also create outbursts.  \cite{beasok2015} proposed that young, 
hot WDs should ionize nebulae formed from an episode of AGB 
reincarnation due to engulfment of planetary hydrogen.
\cite{fosetal2012} suggested that polarimetry can be used
to detect non-transiting WD habitable zone planets because the signal 
would be 2-5 orders of magnitude larger than for habitable planets 
around typical M-dwarf to Sun-like stars.  
Finally, exomoons in MS systems are on the verge of discovery; 
\cite{pasass2014} calculated the time-of-arrival signal that a hypothetical 
exomoon on an inclined orbit around a circumpulsar planet would induce.

\subsection{Theoretical endeavours}

Theoretical progress is lagging behind both the mounting 
observations of post-MS systems and the multitude of theory papers about
MS planetary systems.  Post-MS systems includes rich dynamics that is not
found in MS systems (Fig. \ref{FigForces}), and which needs to be explored in greater detail
in order to understand physics and improve models.

{\bf EFFECTS OF RADIATION}
\ \ \ \
Firstly, no study has yet self-consistently integrated the set of Eqs. 
\ref{eqyark}, \ref{eqyorp}, \ref{eqsubl}, and \ref{eqtemp}, or a
similar set.
Also, the applicability of the Yarkovsky geometry in Eq. \ref{eqyark} 
for highly eccentric objects is limited.  In general, $w$ may be a function
of time, even for spherical SBs.  We also know that asteroid-sized SBs
are generally not spherical, and for those objects Eq. \ref{eqyark} should 
be modified.  Further, Eq. \ref{eqyorp} provides an averaged quantity.  
The high luminosity of GB stars and the duration of the SB orbit may 
require one to model spin changes on orbital rather than just secular timescales.
We also need to characterise the debris fields produced from YORP spin-up
of SBs so that we understand from where WD pollution arises, and can
make a chemical link with first-generation formation.
Eq. \ref{eqsubl} applies only for a single-species homogeneous sphere.
We know that real asteroids are multi-layered, aspherical and contain many
chemical species.  In order to quantify the amount of water remaining in 
these SBs, more sophisticated models are required.  Finally, the 
equations of motion for a moon (which orbits some SB) will be different
than what is presented here, partly because of shadowing.

How do atmospheres change from post-MS stellar radiation?  An important
follow-up to the studies of \cite{villiv2007}, \cite{beasok2011} and
\cite{spimad2012} is to solve the hydrodynamical 
equations for atmospheric escape.  Further, these studies do not 
necessarily agree about the extent of Jupiter's atmosphere that will
escape.  How much will Saturn, Uranus and Neptune be ablated?  We know
that the four giant planets in the Solar system will remain on stable orbits
during the Solar GB phase, but how will they physically evolve?

{\bf EFFECTS OF TIDES}
\ \ \ \
As is also true for MS tidal studies, tidal investigations for GB, WD and NSs
can be improved by sampling tidal dissipation prescriptions for different 
planetary rheologies and stellar structures.  Doing so will better constrain
the timescales for SB destruction and the critical engulfment radius.  
Treating a population of known exoplanets
with similar prescriptions (as in Fig. 4 of \cite{norspi2013}) 
might yield a very different result than a more individualised treatment of each planet
(Fig. 15 and Table 2 of \cite{henhur2014}).  For GB systems, 
Eqs. \ref{zahna}-\ref{zahne} need to be extended to higher SB eccentricities,
and potentially coupled with stellar spin dynamics \cite{norspi2013} at those 
higher eccentricities.  For stars with similar radial extensions during both 
the RGB and AGB phases, different tidal prescriptions may 
apply \cite{musvil2012,viletal2014}
and might need to be treated together self-consistently within a single
simulation.

{\bf EFFECTS OF MASS LOSS}
\ \ \ \
Compared to radiation and tides, our understanding of how SBs evolve due to
stellar mass loss is better.  Nevertheless, the consequences of
accretion onto SBs has largely been ignored, as has the potential for
GB stars to experience kicks during non-instantaneous mass loss.
The effects of ram pressure and entrainment on smaller SBs such as boulders 
and pebbles need better quantification.  How these effects, along with
frictional drag, affect the transition between adiabatic and non-adiabatic
motion is not yet clear.  The transition itself, even in the non-accreting
case, needs further exploration in order to explain e.g. the nonmonotonic
behaviour in Fig. 10 of \cite{verwya2012}.  How resonances are formed,
broken and maintained \cite{voyetal2013} may crucially affect WD pollution 
rates \cite{debetal2012} and need further analysis on a 
commensurability-by-commensurability basis.  Finally, simulations
of the long-term MS future of the Solar system typically do not
include effects from MS Solar mass loss.  Although small, 
mass loss during the MS will cause a shift much greater than the 
instability-inducing 0.38 mm shift suggested by \cite{lasgas2009}.

{\bf DIFFERENT PLANETARY ARCHITECTURES}
\ \ \ \
So far, self-consistent full-lifetime numerical simulations of planetary systems
have been limited in scope in terms of number of bodies, amount of time simulated
and number of simulations.  Room for improvement is significant, but must overcome
computational hurdles.  We need post-MS multi-planet
simulations with an asteroid or Kuiper belt,
a situation that will mirror the fate of our own Solar system.
The planetary mass which maximizes the accretion 
rate onto the eventual WD is still unknown \cite{frehan2014}.  Where is the sweet 
spot for one-planet and other configurations?
We need to determine the orbital and physical evolution of post-MS
exomoons, Trojan asteroids, and exorings, and how they will be subject
to close encounters from other objects.  We need to determine how fragments
from SB-SB collisions evolve and are subject to scattering events.
\cite{musetal2014} estimated through simulations of three planets only that no more
than 1\% of WDs should eventually host close-in giants planets.  We must 
improve that estimate with (i) terrestrial-mass planets, (ii) unequal-mass planets, and
(iii) SBs smaller than planets.  Also, further exploration of binary and triple stellar
planet-hosting systems \cite{shatho2013,micper2014} will provide us with a more
representative, population-based perspective on the evolution of planetary systems in 
the Milky Way.

Investigations detailing the fates of specific known exoplanetary systems are lacking.  
The famous HR 8799 system has a parent star with a progenitor mass that is typical of 
the currently observed WD population (see Fig. \ref{FigSpec}) and represents a potential 
exemplar for polluted WD systems.  The fates of other wide-orbit multi-planet systems 
on the current edge of MS stability (see e.g. Sec 6.3 of \cite{veretal2013})
are not yet clear.

{\bf SECOND-GENERATION FORMATION}
\ \ \ \
\cite{curhan2007} called for more sophisticated extensions of their
models of second-generation circumpulsar disc formation.  In their 
Sec. 7, they list several simplifications and assumptions adopted,
which have not yet been lifted nearly a decade after that paper
has been published.  Many of the recent advancements in first-generation
planet formation theory would also be applicable to second-generation
SB formation.

{\bf WD DEBRIS DISC FORMATION AND EVOLUTION}
\ \ \ \
The details of tidal disruption of an SB entering the WD Roche radius (and forming
a disc) should
be improved, particularly given the fine detail of some observations like those
of \cite{manetal2015} and in Fig. \ref{FigDisc}.  As mentioned by 
\cite{verleibongae2014}, improved $N$-body
rubble pile simulations may include ``soft spheres'', with rolling and twisting 
friction between particles at multiple points of contact \cite{schetal2012}.  
Alternatively, the spheres could have internal strength, or not be in the shapes
of spheres at all.  \cite{movetal2012} instead used polyhedral regular or
or irregular grains to represent the building blocks of SBs.  The SBs can be 
different shapes and spins,
and when appropriate, harbour different equations of state \cite{guietal2011}.
Numerical modeling of the disruption of an incoming SB at an already existing
disc would help characterise this likely scenario.

The evolution of the disc requires coupled modeling of the dust and gas, as in
\cite{metetal2012}.  They provided many useful relations which may be utilized
for further modeling.  Considering the simpler case of gas only, 
with potential self-generation of eccentricity
and precession, is another approach which may help us disentangle different
effects in these complex systems.  Further work on spectral modeling
of these discs \cite{haretal2011,haretal2014} can link their composition
with orbital evolution.

{\bf WD ACCRETION}
\ \ \ \
Improvements in atmospheric modeling will allow us to better distinguish
accreted material from radiatively levitated material in hot WDs
and from dredged-up carbon in cool WDs.  In concert with observations, 
modeling the onset and tail-off of accretion events with different 
functional forms than Eq. \ref{eq3states} and as a function of chemical
elements may help us better constrain progenitor masses and evaluate the
steady-state assumption in DBZ WDs.  Accretion from direct collisions between
an SB and WD needs to be explored in more detail, to distinguish fragmentation,
sublimation and explosion regimes.

{\bf FINALLY}
\ \ \ \
Will the Sun become a polluted WD?  The chances are good,
with multiple surviving planets potentially perturbing Kuiper belt and
asteroid belt fragments onto the Solar WD. However, other polluted WDs
will have harboured different planetary architectures on the MS.  Extrasolar planets, first
unknowingly seen inside of a WD \cite{vanmaanen1917} and then by their 
own right orbiting a pulsar \cite{wolfra1992}, have taught us that
``their diversity cannot be easily foreseen from extrapolations of our 
knowledge of the Solar System'' \cite{wolszczan2012}.

\section*{Acknowledgments}

I solicited input on this manuscript from many researchers, from graduate students to senior professors, and was rewarded with valuable comments and perspectives that improved the final product. William F. McDonough provided initial sage advice, and later the following individuals provided explicit suggestions for text, equations and physics:  Siegfried Eggl, Jay Farihi, Boris T. Gänsicke, Nikolaos Georgakarakos, Mark A. Hollands, Michael Jura, Detlev Koester, Christopher J. Manser, Thomas R. Marsh, Noam Soker, Nicholas C. Stone, Pier-Emmanuel Tremblay, Andrew Vanderburg, David J. Wilson and Siyi Xu.  Boris T. Gänsicke and Jay Farihi kindly provided updated data for, respectively, Fig. \ref{FigAccr} and Fig. \ref{FigConv}.  I also appreciate the referee's comments, which helped improve the narrative.

This research benefited by support from both the European Union through ERC grant number 320964 and the Royal Society through grant number IE140641.

\end{document}